\documentclass[11pt,a4paper]{scrartcl}
\pdfoutput=1

\usepackage{jheppub}
\usepackage{amsmath,amstext,amssymb}
\usepackage{psfrag,graphicx,subfigure}
\usepackage{relsize}
\usepackage[abs]{overpic}
\addtocontents{toc}{\protect\setcounter{tocdepth}{3}}

\def\calt{{\cal T}}
\def\calj{{\cal J}}
\def\ie{{i.\,e.\ }}

\def\m{\mathfrak{m}}

\def\call{{\cal L}}

\def\re{\text{Re}}
\def\im{\text{Im}}
\def\k{\kappa}

\def\tmI{\text{{\tiny{I}}}}
\def\tmJ{\text{{\tiny{J}}}}
\def\tmK{\text{{\tiny{K}}}}
\def\p{\text{{\tiny +}}}
\def\m{\text{{\small -}}}
\def\del{\partial}

\def\ee{{\mathrm e}}
\def\ii{{\mathrm i}}

\newcommand{\dd}{\mathrm{d}}

\title{New Transport Properties of Anisotropic Holographic Superfluids}
\author[a]{Johanna Erdmenger,}
\author[b]{Daniel Fern\'andez}
\author[a]{and Hansj\"org Zeller}
\affiliation[a]{Max-Planck-Institut f\"ur Physik (Werner-Heisenberg-Institut),\\
F\"ohringer Ring 6, 80805 M\"unchen, Germany}
\affiliation[b]{Departament de F\'\i sica Fonamental \&  Institut de
Ci\`encies del Cosmos (ICC), Universitat de Barcelona (UB), Mart\'{\i}  i
Franqu\`es 1, E-08028 Barcelona, Spain}
\emailAdd{jke@mppmu.mpg.de}
\emailAdd{daniel@ffn.ub.edu}
\emailAdd{zeller@mppmu.mpg.de}

\abstract{We complete the analysis of transport phenomena in p-wave superfluids within gauge/gravity duality, using the $SU(2)$ Einstein-Yang-Mills model with backreaction. In particular, we analyze the fluctuation modes of helicity zero in addition to the helicity one and two modes studied earlier. We compute a further transport coefficient, associated to the first normal stress difference, not previously considered in the holographic context. In the unbroken phase this is related to a minimally coupled scalar on the gravity side. Moreover we find transport phenomena related to the thermoelectric and piezoelectric effects, in particular in the direction of the condensate, as well as the flexoelectric effect. These are similar to phenomena observed in condensed matter systems.}

\date{\today}


\preprint{MPP-2012-156, ICCUB-12-473}

\begin{document}

 \maketitle


\section{Introduction}
\label{sec:Introduction}
Gauge/gravity duality has become a valuable tool for gaining insight into the physics of many different strongly coupled theories and, in particular, is being used to successfully describe their hydrodynamical behavior, with the prospect of making contact with systems found in nature. Recently, many new features within hydrodynamics have been discovered using gauge/gravity duality: For instance, the importance of anomalies for relativistic hydrodynamics as applied to quark-gluon plasma first appeared in the context of gauge/gravity duality~\cite{Erdmenger:2008rm,Banerjee:2008th}. Subsequently, in~\cite{Loganayagam:2012zg,Loganayagam:2012pz,Son:2012zy,Manes:2012hf,Son:2009tf} and~\cite{Jensen:2012kj,Bhattacharyya:2012xi} it has been realized by thermal field theory computations and from general hydrodynamics arguments that anomalies induce modifications in the constitutive relations of relativistic hydrodynamics. Moreover, in~\cite{Landsteiner:2012kd,Kalaydzhyan:2012ut,Gahramanov:2012wz} chiral anomalies have been shown to give rise to non-dissipative transport coefficients. Anisotropy has been included by considering the backreacted holographic p-wave superfluid (see below) and by means of a position-dependent theta-term, leading to several interesting effects~\cite{Gynther:2012mw,Mateos:2011tv,Chernicoff:2012bu}. A common feature of these systems is that the breaking of symmetries brings about a richer structure to the theory, so that new phenomena are unveiled.

A very suitable system to study anisotropic hydrodynamics is the holographic backreacted p-wave superfluid, in which the rotational symmetry is broken by a vector condensate which may be interpreted as a vector meson. This system has been studied in~\cite{Ammon:2009xh,Natsuume:2010ky,Erdmenger:2010xm,Erdmenger:2011tj,Basu:2011tt,Basu:2009vv}. It involves a finite $SU(2)$ charge density or isospin density. In the present article we present the study of the remaining hydrodynamic modes that were not accounted for in \cite{Erdmenger:2011tj} and describe the corresponding new transport properties.

In~\cite{Erdmenger:2010xm,Erdmenger:2011tj}, the helicity two and one fluctuations have been analyzed. It has been found that the helicity one modes lead to contributions to the viscosity tensor whose ratio with the entropy density is non-universal at leading order in the 't Hooft coupling and $N$. These contributions are temperature dependent and satisfy the viscosity bound, $\eta/s\geq1/4\pi$. This is in contrast to the $\theta$-term model of~\cite{Mateos:2011tv} where the usual viscosity bound~\cite{Kovtun:2004de,Buchel:2003tz,Iqbal:2008by} is violated~\cite{Rebhan:2011vd}. This happens already for Einstein gravity, violations of the bound by Gauss-Bonnet terms have been studied in~\cite{Buchel:2008vz}.

The Einstein-Yang-Mills model may also be used as a starting point to derive similar universal relations, such as the holographic realization of Homes' law~\cite{Erdmenger:2012ik} of condensed matter physics. Furthermore when considering finite $SU(2)$ magnetic fields, the system admits more than one possible solution (or state), but similarly to the holographic superfluid at finite $SU(2)$ density, only one is physically realized, determined by the lowest free energy. A magnetic field generates an Abrikosov lattice~\cite{Bu:2012mq} in a superconductor, which becomes the preferred state if the magnetic field is sufficiently large.

The Einstein-Yang-Mills system we consider in this publication is motivated by the D3/D7 setup~\cite{Karch:2002sh}, which allows for temperature and matter in the fundamental representation to be added to the system. Holographically, the Hawking temperature of a black hole geometry coincides with the temperature $T$ of the dual thermal field theory. A chemical potential $\mu$ can also be introduced by placing a non-vanishing boundary condition upon the bulk gauge field. Given these ingredients, it is possible to do thermodynamics, since each solution labeled with $T/\mu$ describes a different thermal state of the dual field theory.

In this paper we consider a superfluid generated by a finite $SU(2)$ density (for an extensive study of this background see~\cite{Ammon:2009xh,Erdmenger:2011tj}). In this case, the temperature determines the preferred state, i.e. at some critical value $T_c$ the system undergoes a phase transition between the normal ($T>T_c$) and superfluid ($T<T_c$) states. We will be interested in the superfluid phase and study its transport properties. To do so, we consider fluctuations in a backreacted holographic p-wave superfluid theory defined in an AdS$_5$ geometry with an $SU(2)$ Yang-Mills gauge field. The boundary condition that fixes the chemical potential,
\begin{equation}
\lim_{r \to \infty} A_t=\mu\; ,
\end{equation}
breaks explicitly this $SU(2)$ symmetry, leaving a $U(1)_3$ gauge symmetry. Depending on the values of $\mu$, the system may present a superfluid or a normal phase, with the order of the phase transition being controlled by a parameter $\alpha$, which on the gravity side measures the effect of the gauge fields onto the geometry. The superfluid state is thermodynamically preferred at low temperatures compared to the chemical potential, and the transition to this phase is characterised by the formation of a vector condensate $\langle \calj^x_1 \rangle$, as opposed to the case of an s-wave superfluid, in which a scalar field condenses. The vector condensate designates a particular direction both in momentum and flavor space, and as a consequence the spatial rotational $SO(3)$ symmetry and the $U(1)_3$ symmetry are spontaneously broken. Schematically, this process can be represented as
\begin{equation}
\begin{split}
SU(2) \xrightarrow[\text{{\tiny Expl.B}}]{} U(1)_3 &\xrightarrow[\text{{\tiny SSB}}]{} \mathbb{Z}_2 \; ,\\
SO(3) &\xrightarrow[\text{{\tiny SSB}}]{} SO(2) \; .
\end{split}
\end{equation}

This is an example of spontaneous breaking of continuous symmetries in gauge/gravity duality (first achieved in~\cite{Gubser:2008px}) to construct holographic superfluids or superconductors. This technique was initially developed by breaking Abelian symmetries~\cite{Hartnoll:2008vx,Hartnoll:2008kx} and later adapted to p-wave superconductors/superfluids~\cite{Gubser:2008wv} as in the case at hand, giving rise to the first string theory embeddings of these constructions~\cite{Ammon:2008fc,Basu:2008bh,Ammon:2009fe}.

We present an analysis of the perturbations of the spacetime metric, $h_{\mu\nu}$, and of the Yang-Mills field, $a^a_\mu$, about the Einstein-Yang-Mills model in $AdS_5$. Due to the breaking of the spatial rotational symmetry, these fluctuations can be grouped according to their transformation behavior under the remaining $SO(2)$ rotational symmetry around the $x$-axis. In this paper, we present the fluctuations which transform as scalars under this group. Vector and tensor fluctuations have been studied in~\cite{Erdmenger:2010xm,Erdmenger:2011tj}. To make the equations tractable, we set the spatial momentum $\vec k =0$. This simplification leads to an additional $\mathbb{Z}_2$ symmetry under which the scalar fluctuations can be characterized further. We end up with two distinct blocks, the first of which contains, among others, the gauge field fluctuation $a^3_x$ and the metric fluctuation $h_{tx}$, and the second one the diagonal metric fluctuations, $a^1_x$ and $a^2_x$.

From the field theory point of view, the corresponding correlation functions are related to the thermoelectric effect which correlates charge and heat transport in the direction of the condensate, since $a^3_x$ can be identified with an electric field in the $x$ direction and $h_{tx}$ with a temperature gradient in the $x$ direction. This effect was studied for holographic s-wave superfluids~\cite{Hartnoll:2009sz,Hartnoll:2008kx} and for p-wave superfluids in the transverse directions~\cite{Erdmenger:2011tj}, but to our knowledge, this is the first time this effect has been calculated with backreaction and in the direction of the condensate.

The second block contains, among others, the diagonal metric fluctuations $h_{xx}-h_{yy}$ and the gauge field fluctuations $a^1_x$ and $a^2_x$. A field theoretic description of the corresponding Green's functions is not fully addressed in this paper and is left for future work. However, we know that some of the modes in this block are related to the transport coefficients in the viscosity tensor $\eta^{ijkl}$. In general, the viscosities of a system are encoded in a rank four tensor $\eta^{ijkl}$ which in the most general case has 21 independent components. Due to the symmetries of the system at hand, we are left with five independent components of the tensor $\eta^{ijkl}$~\cite{Landau:1959te,Gennes:1974lc}, two of which are shear viscosities, $\eta_{xy}$ and $\eta_{yz}$, that were addressed in \cite{Erdmenger:2011tj}. Two of the remaining components are bulk viscosities and can be set to zero using the tracelessness condition for the conformal energy-momentum tensor, leaving one free transport coefficient, denoted by $\lambda$. While $\eta$ and $\zeta$ measure the response of the system to deformations due to shear or normal stress, $\lambda$ is related to the normal stress difference that is induced by an anisotropic strain. Our holographic computation shows that in the zero frequency limit, the ratio of $\lambda$ to the entropy density is finite. Moreover, in the normal phase it turns into the shear viscosity of the isotropic fluid, which we simply denote by $\eta$. Therefore it acquires a fixed value given by the well-known
\begin{equation}
\frac{\eta}{s} = \frac{1}{4\pi}\,.
\end{equation}
Note that we normalized $\lambda$ in a way that at the phase transition it matches $\eta$. In the broken phase we see a temperature dependence and the resulting curve does not fall below $1/(4\pi)$ for any backreaction parameter $\alpha$ and for any temperature.

Since we have completed our analysis of all fluctuation modes in the p-wave system, let us now summarize them, as well as the transport phenomena they correspond to:
\begin{enumerate}
\item $h_{yz}$ (helicity two) is related to the shear viscosity $\eta_{yz}$ which for all values of $T$ takes the universal value $\eta/s=1/4\pi$ (see~\cite{Erdmenger:2011tj}),
\item $h_{x\perp}$ (helicity one) is related to the shear viscosity $\eta_{x\perp}$ which shows a temperature dependence in the broken phase (see~\cite{Erdmenger:2011tj}),
\item The coupling between $a^\pm_\perp=a^1_\perp \pm i a^2_\perp$ (helicity one) and $h_{x\perp}$ leads to an effect which is similar to the flexoelectric effect known from crystals (see~\cite{Erdmenger:2011tj}),
\item $a^3_\perp$ is related to the ``electrical'' conductivity $\sigma^{\perp\perp}$ (helicity one), and its coupling to $h_{t\perp}$ (helicity one) is related to the so called thermoelectric effect transverse to the condensate (see~\cite{Erdmenger:2011tj}),
\item $\Phi_4 \sim a^3_x$ (helicity zero) is related to the ``electrical'' conductivity $\sigma^{xx}$, and its coupling to $h_{tx}$ (helicity zero) gives the thermoelectric effect in the direction of the condensate (see section \ref{sec:Transport-Properties}),
\item $\Phi_3 \sim h_{xx}-h_{yy}$ (helicity zero) is related to the transport coefficient $\lambda$ found in the viscosity tensor $\eta^{ijkl}$ and its coupling to $\Phi_\pm \sim a^\pm_x$ (helicity zero) shows a behaviour similar to the piezoelectric effect (see section \ref{sec:Transport-Properties}).
\end{enumerate}

We see that the study of fluctuations in a backreacted holographic p-wave superfluid provides a rich structure of different effects which by using the fluctuation-dissipation theorem may be related to well-known transport phenomena in other areas of physics.

The paper is organized as follows: In section \ref{sec:Holographic-Setup-and}, we recapitulate the backreacted holographic p-wave superfluid. In section \ref{sec:Perturbations-about-the}, the scalar fluctuations (helicity zero) and the corresponding Green's functions on the gravity side are presented. The following section \ref{sec:Transport-Properties} contains our results for the transport properties of the superfluid and an approach to interpreting them from a hydrodynamical point of view. In \ref{sec:Conclusion} we present our conclusions. Many of the technical details are collected in the Appendices: In \ref{sec:Holographic-Renormalization}, we discuss the necessary holographic renormalization, in \ref{sec:Constructing-the-Gauge} we specify the physical fields of the system, in \ref{sec:Numerical-Evaluation-of} we review the numerical procedure to deal with the coupled equations of motion, and finally some general remarks on anisotropic fluids are given in \ref{sec:General-Remarks-on}.


\section{Holographic Setup and Equilibrium}
\label{sec:Holographic-Setup-and}
The setup used in this paper was already described in \cite{Ammon:2009xh,Erdmenger:2011tj}. Therefore, here we give a brief review of its most important properties.
We consider $SU(2)$ Einstein-Yang-Mills theory in $(4+1)$-dimensional asymptotically AdS space. The action is
\begin{equation}
\label{eq:action}
S = \int\!\dd^5x\,\sqrt{-g} \, \left [ \frac{1}{2\k_5^2} \left( R -\Lambda\right) - \frac{1}{4\hat g^2} \, F^a_{MN} F^{aMN} \right] + S_{\text{bdy}}\,, 
\end{equation}
where $\k_5$ is the five-dimensional gravitational constant, $\Lambda = - \frac{12}{L^2}$ is the cosmological constant (with $L$ being the $AdS$ radius), and $\hat g$ is the Yang-Mills coupling constant. It is convenient to define
\begin{equation} \label{eq:alpha}
\alpha \equiv \frac{\kappa_5}{\hat g} \, ,
\end{equation}
which measures the strength of the backreaction. The $SU(2)$ field strength $F^a_{MN}$ is defined by
\begin{equation}
F^a_{MN}=\del_M A^a_N -\del_N A^a_M + \epsilon^{abc}A^b_M A^c_N \,,
\end{equation}
where capital Latin letter indices run over $\{t,x,y,z,r\}$, with $r$ being the $AdS$ radial coordinate, and $\epsilon^{abc}$ is the totally antisymmetric tensor with $\epsilon^{123}=+1$. The $A^a_M$ are the components of the matrix-valued Yang-Mills gauge field $A=A^a_M\tau^a dx^M$,  where the $\tau^a$ are the $SU(2)$ generators, related to the Pauli matrices by $\tau^a=\sigma^a/2\ii$. Finally, the $S_{\text{bdy}}$ term includes boundary terms, namely the Gibbons-Hawking boundary term as well as counterterms required for the on-shell action to be finite, that will be discussed below. It does not affect the equations of motion.

The Einstein and Yang-Mills equations derived from the above action are
\begin{align}
\label{eq:einsteinEOM}
R_{M N}+\frac{4}{L^2}g_{M N}&=\k_5^2\left(T_{MN}-\frac{1}{3}{T_{P}}^{P}g_{MN}\right)\,, \\
\label{eq:YangMillsEOM}
\nabla_M F^{aMN}&=-\epsilon^{abc}A^b_M F^{cMN} \,,
\end{align}
where the Yang-Mills stress-energy tensor $T_{MN}$ is
\begin{equation}
\label{eq:energymomentumtensor}
T_{M N}=\frac{1}{\hat{g}^2}\left(F^a_{PM}{F^{aP}}_{N}-\frac{1}{4}g_{MN} F^a_{PQ}F^{aPQ}\right)\,.
\end{equation}

To solve this equations, we use the following ans\"atze for the gauge field and the metric, which can be motivated from symmetry considerations~\cite{Gubser:2008wv,Ammon:2009xh}
\begin{gather}
\label{eq:gaugefieldmetricansatz}
A=\phi(r)\tau^3\dd t+w(r)\tau^1\dd x\,, \\
\dd s^2 = -N(r)\sigma(r)^2\dd t^2 + \frac{1}{N(r)}\dd r^2 +r^2 f(r)^{-4}\dd x^2 + r^2f(r)^2\left(\dd y^2 + \dd z^2\right)\,,
\end{gather}
where $N(r)\equiv-\frac{2m(r)}{r^2}+\frac{r^2}{L^2}$. The $AdS$ boundary is at $r\to\infty$ and for our black hole solutions we denote the position of the horizon as $r_h$.

This ansatz is compatible with the well-known $AdS$ Reissner-Nordstr\"om solution, where $w(r)=0$ for all values of $r$. This solution features $w(r)=0$, so it preserves the $SO(3)$ symmetry and corresponds to the normal phase of the system. There is a second solution with non-vanishing $w(r)$, which can only be computed numerically. The second solution breaks the rotational $SO(3)$ symmetry and describes the condensed superfluid phase. Due to our choice of boundary conditions, this breaking occurs spontaneously. For completeness, we state here the coefficients of the expansion at the horizon (in terms of $(r/r_h-1)^n$),
\begin{equation}
\left\{\phi^h_1, \sigma^h_0, w^h_0, f^h_0\right\} \,,
\end{equation}
being $\phi^h_0=0$ in order for $A$ to be well defined as a one-form~\cite{Kobayashi:2006sb}, and of the expansion at the boundary (in terms of $(r_h/r)^{2n}$),
\begin{equation}
\label{eq:coeffb}
\left\{\mu, \phi^b_1, m^b_0, w^b_1, f^b_2\right\} \,.
\end{equation}
Note that $w^b_0=0$, otherwise the $SO(3)$ would be broken explicitly instead of spontaneously. Besides, we can fix the metric to have asymptotic $AdS$ boundary conditions, so that $\sigma^b_0=f^b_0=1$. The fields can be made dimensionless through $m(r)\to r_h^4 m(r)$, $\phi(r)\to r_h \phi(r)$ and $w(r)\to r_h w(r)$, while $f(r)$ and $\sigma(r)$ are already dimensionless. 

In terms of these coefficients we can express the different field theory quantities, such as temperature and entropy density, given by
\begin{equation}
  \label{eq:temperature}
  T=\frac{\sigma^h_0}{12\pi}\left(12-\alpha^2 \frac{{(\phi^h_1)}^2}{{\sigma^h_0}^2}\right)\,r_h\,, \qquad   s=\frac{2\pi}{\k_5^2}r_h^3\,.
\end{equation}
The field theory expectation values of the dual operators of the different fields are directly related to the expansion coefficients. For the charge density and the condensate we have
\begin{equation}
\label{eq:density}
 \langle \calj^t_3\rangle= -\frac{2\alpha^2}{\k_5^2}r_h^3 \, \phi^b_1\,, \qquad \langle \calj^x_1\rangle= \frac{2\alpha^2}{\k_5^2}r_h^3 \, w^b_1\,,
\end{equation}
and for the energy-momentum tensor~\cite{Balasubramanian:1999re,deHaro:2000xn} they are
\begin{equation}
\label{eq:cftstressenergytensor}
\langle \calt_{tt} \rangle=\frac{3r_h^4}{\k_5^2}\,m^b_0\,, \quad  \langle \calt_{xx} \rangle= \frac{r_h^4}{\k_5^2}\left(m^b_0-8f_2^b\right)\,, \quad   \langle \calt_{yy} \rangle = \langle \calt_{zz} \rangle= \frac{r_h^4}{\k_5^2}\left(m^b_0+4f_2^b\right)\,.
\end{equation}

\begin{figure}[t]
\centering
\includegraphics[width=0.8\textwidth]{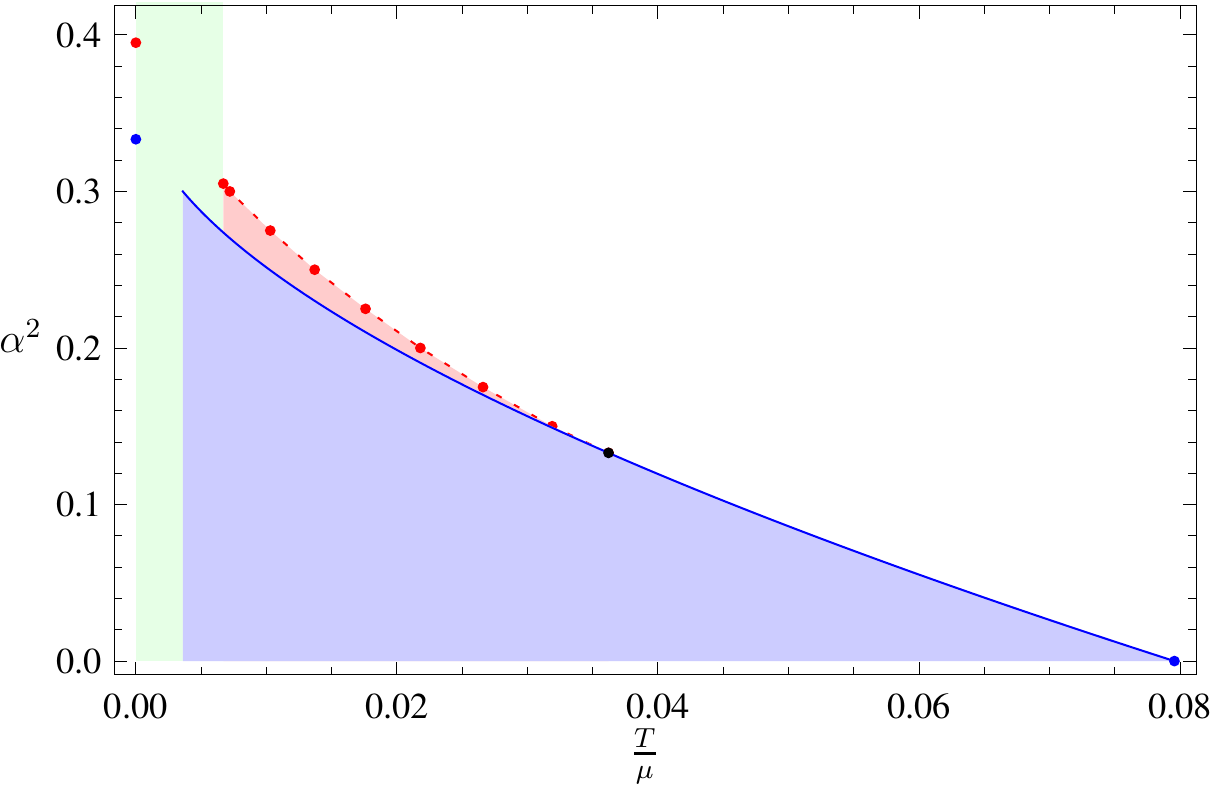}
\caption{This diagram shows the dependence of the order of the phase transition on $\alpha^2$ and $T_c/\mu$. For a description of the plot see the text. This figure is taken from~\cite{Erdmenger:2011tj}.}
\label{fig:phasediag}
\end{figure}

In \cite{Ammon:2009xh} it was found that the value of the Yang-Mills coupling constant $\alpha$ determines if the phase transition is second order ($\alpha\le \alpha_c=0.365$) or first order ($\alpha > \alpha_c=0.365$). The quantitative dependence of the critical temperature on the parameter $\alpha$ is given in figure~\ref{fig:phasediag}. The broken phase is thermodynamically preferred in the blue and red regions, while in the white region the ground state is the Reissner-Nordstr\"om black hole. The phase transition from the white to the blue region is second order, while the one from the white to the red region is first order. The black dot determines the critical point where the order of the phase transition changes. In the green region the numerics are unstable. At zero temperature, the data may be obtained analytically as described in \cite{Basu:2009vv,Gubser:2010dm}.


\section{Perturbations about Equilibrium}
\label{sec:Perturbations-about-the}
In this section we study the response of the holographic p-wave superfluid under small perturbations. This analysis is necessary to ultimately compute the transport coefficients of the system. On the gravity side, the perturbations are given by fluctuations of the metric $h_{MN}(x^\mu,r)$ and the gauge field $a^a_{M}(x^\mu,r)$. Thus we are studying a total of $14$ physical modes: $5$ coming from the massless graviton in 5 dimensions and  $3\times 3$ from the massless vectors in five dimensions.  Due to time and spatial translation invariance in the Minkowski directions, the fluctuations can be decomposed in a Fourier decomposition as
\begin{equation}
\label{eq:fourier}
\begin{split}
h_{MN}(x^\mu,r)&=\int\!\frac{\dd^4k}{(2\pi)^4}\ee^{\ii k_\mu x^\mu}\,\hat{h}_{MN}(k^\mu,r)\,,\\
a^a_{M}(x^\mu,r)&=\int\!\frac{\dd^4k}{(2\pi)^4}\ee^{\ii k_\mu x^\mu}\,\hat{a}^a_{M}(k^\mu,r)\,.
\end{split}
\end{equation}
To simplify notations, we drop the hat on the transformed fields which we use from now on if not stated otherwise.

\subsection{Characterization of Fluctuations and Gauge Fixing}
\label{sec:Characterization-of-Fluctuations}
In general, we would have to introduce two spatial momenta: one longitudinal to the direction of the condensate, $k_{\|}$, and another one perpendicular to it, $k_\perp$. Thus, $k^\mu=(\omega,k_\|,k_\perp,0)$. But introducing a momentum perpendicular to the condensate breaks the remaining rotational symmetry $SO(2)$ down the discrete $\mathbb{Z}_2$ parity transformation $P_\perp$: $k_\perp\to -k_\perp$, $x_\perp\to -x_\perp$. This leads to a mixing of most of the fields making the problem of solving the corresponding differential equations unmanageable. Thus we do not study this case further in this paper. However, a momentum exclusively in the direction longitudinal to the condensate, or zero spatial momentum, preserves the $SO(2)$ rotational symmetry such that we can classify the fluctuations according to their transformation under the $SO(2)$ symmetry (see table~\ref{tab:classification}). The modes of different helicity decouple from each other. The momentum longitudinal to the condensate, however, breaks the longitudinal parity invariance $P_\|$. In this paper we will set this spatial momentum to zero as well. Therefore, we can classify the modes further by their behaviour under the longitudinal parity transformation $P_\|$. Under this transformation the helicity 0 fields are divided into two blocks, the first block contains $h_{tx},\, a^3_x,\, a^1_t$ and $a^2_t$ and the second one $h_{tt},\, h_{xx},\, h_{yy}+h_{zz},\, a^1_x,\, a^2_x$ and $a^3_t$.

In order to obtain the physical modes of the system we have to fix the gauge freedom. 
We choose a gauge where $a_r^a\equiv 0$ and $h_{Mr}\equiv 0$ such that the equations of motion for 
these fields become constraints. These constraints fix the unphysical fluctuations in each helicity sector and allow only the physical modes to fluctuate. The physical modes may be constructed by enforcing them to be invariant under the residual gauge transformations, $\delta a_r^a=0$ and $\delta h_{Mr}=0$ (see appendix~\ref{sec:Constructing-the-Gauge}). Thus, the physical fields we define are given in terms of the fluctuations, and classified as (from here on we set $k_\|=k$)
\begin{equation}
\label{eq:physicalmodes}
\begin{aligned}
&\text{helicity two:}\; && \Xi=g^{yy}h_{yz},~h_{yy}-h_{zz}\,,\\
&\text{helicity one:} && \Psi=g^{yy}(\omega h_{xy}+k h_{ty});~a^a_y\,,
\end{aligned}
\end{equation}
and for helicity zero:
\begin{equation}
\label{eq:physicalmodeshelicityzero}
     \begin{split}
      \Phi_1 =&a^1_x-\frac{\ii k}{\phi}a^2_t+\frac{k^2}{w \phi}a^3_t+\frac{k \omega}{w \phi}a^3_x+\frac{k w}{\omega}\xi_{tx}-\\
&-\frac{k^2 f^4 N w \sigma^2}{2 r^2 \omega ^2}\xi_t+\frac{k^2 f^5 w^2 \sigma \phi \left(\sigma N'+2 N \sigma' \right)-2 r^2 \omega ^2 f\left(w \phi w'+k^2 \phi ' \right)}{4 r \omega ^2 w \phi \left(f+r f'\right)}\xi_y,\\
      \Phi_2 =&a^2_x+\frac{\ii \left(-k^2+w^2\right)}{\omega  w}a^3_t-\\
&-\frac{\ii k}{w}a^3_x-\frac{\ii w \phi}{2 \omega }\xi_t+\frac{\ii r f \left(w^2 \phi \left (\sigma N' + 2 N\sigma' \right) + 
 2 N \left (k^2 - w^2 \right)\sigma \phi'\right)}{4 \omega  N w \sigma \left(f+r f'\right)}\xi_y,\\
    \Phi_3 =&\xi_x+\frac{2 k}{\omega }\xi_{tx}-\frac{k^2 f^4 N\sigma^2}{r^2 \omega ^2}\xi_t+\frac{4 r^2 \omega ^2 f'-2 r \omega ^2 f+k^2 f^5 \sigma \left(\sigma N'+2 N \sigma '\right)}{2 r \omega ^2 \left(f+r f'\right)}\xi_y,\\
      \Phi_4 =&a^3_x+\frac{k}{\omega }a^3_t-\frac{w \phi}{\omega ^2-\phi^2}a^1_t-\frac{\ii \omega w}{\omega ^2-\phi^2}a^2_t+\frac{w^2 \phi}{\omega ^2-\phi^2}\xi_{tx}-\\
&-\frac{k f^4 N w^2 \sigma^2 \phi}{2 r^2 \omega \left(\omega^2- \phi^2\right)}\xi_t+\frac{k f \left(f^4 w^2 \sigma \phi \left(\sigma N'+2 N \sigma '\right)+2 r^2 \left(-\omega ^2+\phi^2\right) \phi '\right)}{4 r \omega  \left(\omega ^2-\phi^2\right) \left(f+r f'\right)}\xi_y ,
     \end{split}
    \end{equation}
with
\begin{equation}
\label{eq:defxi}
  \xi_y = g^{yy}h_{yy}, \qquad \xi_x = g^{xx}h_{xx}, \qquad \xi_t = g^{tt}h_{tt}, \qquad \xi_{tx} = g^{xx}h_{tx}.
\end{equation}

\begin{table}
\centering
\begin{tabular}{l|ccc}
& dynamical fields & constraints &  \# physical modes\\
\hline
helicity 2 & $h_{yz},h_{yy}-h_{zz}$ & none & 2\\
helicity 1 & $h_{ty},h_{xy};a^a_{y}$ & $h_{yr}$ & 4\\
	        & $h_{tz},h_{xz};a^a_{z}$ & $h_{zr}$ & 4\\
helicity 0 & $h_{tt},h_{xx},h_{yy}+h_{zz},h_{xt};a^a_t,a^a_x$ & $h_{tr},h_{xr},h_{rr};a^a_r$ & 4
\end{tabular}
\caption{Classifications of the fluctuations according to their transformation under the little group $SO(2)$. The constraints are given by the equations of motion for the fields which are  set to zero due the fixing of the gauge freedom:  $a_r^a\equiv 0$ and $h_{rM}\equiv 0$. The number of physical modes is obtained by the number of dynamical fields minus the number of constraints. Due to $SO(2)$ invariance the fields in the first and second line of the helicity one fields can be identified.}
\label{tab:classification}
\end{table}

First we look at the asymptotic behavior of the helicity zero physical fields in terms of the asymptotic form of the background \eqref{eq:coeffb} and the fluctuation fields~\eqref{eq:expflucbdybl2}. The physical fields are chosen so that each one can be identified at the boundary with a fluctuation field, or a combination of them. In fact, in this limit they asymptote to
   \begin{equation}
    \label{eq:physfieldsasympt}
    \begin{split}
\Phi_1(\omega,r) &\longrightarrow \left(a^1_x\right)^b_0, \\
\Phi_2(\omega,r) &\longrightarrow \left(a^2_x\right)^b_0, \\
\Phi_3(\omega,r) &\longrightarrow \left(\xi_x\right)^b_0-\left(\xi_y\right)^b_0, \\
\Phi_4(\omega,r) &\longrightarrow \left(a^3_x\right)^b_0\,.
    \end{split}
   \end{equation}
Note that this computation was done in the $\vec k = 0$ limit, since this is the relevant limit for this paper. The resulting correlators from the helicity zero modes will be written in terms of this physical fields.

\subsection{Equations of Motion, On-shell Action and Correlators}
\label{sec:Equation-of-Motion}
In the following we will focus on the response exclusively due to time dependent perturbations, \ie $k^\mu=(\omega,0,0,0)$. In this case in addition to the $SO(2)$ symmetry, $P_\|$ parity is conserved which allows us to decouple some of the physical modes in the different helicity blocks. In this section we obtain the retarded Green's functions $G$ of the gauge theory corresponding to the stress-energy tensor $T^{\mu\nu}$ and the currents $J^\mu_a$, defined as two point functions, as in
\begin{equation}
\label{eq:correlators}
\begin{split}
G^{\mu\nu,\rho\sigma}(k)&=-\ii\int\!\dd t \dd^3 x\,\ee^{-\ii k_\mu x^\mu}\;\theta(t)\langle[T^{\mu\nu}(t,\vec{x}),T^{\rho\sigma}(0,0)]\rangle\,,\\
G_{a,b}^{\mu,\nu}(k)&=-\ii\int\!\dd t\dd^3 x\,\ee^{-\ii k_\mu x^\mu}\;\theta(t)\langle[J_a^\mu(t,\vec{x}),J_b^\nu(0,0)]\rangle\,,\\
{G^{\mu\nu}}^\rho_a(k)&=-\ii\int\!\dd t\dd^3 x\,\ee^{-\ii k_\mu x^\mu}\;\theta(t)\langle[T^{\mu\nu}(t,\vec{x}),J_a^\rho(0,0)]\rangle\,,\\
{G_a^\rho}^{\mu\nu}(k)&=-\ii\int\!\dd t\dd^3 x\,\ee^{-\ii k_\mu x^\mu}\;\theta(t)\langle[J_a^\rho(t,\vec{x}),T^{\mu\nu}(0,0)]\rangle\,.
\end{split}
\end{equation}
Here  $T^{\mu\nu}$ and $J^\mu_a$ are respectively the full stress-energy tensor and current, which include the equilibrium parts of Sec.~\ref{sec:Holographic-Setup-and}, $\langle \calt^{\mu\nu}\rangle$ and $\langle \calj^\mu_a\rangle$, as well as the corresponding dissipative parts which arise due to the introduction of fluctuations in our model.

We use the methods developed in the context of gauge-string duality to extract these Green's functions. First we determine the on-shell action at the boundary of the asymptotically AdS space from which we can easily read of the Green's functions using the recipe described in the seminal paper \cite{Son:2002sd} and its generalisation to the case of operator mixing (c.f.~\cite{Kaminski:2009dh}).

We refer the reader to \cite{Erdmenger:2011tj} for the treatment of the helicity one and two modes. Here we present the analysis of the helicity zero fluctuations. 

\subsubsection{Helicity zero modes}
\label{sec:fluchel0}
The equations of motion corresponding to these fluctuations are very lengthy, therefore, to guarantee readability, we omit them here. They can be derived by expanding the action (\ref{eq:action}) up to second order in the fluctuations and varying it with respect to the corresponding fields.

Due to the parity symmetry $P_\|$ in the $k=0$ case the modes split into two blocks, one transforming oddly (block 1) the other evenly (block 2) under $P_\|$.

\vspace{0.3cm}
 {\bf \emph{Block 1 - Parity odd}}
\vspace{0.2cm}

The first block is composed by the modes $\{ a_t^1, a_t^2, a_x^3, \xi_{tx} \}$.
The contribution of these modes to the on-shell action is\footnote{Here and in other similar expressions ahead, the products are to be understood as evaluated on opposite values of the frequency, as is natural for a Lagrangian written in Fourier space. For instance, $\xi_{tx}\;a^1_t$ would actually be $\xi_{tx}(-\omega, r)\; a^1_t(\omega,r)$.}
\begin{equation}
   \label{eq:hel0actionbl1}
\begin{split}
\tilde S^{\text{on-shell}}_{\text{hel.0, bl.1}} =& \frac{1}{\kappa^2_5} \int \frac{\dd^4 k}{{(2\pi)}^4} \left\lbrace \frac{r^5}{4 f^4 \sigma} \xi_{tx} {\xi_{tx}}'+\frac{r^3 \alpha^2}{2 \sigma} \left(a_t^1 {a_t^1}'+ a_t^2 {a_t^2}' \right)-\frac{r \alpha^2 f^4 N \sigma}{2} a_x^3 {a_x^3}' \right.\\
&\left.\left. \frac{3 r^4}{2 f^4 \sigma}{\xi_{tx}}^2 -\frac{r^3 \alpha^2}{2 \sigma}  \xi_{tx} \left(w' a_t^1 + \phi ' a_x^3 \right) \right\rbrace \right|_{r=r_{\text{bdy}}}\, ,
\end{split}
\end{equation}
which is divergent as we send $r_{\text{bdy}} \to \infty$. The divergence can be cured via holographic renormalization, i.e. the addition of covariant boundary counterterms that cancel the divergences without affecting the equations of motion (see appendix \ref{sec:Holographic-Renormalization}). To obtain the boundary action we plug the field expansions at $r_\text{bdy}$ into equation \eqref{eq:hel0actionbl1}. Since we have four fields satisfying four second order differential equations and three constraints (coming from setting $h_{xr},\, a^1_r,\, a^2_r$ to zero) we are left with a total of five ($8-3=5$) independent parameters at the boundary, $\{ \left(a_t^1\right)_0^b, \left(a_t^2\right)_0^b, \left(a_x^3\right)_0^b, \left(a_x^3\right)_1^b, \left(\xi_{tx}\right)_0^b \}$ (see~\eqref{eq:expflucbdybl1}). There is some freedom in choosing the undetermined coefficients, however the present choice is convenient for the later use of the gauge/gravity dictionary. We express the renormalized on-shell action at the boundary in terms of these coefficients\footnote{All fields in the following boundary action are dimensionless, i.e. we pulled out $r_h$. Wherever the context allows, we are sloppy with the notation and do not give the dimensionless fields new names.},
\begin{equation}
   \label{eq:hel0actionrenbl1}
\begin{split}
S^{\text{on-shell}}_{\text{hel.0, bl.1}} =& \frac{r_h^4}{\kappa^2_5} \int \frac{\dd^4 k}{{(2\pi)}^4} \left\lbrace \frac{\alpha^2 \mu \phi^b_1}{\omega^2-\mu^2} {\left(a_t^1\right)_0^b}^2 + \frac{\alpha^2 \mu \phi^b_1}{\omega^2-\mu^2}{\left(a_t^2\right)_0^b}^2 - \frac{\alpha^2 \omega^2}{4}{\left(a_x^3\right)_0^b}^2 \right.\\
& -\frac{3}{2} m^b_0 {\left(\xi_{tx}\right)_0^b}^2 + \frac{2 \ii \alpha^2 \omega  \phi^b_1}{\omega^2-\mu^2}\left(a_t^1\right)_0^b \left(a_t^2\right)_0^b + \frac{\alpha^2 \mu  w^b_1}{\omega^2-\mu^2}\left(a_t^1\right)_0^b \left(a_x^3\right)_0^b \\
& \left. - \frac{\ii \alpha ^2 \omega  w^b_1}{\omega^2-\mu^2}\left(a_t^2\right)_0^b \left(a_x^3\right)_0^b +\alpha ^2 \left(a_x^3\right)_0^b \left(a_x^3\right)_1^b +2 \alpha ^2 \phi^b_1 \left(a_x^3\right)_0^b \left(\xi_{tx}\right)_0^b  \right\rbrace \, .
\end{split}
\end{equation}

As we discuss in appendix \ref{sec:Constructing-the-Gauge}, there is a residual gauge freedom left, which has to be taken into account to obtain physically sensible observables. Using the gauge transformations given in (\ref{eq:gengaugetr}) for $\vec k=0$ and setting $K_t=K_r=\Lambda^3_0=0$, since they do not affect the fields discussed in this block, we obtain the unique linear combination (up to an overall scaling discussed in the previous paragraph)
\begin{equation}
\Phi_4 =a^3_x+w\frac{\ii \omega a^2_t+\phi a^1_t-w \phi \xi_{tx}}{\phi^2-\omega^2}
\end{equation}
Following \cite{Gubser:2008wv}, we rewrite the boundary action \eqref{eq:hel0actionrenbl1} in terms of gauge-equivalent fields, which guarantees that our solutions are gauge invariant. The set of allowed transformations is parametrized by three coefficients. The gauge equivalents to the fields which solve the equations of motion and constraints are
\begin{equation}
   \label{eq:gaugeclass}
\begin{split}
a_t^1 &\to \alpha_0 a_t^1 -\ii\omega\Lambda^1_0-\phi \Lambda^2_0 - \ii\omega  w K_x \, ,\\
a_t^2 &\to \alpha_0 a_t^2  -\ii\omega \Lambda^2_0+\phi\Lambda^1_0 \, ,\\
a_x^3 &\to \alpha_0 a_x^3 +w\Lambda^2_0 \, ,\\
\xi_{tx} &\to \alpha_0 \xi_{tx} -\ii\omega K_x \, .
\end{split}
\end{equation}
Note that we also took an overall multiplicative scaling factor into account. This can be included because different solutions of the equations of motion are related by a rescaling of the fields. These expressions give a relation, parametrized by four coefficients $\{ \alpha_0, \Lambda^1_0, \Lambda^2_0, K_x \}$, between different sets of classical solutions which are equivalent.

In order to compute the two-point functions as derivatives of the classical action, we will follow the directions given in \cite{Gubser:2008wv}, which instructs us to prescribe the value of the perturbations at the boundary, respectively defined as $\{ \beta_t^1, \beta_t^2, \beta_x^3, \beta_{tx} \}$, in terms of the gauge-equivalent quantities defined in \eqref{eq:gaugeclass}. Those are
\begin{equation}
   \label{eq:betadef}
\begin{split}
\beta_t^1 &= \alpha_0 \left(a_t^1\right)_0^b - \ii\omega \Lambda^1_0 - \phi^b_0 \Lambda^2_0 - \ii\omega  w K_x \, ,\\
\beta_t^2 &= \alpha_0 \left(a_t^2\right)_0^b - \ii\omega \Lambda^2_0 + \phi^b_0 \Lambda^1_0 \, ,\\
\beta_x^3 &= \alpha_0 \left(a_x^3\right)_0^b \, ,\\
\beta_{tx} &= \alpha_0 \left(\xi_{tx}\right)_0^b - \ii\omega K_x \, .
\end{split}
\end{equation}
The four coefficients of the gauge transformation can be chosen so that the fields asymptote to these vales. Thus, we are effectively fixing the gauge, because the gauge freedom is ``absorbed" in the freedom of choosing the boundary values. Then, we rewrite the boundary action \eqref{eq:hel0actionrenbl1} in terms of the $\beta_i$ and obtain 
\begin{equation}
S^{\text{on-shell}}_{\text{hel.0, bl.1}} =  \frac{r_h^4}{\kappa^2_5} \int \frac{\dd^4 k}{{(2\pi)}^4} \begin{pmatrix} \beta_t^{1*} & \beta_t^{2*} & \beta_x^{3*} & \beta_{tx}^* \end{pmatrix} \mathcal{G}_{(1)}(\omega) \begin{pmatrix} \beta_t^1 \\ \beta_t^2 \\ \beta_x^3 \\ \beta_{tx} \end{pmatrix}\, , 
\end{equation}
where $\mathcal{G}_{(1)}$ is the Green's function matrix of this block, which relates the response of the system to field fluctuations $a_t^1$, $a_t^2$, $a_x^3$ and $h_{tx}$. Note that, following our convention, the fields on the left row vector are evaluated in $-\omega$ and the fields on the right column vector are in $\omega$. Next, by taking derivatives $\partial^2/\partial \beta^*(-\omega)\partial\beta(\omega)$ of the action above we obtain
\begin{equation*}
\begin{pmatrix}\langle J^{t}_{1} \rangle(\omega)\\[1ex] \langle J^{t}_{2} \rangle(\omega)\\[1ex] \langle J^{x}_{3} \rangle(\omega)\\[1ex] \langle T^{tx} \rangle(\omega)\end{pmatrix}=\begin{pmatrix}\frac{\delta S^\text{on-shell}_{\text{helicity 0}}}{\delta {\left(a_t^1\right)_0^b}(-\omega)}\\[2ex] \frac{\delta S^\text{on-shell}_{\text{helicity 0}}}{\delta {\left(a_t^2\right)_0^b}(-\omega)}\\[2ex] \frac{\delta S^\text{on-shell}_{\text{helicity 0}}}{\delta {\left(\Phi_4\right)_0^b}(-\omega)}\\[2ex] \frac{\delta S^\text{on-shell}_{\text{helicity 0}}}{\delta {\big(\xi_{tx}\big)_0^b}(-\omega)}\end{pmatrix}=\begin{pmatrix}G_{1,1}^{t,t}(\omega) & G_{1,2}^{t,t}(\omega) & G_{1,3}^{t,x}(\omega) & {G_1^t}^{tx}(\omega)\\[1ex] G_{2,1}^{t,t}(\omega) & G_{2,2}^{t,t}(\omega) & G_{2,3}^{t,x}(\omega) & {G_2^t}^{tx}(\omega)\\[1ex] G_{3,1}^{x,t}(\omega) & G_{3,2}^{x,t}(\omega) & G_{3,3}^{x,x}(\omega) & {G_3^x}^{tx}(\omega)\\[1ex] {G^{tx}}_1^t(\omega) & {G^{tx}}_2^t(\omega) & {G^{tx}}_3^x(\omega) & G^{tx,tx}(\omega)\end{pmatrix}
\begin{pmatrix} {\left(a_t^1\right)_0^b}(\omega)\\[1ex] {\left(a_t^2\right)_0^b}(\omega)\\[1ex] {\left(\Phi_4\right)_0^b}(\omega)\\[1ex]  {\big(\xi_{tx}\big)_0^b}(\omega)\end{pmatrix}\,,
\end{equation*}
which explicitly written in terms of the field theory expectation values is
\begin{equation}
\label{eq:hel0responseb1}
\mathcal{G}_{(1)}(\omega) = \begin{pmatrix}  \frac{\mu}{\mu^2-\omega^2}\langle \mathcal{J}^t_3\rangle & \frac{\ii\omega}{\mu^2-\omega^2}\langle \mathcal{J}^t_3\rangle & \frac{-\mu}{\mu^2-\omega^2}\langle \mathcal{J}^x_1\rangle & 0 \\[1ex] \frac{-\ii\omega}{\mu^2-\omega^2}\langle \mathcal{J}^t_3\rangle & \frac{\mu}{\mu^2-\omega^2}\langle \mathcal{J}^t_3\rangle & \frac{\ii\omega}{\mu^2-\omega^2}\langle \mathcal{J}^x_1\rangle & 0 \\[1ex] \frac{-\mu}{\mu^2-\omega^2}\langle \mathcal{J}^x_1\rangle  & \frac{-\ii\omega}{\mu^2-\omega^2}\langle \mathcal{J}^x_1\rangle & G_{3,3}^{x,x}(\omega) & -\langle \mathcal{J}^t_3\rangle \\[1ex] 0 & 0 & -\langle \mathcal{J}^t_3\rangle & -\langle \mathcal{T}_{tt}\rangle \end{pmatrix}\, ,
\end{equation}
where we already included a factor of 2 coming from the prescription developed in~\cite{Son:2002sd} for real-time correlators. Note that using the prescription above we automatically get the correlator which includes the physical field $\Phi_4$ instead of $a^3_x$ as it is pointed out in~\cite{Gubser:2008wv}.

The matrix is completely determined by the background solution near the boundary, except for one entry, the one corresponding to the two-point correlator of $\Phi_4$, which in terms of the parity odd helicity zero modes reads
\begin{equation}
\label{eq:g33exp}
G_{3,3}^{x,x}(\omega) = -\frac{1}{2}\alpha^2\omega^2 + \frac{2\,\alpha^2}{\left(a_x^3\right)_0^b} \left[ \left(a_x^3\right)_1^b + w^b_1 \frac{\mu \left(a_t^1\right)_0^b +i\omega \left(a_t^2\right)_0^b}{\mu^2-\omega^2}  \right] \,.
\end{equation}

Rewriting this correlator in terms of the physical field we obtain
\begin{equation}
\label{eq:G33bl1}
G_{3,3}^{x,x}(\omega) = -\alpha^2 \left( \left. r^3 \frac{\Phi'_4(r)}{\Phi_4(r)}  \right|_{r=r_{\text{bdy}}} + \text{ counter terms } \right)\,
\end{equation}
showing that all entries of $\mathcal{G}_{(1)}$ are gauge invariant.

To compute this correlator we have to numerically integrate the equations of motion and constraint equations of this block. Since we choose infalling conditions at the horizon, we fix four of the eight independent coefficients. And out of the remaining four coefficients, three are fixed by the constraint equations, leaving us with one free parameter. This parameter corresponds to the overall scaling of the physical field and is related to $\alpha_0$ in~\eqref{eq:gaugeclass} and~\eqref{eq:betadef}. Since the correlators are defined by ratios of the boundary values, this parameter is scaled out and we can just set it to one. From the solution of the numerical integration, we read off the boundary values of the fields and plug them into~\eqref{eq:g33exp} to obtain the Green's functions. The results are presented in section \ref{sec:Transport-Properties}, together with a qualitative analysis of the thermoelectric effect associated to these correlators.

\vspace{0.3cm}
{\bf \emph{Block 2 - Parity even}}
\vspace{0.2cm}

The second block is composed by the modes $\{ a_t^3, a_x^2, a_x^1, \xi_t, \xi_y, \xi_x \}$, which combine to form three physical fields. The combinations we chose were defined in~\eqref{eq:physicalmodeshelicityzero}, and in this section we are taking $\vec k=0$, in which case they reduce to
\begin{equation}
   \label{eq:physf2}
\begin{split}
\Phi_1 &= a_x^1 - \frac{f w'}{2\left(f+rf'\right)}\xi_y\, ,\\
\Phi_2 &= a_x^2 +\frac{\ii w}{\omega}a_t^3 - \frac{\ii w \phi}{2\omega}\xi_t + \frac{\ii rfw\left( 2\phi N\sigma'+\phi N'\sigma -2\phi' N\sigma \right)}{4\omega N\sigma \left( f+rf' \right)}\xi_y\, \\
\Phi_3 &= \xi_x + \frac{2rf'-f}{f+rf'}\xi_y\, ,
\end{split}
\end{equation}

The contribution of this second block of helicity zero modes to the on-shell action is
\begin{equation}
   \label{eq:hel0actionbl2}
\begin{split}
\tilde S^{\text{on-shell}}_{\text{hel.0, bl.2}} =& \frac{1}{\kappa^2_5} \int \frac{\dd^4 k}{{(2\pi)}^4} \left\lbrace \frac{r^3 N \sigma}{4} \xi_y {\xi_y}' -\frac{r \alpha^2  f^4 N \sigma}{2} \left( a_x^1 {a_x^1}' + a_x^2 {a_x^2}' \right) + \frac{r^3 \alpha ^2}{2 \sigma} a_t^3 {a_t^3}'  \right. \\
&-\frac{3 r^2 N \sigma}{8} {\xi_t}^2 -\frac{r^2}{8 f} \left( 2 f N \sigma +\frac{ f r \sigma N' }{2} +f N r \sigma' +2 N r \sigma  f' \right) {\xi_x}^2 \\
& +\frac{r^3 N \sigma}{4}  \xi_y  \left( {\xi_t}'+{\xi_x}'\right) + \frac{r^2}{4f} \left( 5 f N \sigma -r N \sigma f' +\frac{r f \sigma N'}{2} +r f N \sigma' \right) \xi_y \xi_t \\
& + \frac{r^3 N \sigma}{4} \left( \xi_t + \xi_x \right) {\xi_y}' +\frac{r^2}{2 f} \left( 2 f N \sigma+\frac{r N \sigma f'}{2} +\frac{r f \sigma N'}{2} +r f N \sigma' \right) \xi_y \xi_x \\
& + \frac{r^3 N \sigma}{8} \left(\xi_t {\xi_x}'+\xi_x {\xi_t}' \right) +\frac{r^2}{8 f} \left( 5 f N \sigma+ 2r N \sigma f' +\frac{r f \sigma N'}{2} +r f N \sigma' \right) \xi_t \xi_x \\
&\left.\left. -\frac{r \alpha^2 f^4 N \sigma  w'}{4} a_x^1 \left( \xi_t- \xi_x+2 \xi_y \right) - \frac{r^3 \alpha^2 \phi'}{4 \sigma} a_t^3 \left( \xi_t- \xi_x -2 \xi_y \right) \right\rbrace \right|_{r=r_{\text{bdy}}}\, ,
\end{split}
\end{equation}
which again is divergent. The renormalized on-shell action is derived and presented in the appendix, see \eqref{eq:hel0actionrenbl2}.

Since we have six fields determined by second order differential equations and three constraints, we end up with twelve ($12-3=9$) undetermined coefficients of the boundary expansion, in terms of which the expression above is written. They are
\begin{equation}
\label{eq:bicoeff}
\left\{ \left(a_t^3\right)_0^b, \left(a_x^2\right)_0^b, \left(a_x^2\right)_1^b, \left(a_x^1\right)_0^b, \left(a_x^1\right)_1^b, \left(\xi_t\right)_0^b, \left(\xi_y\right)_0^b, \left(\xi_y\right)_2^b, \left(\xi_x\right)_0^b \right\}\, .
\end{equation}
Notice that six of them (the ones with subscript 0) coincide with the boundary values of the fields. The other three are higher-order coefficients. They are undetermined since the boundary expansion does not know about the boundary conditions set on the horizon (\ie that they must satisfy an infalling condition at the horizon). Actually when integrating the equations these coefficients are fixed by the boundary values we choose at the horizon. However, how the expansion coefficients at the boundary depend on the coefficients at the horizon cannot be addressed analytically, since in the bulk we can only solve the equations of motion numerically and this dependence is precisely determined by the behaviour of the fields in the bulk. 

As in the previous case, the Green's functions cannot be extracted directly from \eqref{eq:hel0actionbl2} because there is a residual gauge freedom left under which the fluctuation fields are not invariant. To fix the gauge freedom, we can apply again the formalism used before to derive the gauge-equivalent solutions. In this case, we have to look for the restricted set of gauge transformations and rescalings that keeps unaffected the perturbations of the first block. This set is parametrized by six coefficients $\{ \alpha_0^i, \Lambda^3_0, K_t, K_r  \}$, with $i=1,2,3$, and gives the gauge-equivalents of a solution, which are
\begin{equation}
   \label{eq:gaugeclass2}
\begin{split}
a_t^3 &\to \alpha_0 a_t^3 -\ii \omega\Lambda^3_0 + \ii\omega \phi K_t + \left(\sqrt{N}\phi' -\omega^2 (\phi A- C_\phi) \right) K_r \, ,\\
a_x^2 &\to \alpha_0 a_x^2 - w \Lambda^3_0 - \ii\omega w C_\phi K_r \, ,\\
a_x^1 &\to \alpha_0 a_x^1 + \sqrt{N}w' K_r \, ,\\
\xi_t &\to \alpha_0 \xi_t + 2\ii\omega K_t + \left(\frac{\sigma N' + 2 N\sigma' }{\sqrt{N} \sigma} - 2\omega^2 A \right) K_r \, ,\\
\xi_y &\to \alpha_0 \xi_y +\frac{2 \sqrt{N} \left( f+r f'\right)}{r f}K_r \, ,\\
\xi_x &\to \alpha_0 \xi_x +\frac{2 \sqrt{N} \left(f-2 r f'\right)}{r f}K_r \, ;
\end{split}
\end{equation}
where $A$, $C_\phi$ are defined in~\eqref{eq:transfdeltar}.
We do not explicitly write the 3 independent scale factors out, rather we use a general scaling $\alpha_0$. As will be explained later, the 3 independent scale parameters are related to the freedom of choosing the value for 3 of the fields at the horizon. However, due to the complicated mixing of the fields in the bulk it is not known how this translates into the scaling at the boundary.  Following the steps of \cite{Gubser:2008wv}, we would now proceed by prescribing the values of the perturbations at the boundary $\{ \beta_t^3, \beta_x^2, \beta_x^1, \beta_t, \beta_y, \beta_x \}$ by evaluating the asymptotic behavior of the gauge transformations, that we find is given by
\begin{equation}
   \label{eq:betadef2}
\begin{split}
\beta_t^3 &=  \left(a_t^3\right)_0^b - \ii\omega \Lambda^3_0 +\ii \omega \phi^b_0 K_t \, ,\\
\beta_x^2 &=  \left(a_x^2\right)_0^b \, ,\\
\beta_x^1 &=  \left(a_x^1\right)_0^b \, ,\\
\beta_t &=  \left(\xi_t\right)_0^b +2K_r+2\ii\omega K_t \, ,\\
\beta_y &=  \left(\xi_y\right)_0^b +2K_r \, ,\\
\beta_x &=  \left(\xi_x\right)_0^b +2K_r \, .
\end{split}
\end{equation}
Since we do not know how the scale parameters enter the above equations, we have to alter our approach in deriving the action in terms of the physical fields.

We parametrize the fluctuations in such a way that each physical mode asymptotes (see equation \eqref{eq:physfieldsasympt}) to the boundary value of a fluctuation field. For this reason, from here on we will work with
\begin{equation}
\label{eq:mphel0}
\xi_p(\omega,r)= \xi_x(\omega,r)+\xi_y(\omega,r)\; , \qquad \xi_m(\omega,r)= \xi_x(\omega,r)-\xi_y(\omega,r)\; .
\end{equation}
In addition, we perform a rotation of the $a^1_x,a^2_x$ into
\begin{equation}
\label{eq:plusminushel0}
a^+_x(\omega,r)= a^1_x(\omega,r)+\ii\;a^2_x(\omega,r)\; , \qquad a^-_x(\omega,r)= a^1_x(\omega,r)-\ii\;a^2_x(\omega,r)\; .
\end{equation}
Accordingly, we rotate the corresponding physical fields into $\Phi_\pm=\Phi_1\pm \ii\;\Phi_2$, so that their respective boundary values coincide with those of $a^\pm_x$. This parametrization is more convenient, since the $a^+_x$ and $a^-_x$ fields transform under the fundamental representation of the unbroken $U(1)_3$. That is, they behave in a similar fashion as electrically charged vector mesons do under the $U(1)_\text{em}$. To make contact with the unbroken phase, we keep the parametrization also in the broken phase. Notice that these fields are conjugate of one another: $\left(a^\pm (\omega)\right)^*=a^\mp(-\omega)$.

Next, we will invert the definitions~\eqref{eq:physf2} and solve for the selected fluctuation fields $\varphi_\tmI=\{ a^\pm_x, \xi_m \}$ in terms of the corresponding physical fields $\Phi_\tmI=\{\Phi_\pm,\Phi_3\}$. The idea is to replace these three fields and write the on-shell action in terms of the physical fields of this block along with the remaining fluctuations $\varphi_i=\{a^3_t,\xi_t,\xi_p\}$. This can be seen as a change to a more convenient basis, which guarantees that the resulting correlators are free of gauge ambiguity.

We perform the replacement in~\eqref{eq:hel0actionbl2} and in the corresponding counterterms (App.~\ref{sec:Counterterms}). In terms of the expansion coefficients of the physical fields at the boundary we obtain the on-shell action
\begin{equation}
\label{eq:hel0actionrewritten}
\begin{split}
S^{\text{on-shell}}_{\text{hel.0, bl.2}} &= \frac{r_h^4}{\kappa^2_5} \int \frac{\dd^4 k}{{(2\pi)}^4} \left\lbrace \frac{\alpha^2}{2}\left[ \left(\Phi_\p\right)^b_0 \left(\Phi_\m\right)^b_1 + \left(\Phi_\m\right)^b_0 \left(\Phi_\p\right)^b_1 \right] \right. \\
+&\frac{1}{3}\left( \Phi_3\right)^b_0 \left(\Phi_3 \right)^b_2-\frac{1}{4}\alpha^2\left(\mu+\omega\right)^2 \left(\Phi_\p\right)^b_0 \left(\Phi_\m\right)^b_0 \\
+&\frac{2\mu-\omega}{12\omega} \alpha^2 w^b_1 \left[\left(\Phi_\p\right)^b_0\left(\Phi_3\right)^b_0+\left(\Phi_3\right)^b_0\left(\Phi_\m\right)^b_0 \right] -\left(\frac{\omega^4}{64}+\frac{5f^b_2}{3}+\frac{19m^b_0}{96}\right){\left(\Phi_3\right)^b_0}^2 \\
+&\frac{\alpha^2w^b_1}{\omega}\left[ \left(\Phi_\m\right)^b_0-\left(\Phi_\p\right)^b_0 \right]\left(a^3_t\right)^b_0+ \left(f^b_2+\frac{m^b_0}{16}\right)\left(\Phi_3\right)^b_0\left[3 \left(\xi_p\right)^b_0-2 \left(\xi_t\right)^b_0  \right] \\
+&\frac{\mu-\omega}{4\omega}\alpha^2 w^b_1 \left[ \left( \left(\xi_p\right)^b_0-2\left(\xi_t\right)^b_0 \right)\left(\Phi_\p\right)^b_0 + \left(\Phi_\m\right)^b_0\left( \left(\xi_p\right)^b_0-2\left(\xi_t\right)^b_0 \right) \right] \\
+&\left. \left. \frac{m^b_0}{32}\left[ 12{\left(\xi_t\right)^b_0}^2 -9{\left(\xi_p\right)^b_0}^2 +12\left(\xi_p\right)^b_0\left(\xi_t\right)^b_0\right] \right\rbrace \right|_{r=r_{\text{bdy}}}
\end{split}
\end{equation}
The fields in this action are defined by equation \eqref{eq:physfieldsasympt} and \eqref{eq:mphel0} and below equation \eqref{eq:plusminushel0}. This new action (including the $\varphi_i$ part), when written in terms of the block 2 perturbation modes, coincides exactly with what we have in~\eqref{eq:hel0actionrenbl2}. 

The part involving the physical fields only can schematically be written as
\begin{equation*}
\label{eq:actionphyssector}
S^{\text{on-shell}}_{\text{hel.0, bl.2}} = \frac{r_h^4}{\kappa^2_5} \int \frac{\dd^4 k}{{(2\pi)}^4}\left[\Phi^*_\tmI (-\omega,r) A(k,r)_{\tmI\tmJ}\partial_r \Phi_\tmJ(\omega,r)+\Phi^*_\tmI(-\omega,r) B(k,r)_{\tmI\tmJ}\Phi_\tmJ(\omega,r)  \right]_{r=r_b} \,,
\end{equation*}
where the derivatives $\partial_r \Phi_\tmI$ evaluated at the boundary absorb the higher-order coefficients of the expansions (see~\eqref{eq:bicoeff}), in the same way that the $\Phi_\tmI(r_b)$ absorb the boundary values of the replaced fields. Of the matrices $A$, $B$; we only need to know their asymptotic values at the cutoff $r_\text{bdy}$, which are given by
   \begin{align}
    A(\omega,r_\text{bdy}) = \left(\begin{array}{ccc}
     -\frac{1}{4}\alpha^2r_{\text{bdy}}^3 & 0 & 0 \\
     0 & -\frac{1}{4}\alpha^2r_{\text{bdy}}^3 &0 \\
      0 & 0 & -\frac{1}{12}r_{\text{bdy}}^5
    \end{array}\right),\,
\end{align}
and
   \begin{align}
\label{eq:Bmatrix}
    B(\omega,r_{\text{bdy}}) = \left(\begin{array}{ccc}
      \frac{1}{4}\alpha^2 \left(\mu-\omega\right)^2 \log\left(\frac{r_h}{r}\right) & 0 &- \frac{2\mu+\omega}{24\omega}\alpha^2 w_1^b \\
     0 & \frac{1}{4}\alpha^2 \left(\mu+\omega\right)^2 \log\left(\frac{r_h}{r}\right) & \frac{2\mu-\omega}{24\omega} \alpha^2 w_1^b \\
      - \frac{2\mu+\omega}{24\omega}\alpha^2 w_1^b & \frac{2\mu-\omega}{24\omega}\alpha^2 w_1^b & B_{33}(\omega,r_{\text{bdy}})
    \end{array}\right)\; ,
   \end{align}
with $B_{33}(\omega,r)=\frac{1}{96} \left[-4\omega^2 r^2+2\omega^4 \log\left(\frac{r_h}{r}\right)-160f^b_2 -19m^b_0\right]$.

At this point we refer the reader to~\cite{Kaminski:2009dh} for a prescription to calculate the Green's functions in systems where the operators mix. In appendix~\ref{sec:Numerical-Evaluation-of}, we discuss this prescription in more detail and show how it can be generalized to our case.

The matrices $A$, $B$ are the ones used to perform the numerical calculations described there, and the rest of the terms in~\eqref{eq:hel0actionrewritten} give directly the Green's functions. In fact the matrix of Green's functions of this block may be written as
\begin{align}
\label{eq:hel0responseb2}
\begin{pmatrix}\langle J^{t}_{3} \rangle(\omega)\\[1ex] \langle J^{x}_\p \rangle(\omega)\\[1ex] \langle J^{x}_\m \rangle(\omega)\\[1ex] \langle\frac{1}{2}\left(T^{xx}-2T^{\perp\perp}\right) \rangle(\omega) \\[1ex] \langle \frac{1}{2}\left( T^{xx}+2T^{\perp\perp}\right) \rangle(\omega) \\[1ex] \langle T^{tt} \rangle(\omega) \end{pmatrix}=\begin{pmatrix}\frac{\delta S^\text{on-shell}_{\text{helicity 0}}}{\delta {\left(a_t^3\right)_0^b}(-\omega)}\\[2ex] \frac{\delta S^\text{on-shell}_{\text{helicity 0}}}{\delta {\left({\Phi_\p}^*\right)_0^b}(-\omega)}\\[2ex] \frac{\delta S^\text{on-shell}_{\text{helicity 0}}}{\delta {\left({\Phi_\m}^*\right)_0^b}(-\omega)}\\[2ex] \frac{\delta S^\text{on-shell}_{\text{helicity 0}}}{\delta {\big(\Phi_3 \big)_0^b}(-\omega)}\\[2ex] \frac{\delta S^\text{on-shell}_{\text{helicity 0}}}{\delta {\big(\xi_p \big)_0^b}(-\omega)}\\[2ex] \frac{\delta S^\text{on-shell}_{\text{helicity 0}}}{\delta {\big(\xi_t \big)_0^b}(-\omega)}\end{pmatrix}=
\mathlarger{\mathlarger{\mathlarger{\mathcal{G}_{(2)}(\omega)}}}
\begin{pmatrix} {\left(a_t^3\right)_0^b}(\omega)\\[1ex] {\left(\Phi_\p\right)_0^b}(\omega)\\[1ex] {\left(\Phi_\m\right)_0^b}(\omega)\\[1ex]  {\big(\Phi_3\big)_0^b}(\omega)\\[1ex]  {\big(\xi_p\big)_0^b}(\omega)\\[1ex]  {\big(\xi_t\big)_0^b}(\omega)\end{pmatrix}\,,
\end{align}
where the entries are denoted by
\begin{equation}
\mathcal{G}_{(2)}(\omega) = \begin{pmatrix}G_{3,3}^{t,t}(\omega) & G_{3,\p}^{t,x}(\omega) & G_{3,\m}^{t,x}(\omega) & {G_3^t}^{m}(\omega) & {G_3^t}^{p}(\omega) & {G_
3^t}^{t}(\omega)\\[1ex] G_{\p,3}^{x,t}(\omega) & G_{\p,\p}^{x,x}(\omega) & G_{\p,\m}^{x,x}(\omega) & {G_\p^x}^{m}(\omega) & {G_\p^x}^{p}(\omega) & {G_\p^x}^{t}(\omega)\\[1ex] G_{\m,3}^{x,t}(\omega) & G_{\m,\p}^{x,x}(\omega) & G_{\m,\m}^{x,x}(\omega) & {G_\m^x}^{m}(\omega) & {G_\m^x}^{p}(\omega) & {G_\m^x}^{t}(\omega)\\[1ex] {G^{m}}_3^t(\omega) & {G^{m}}_\p^x(\omega) & {G^{m}}_\m^x(\omega) & G^{m,m}(\omega) & G^{m,p}(\omega) & G^{m,t}(\omega) \\[1ex] {G^{p}}_3^t(\omega) & {G^{p}}_\p^x(\omega) & {G^{p}}_\m^x(\omega) & G^{p,m}(\omega) & G^{p,p}(\omega) & G^{p,t}(\omega) \\[1ex] {G^{t}}_3^t(\omega) & {G^{t}}_\p^x(\omega) & {G^{t}}_\m^x(\omega) & G^{t,m}(\omega) & G^{t,p}(\omega) & G^{t,t}(\omega)\end{pmatrix}
\end{equation}
and we find that they are given by $\mathcal{G}_{(2)}(\omega)  = $
\begin{equation*}
\label{eq:greenmathel0b2}
\begin{pmatrix} 0 & -\frac{1}{2\omega}\langle \mathcal{J}^x_1\rangle & \frac{1}{2\omega}\langle \mathcal{J}^x_1\rangle & 0 & 0 & 0 \\[1ex] \frac{1}{2\omega}\langle \mathcal{J}^x_1\rangle & G_{\p,\p}^{x,x}(\omega) & G_{\p,\m}^{x,x}(\omega) & {G_\p^x}^{m}(\omega) & -\frac{\mu+\omega}{8\omega}\langle \mathcal{J}^x_1\rangle & \frac{\mu+\omega}{4\omega}\langle \mathcal{J}^x_1\rangle \\[1ex] -\frac{1}{2\omega}\langle \mathcal{J}^x_1\rangle & G_{\m,\p}^{x,x}(\omega) & G_{\m,\m}^{x,x}(\omega) & {G_\m^x}^{m}(\omega) & \frac{\mu-\omega}{8\omega}\langle \mathcal{J}^x_1\rangle & -\frac{\mu-\omega}{4\omega}\langle \mathcal{J}^x_1\rangle\\[1ex] 0 & {G^{m}}_\p^x(\omega) & {G^{m}}_\m^x(\omega) & G^{m,m}(\omega) & \frac{3}{16}\left(\langle\mathcal{T}_{tt}\rangle-2\langle\mathcal{T}_{xx}\rangle\right) & -\frac{1}{8} \left(\langle\mathcal{T}_{tt}\rangle-2\langle\mathcal{T}_{xx}\rangle\right) \\[1ex] 0 & \frac{\mu-\omega}{8\omega}\langle \mathcal{J}^x_1\rangle & -\frac{\mu+\omega}{8\omega}\langle \mathcal{J}^x_1\rangle & \frac{3}{16} \left(\langle\mathcal{T}_{tt}\rangle-2\langle\mathcal{T}_{xx}\rangle\right) & -\frac{3}{16} \langle\mathcal{T}_{tt}\rangle & \frac{1}{8}\langle\mathcal{T}_{tt}\rangle \\[1ex]  0 & -\frac{\mu-\omega}{4\omega}\langle \mathcal{J}^x_1\rangle & \frac{\mu+\omega}{4\omega}\langle \mathcal{J}^x_1\rangle & -\frac{1}{8} \left(\langle\mathcal{T}_{tt}\rangle-2\langle\mathcal{T}_{xx}\rangle\right) & \frac{1}{8} \langle\mathcal{T}_{tt}\rangle & \frac{1}{4}\langle\mathcal{T}_{tt}\rangle \end{pmatrix} .
\end{equation*}
Now some comments are in order explaining the expectation values of equation \eqref{eq:hel0responseb2} and the notation we use. Regarding the former, we derive $\langle \frac{1}{2}\left( T^{xx}-2T^{\perp\perp}\right) \rangle$ and $\langle \frac{1}{2}\left( T^{xx}+2T^{\perp\perp}\right) \rangle$ in appendix \ref{sec:General-Remarks-on} and furthermore use that $T^{yy}=T^{zz}$. Regarding the notation, note that the index $m$ is related to $\Phi_3$ field, since $\left(\Phi_3\right)^b_0 = \left(\xi_m\right)^b_0$. Furthermore the $^x_\pm$ indices are related to the $\Phi_\p$ and $\Phi_\m$ fields. Here all entries are functions of the background, except for the sector containing the physical fields and their couplings, which has to be computed numerically. As before, to do so we choose infalling boundary conditions at the horizon which leaves us with six of the former twelve free parameters. From the six left, three are fixed by the constraint equations. The value of the three remaining coefficients just scale the solutions and since we compute ratios of the boundary values it does not matter which value we choose for them. Note that a more detailed explanation of the numerical procedure we apply is described in appendix~\ref{sec:Numerical-Evaluation-of}. This part of $\mathcal{G}_{(2)}$ describes the dynamics of this block, and it is related holographically to several interesting properties of the superfluid phase, as we describe in the next section.



\section{Transport Properties}
\label{sec:Transport-Properties}
In this section we extract the transport properties of the holographic p-wave superfluid from the correlation functions presented in the previous section. We split our analysis into distinct transport phenomena.

\subsection{Thermoelectric Effect parallel to the Condensate}
\label{sec:Thermoelectric-Effect-parallel}
We start by presenting the thermoelectric effect parallel to the condensate, i.e.~we look at charge transport and temperature gradients in the $x$ direction. This is related to the first block of helicity zero states we presented in section \ref{sec:fluchel0}. Furthermore our results are in agreement with \cite{Gubser:2008wv} for the non-backreacted case.

The thermoelectric effect describes the simultaneous transport of charge and heat (or energy). This means that an electric field not only leads to a current, but also to a heat flux and, conversely, a temperature gradient leads to an electric current in addition to a heat flux. In holographic systems, this effect was already observed in s-wave superfluids (see e.g.~ \cite{Hartnoll:2008vx,Hartnoll:2009sz,Herzog:2009xv}) or in the p-wave superfluid component transverse to the condensate \cite{Erdmenger:2011tj}. However, in the case at hand, we have a slight complication due to a further coupling of the $a^1_t$ and $a^2_t$ fields to the $a^3_x$ and to the $h_{tx}$ metric component (see section \ref{sec:fluchel0}).

\begin{figure}[t]
\centering
\includegraphics[width=0.9\linewidth]{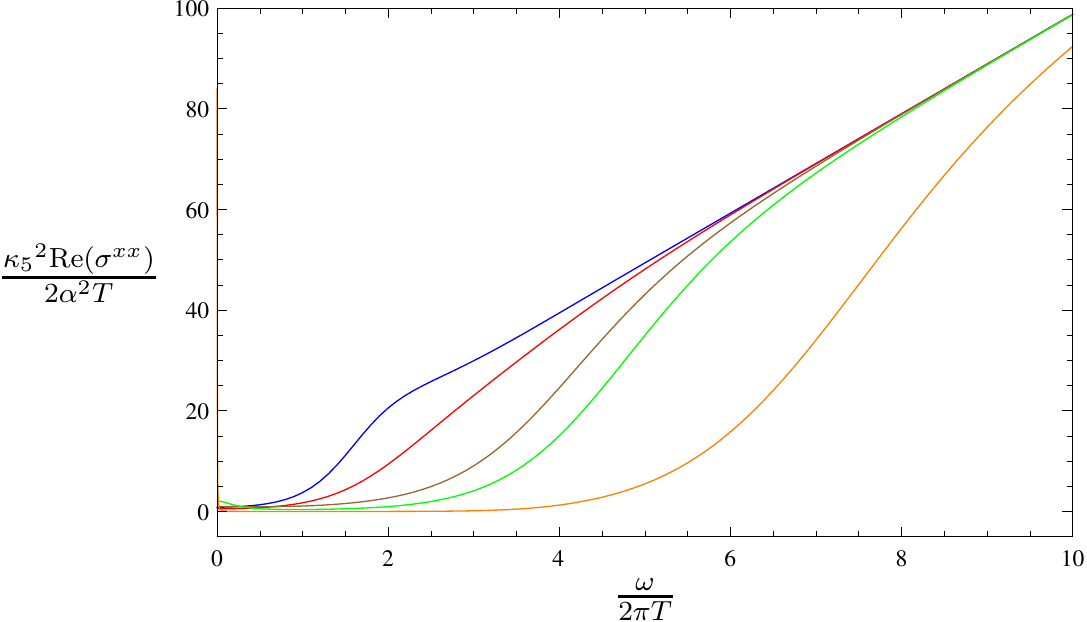}
\put(-300,138){\includegraphics[scale=.43]{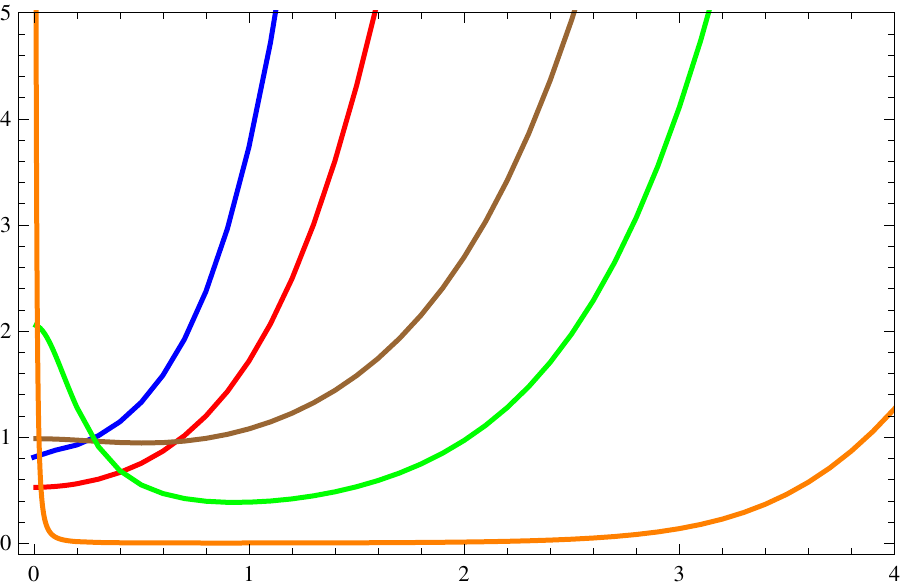}}
\caption{Real part of the conductivity $\re(\sigma^{xx})$ over the frequency $\omega/(2\pi T)$ for $\alpha=0.316$. The color coding is as follows: blue $T=1.63T_c$, red $T=0.98T_c$, brown $T=0.88T_c$, green $T=0.78T_c$, orange $T=0.50T_c$. There is a delta peak at strictly $\omega=0$, not noticeable in this figure, as dictated by the sum rule (the area below the curves has to be the same for any $T$).}
\label{fig:conda3xre_0316}
\end{figure}

\begin{figure}[t]
\centering
\includegraphics[width=0.9\linewidth]{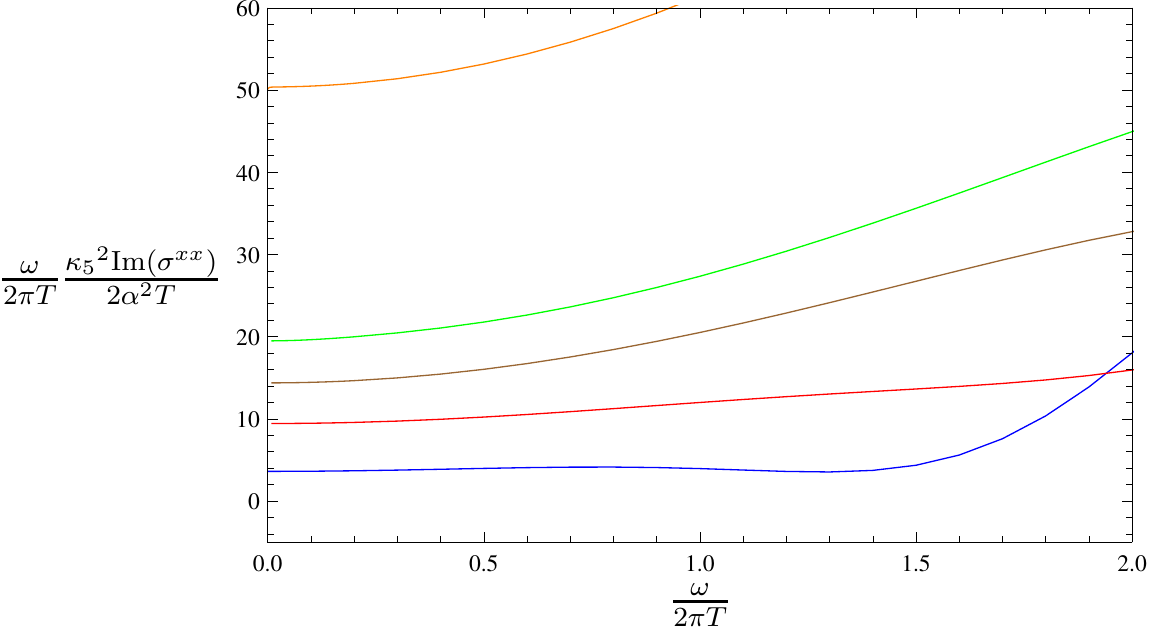}
\caption{Imaginary part of the conductivity times the frequency, $\omega \im(\sigma^{xx})$ over the frequency $\omega/(2\pi T)$ for $\alpha=0.316$. The color coding is as follows: blue $T=1.63T_c$, red $T=0.98T_c$, brown $T=0.88T_c$, green $T=0.78T_c$, orange $T=0.50T_c$. The curves tend to a constant value as $\omega \to 0$, which indicates the presence of a pole at the origin. This is related to the delta peak in the real part of $\sigma^{xx}$.}
\label{fig:conda3xim_0316}
\end{figure}

A straightforward calculation (see~\cite{Hartnoll:2009sz}) shows that $\nabla_x T$ is related to the $g_{tt}$ component of the metric through the change in the period of the Euclidean time.  The change can be done in a way that $\delta g_{tt}$ becomes pure gauge provided a complementary change is done for $A^3_x$ and $g_{tx}$. It is customary to fix the gauge requesting $\delta g_{tt}$ to vanish, and allowing for $\delta A^3_x$ and $\delta g_{tx}$ only. On the other hand, $E_x$ receives an additional contribution from the vector potential, $i\omega \left(\Phi_4\right)^b_0$. The combined effect is so that we define the electric field and temperature gradient
\begin{equation}
\begin{split}
E_x &= \ii \omega \left[ \left(\Phi_4\right)^b_0 +\mu\left( \xi_{tx} \right)^b_0 \right]\, , \\
-\frac{\nabla_x T}{T} &= \ii \omega \left( \xi_{tx} \right)^b_0\,.
\end{split}
\end{equation}
This modes source  the charge current $J^x$ in direction of the condensate and the heat flux $Q^x=T^{tx}-\mu J^x$, respectively. The relation of these currents to the corresponding electrical field and temperature gradient defines the conductivity matrix
\begin{align}
\label{eq:conductdef}
\begin{pmatrix}\langle J^x \rangle\\[1ex] \langle Q^x \rangle \end{pmatrix}=
\begin{pmatrix} \sigma^{xx} & T\alpha^{xx} \\[1ex]  T\alpha^{xx} & T\bar\kappa^{xx} \end{pmatrix}
\begin{pmatrix} E_x \\[1ex]  -(\nabla_x T)/T \end{pmatrix}\,.
\end{align}
Comparing this matrix to the lower right corner of the one in~\eqref{eq:hel0responseb1}, we can identify the electric, thermal and thermoelectric conductivities, which are related to the retarded Green's functions by
\begin{equation}
\begin{split}
\sigma^{xx} &= -\frac{\ii}{\omega} G^{x,x}_{3,3}\, , \\
T\alpha^{xx} &= -\frac{\ii}{\omega} \left( {G^x_3}^{tx} -\mu G^{x,x}_{3,3}  \right) =\frac{\ii}{\omega}\langle\mathcal{J}^t_3\rangle-\mu\sigma^{xx}  \, , \\
T\bar\kappa^{xx} &= -\frac{\ii}{\omega} \left( G^{tx,tx}-2\mu {G^x_3}^{tx} +\mu^2 G^{x,x}_{3,3}   \right)  =\frac{\ii}{\omega}\left( \langle\mathcal{T}_{tt}\rangle -2\mu\langle\mathcal{J}^t_3\rangle \right)+\mu^2\sigma^{xx} \, .
\end{split}
\end{equation}
The conductivity in direction of the condensate $\sigma^{xx}$ has been calculated numerically. The results are shown in figures \ref{fig:conda3xre_0316} and \ref{fig:conda3xim_0316} for $\alpha=0.316<\alpha_c$. The results for other values of $\alpha$ do not show any significant qualitative difference, therefore we do not show them in this paper.

The rest of the matrix~\eqref{eq:hel0responseb1} shows the response of the system due to the $a^1_t$, $a^2_t$ fluctuations. This is a manifestation of the fact that the equations of motion of the gauge field fluctuations are coupled. Therefore, if a temperature gradient excites one of these modes, the other two will respond, and their response is dictated by the coefficients of the matrix. In \cite{Gubser:2008wv}, the $a^1_t$, $a^2_t$ fluctuation fields are interpreted as generating a rotation of the charge density in direction $\langle J^t_1\rangle$ and $\langle J^t_2\rangle$, however without changing its magnitude.

The complete transport matrix of this block then reads
\begin{equation}
\begin{pmatrix}\langle J^{t}_{1} \rangle\\[1ex] \langle J^{t}_{2} \rangle\\[1ex] \langle J^{x} \rangle\\[1ex] \langle Q^x \rangle\end{pmatrix}=\begin{pmatrix}\sigma^{t,t}_{1,1} & \sigma^{t,t}_{1,2} & \sigma^{t,x}_{1,3} & -\mu\sigma^{t,x}_{1,3}\\[1ex] \sigma^{t,t}_{2,1} & \sigma^{t,t}_{2,2} & \sigma^{t,x}_{2,3} & -\mu\sigma^{t,x}_{2,3}\\[1ex] \sigma^{x,t}_{3,1} & \sigma^{x,t}_{3,2} & \sigma^{xx} &  T\alpha^{xx}\\[1ex] -\mu\sigma^{x,t}_{3,1} & -\mu\sigma^{x,t}_{3,2} &  T\alpha^{xx} & T\bar\kappa^{xx} \end{pmatrix}
\begin{pmatrix} i\omega a^1_t \\[1ex] i\omega a^2_t \\[1ex] E_x \\[1ex]  -\frac{\nabla_x T}{T}\end{pmatrix}\,,
\end{equation}
where each of the transport coefficients is simply related to the corresponding Green's function by $\sigma=-\ii G/\omega$. We will now focus on the electric conductivity $\sigma^{xx}$, the others can be obtained from it.

The fact that the longitudinal conductivity $\sigma^{xx}$ has a different behavior than that of the component transverse to the condensate $\sigma^{\perp\perp}$  (c.f.~\cite{Erdmenger:2011tj}) in the broken phase is an effect of the breaking of rotational symmetry.

Let us discuss the similarities and differences between $\sigma^{xx}$ and $\sigma^{\perp\perp}$ (for a discussion of $\sigma^{\perp\perp}$ see~\cite{Erdmenger:2011tj}). The curve of the real part of $\sigma^{xx}$ (fig.~\ref{fig:conda3xre_0316}) shows the correct~\cite{Myers:2007we} asymptotic behaviour for large frequencies, \ie the real part is proportional to the frequency for all temperatures. More precisely, for $\omega\gg T$ we have
\begin{equation}
\frac{{\kappa_5}^2 \text{Re}(\sigma^{xx})}{2\alpha^2 T} \to \pi^2 \frac{\omega}{2\pi T}\,.
\end{equation}
We expect this behavior on dimensional grounds, and as a consequence of the conformal symmetry in our system\footnote{There is no lattice spacing which would spoil the high frequency behavior.}. On the other hand, for decreasing frequencies we see that the conductivity decreases until nearly vanishing. This sharp decrease is a known feature of superconductors. It is present for all temperatures, not only for $T<T_c$. However, for smaller temperatures the decrease takes place at larger values of $\omega$. It is by far not as sharp as in $\sigma^{\perp\perp}$, and furthermore there are some qualitative differences: The bump before decreasing is absent for $\sigma^{xx}$ - the asymptotic value for large frequencies is approached by the curves with smaller temperature from below, rather than from above, as opposed to the perpendicular case. Besides, up to numerical inaccuracy the conductivities do not seem to vanish for any frequency, for temperatures above $0.5T_c$. In comparison, the transverse conductivity has a far stronger temperature suppression in the gapped region. However, below $0.5T_c$ the situation seems to change dramatically as is explained in the next section.

The real part of $\sigma^{xx}$, as opposed to the perpendicular case, increases again for small but finite frequencies and reaches a finite value in the $\omega\to 0$ limit, as seen in the zoomed region of figure~\ref{fig:conda3xre_0316}. This increase in the real part in the zero frequency limit is due to a quasinormal mode which moves up the imaginary axis in the complex frequency plane (see the blue arrow in figure~\ref{fig:polmov}) and seems to reach the origin $\omega=0$ at temperatures slightly above $0.5T_c$. The increase we see towards the $\omega \to 0$ limit comes from the projection of the quasinormal mode onto the real frequency axis. Note that this bump increases with decreasing temperature. Unfortunately it is challenging to compute the exact temperature when the mode arrives at the origin, since we have to rely on numerical calculations. Nevertheless, for temperatures below $0.5T_c$ it appears that a pole is formed and at the same time the real part of the conductivity is more strongly suppressed at finite small frequencies in comparison to cases of temperatures above $0.5T_c$ (see the green and orange curve in zoomed region of figure~\ref{fig:conda3xre_0316}). It seems that somewhere around $0.5T_c$, due to the quasinormal mode at the origin, the conductivity behavior in the direction of the condensate changes. It would be interesting to understand this effect from a field theoretic point of view, we leave this for future work.

Due to the pole in the imaginary part of the conductivity (see fig.~\ref{fig:conda3xim_0316}) and the Kramers-Kronig relation~\cite{Hartnoll:2008kx} we know that at $\omega=0$ the real part must have a delta peak. There are two main contributions to the prefactor of this delta peak, which change with temperature, and they come from the pole at the origin of the imaginary part, expressed as
\begin{equation}
\omega\,\im \left( \sigma^{xx} \right) \simeq A_{\text{{\tiny{D}}}}(\alpha,T) + A^x_s(\alpha)\left(1-\frac{T}{T_c}\right)\,.
\end{equation}
This is reminiscent of the perpendicular case (c.f.~\cite{Erdmenger:2011tj}). In fact, similarly the first contribution $A_{\text{{\tiny{D}}}}$ is a consequence of translational invariance at all temperatures, specifically for temperatures above $T_c$. The other contribution, $A^x_s$, appears when temperatures decrease below $T_c$. This prefactor is expected to be connected to the superfluid density, however it differs from the corresponding factor in the transverse case.

The properties of the two components of the conductivities we state here are very similar to the ones found in the non-backreacted case (see \cite{Gubser:2008wv}). Therefore corrections due to the backreaction seem to be rather small.

\subsection{Viscosities and Flavour Transport Coefficients}
\label{sec:second-block}

The second block of coupled modes transforming as scalars under the $SO(2)$ symmetry includes the fields $a^1_x$, $a^2_x$, $a^3_t$, $\xi_t=g^{tt}h_{tt}$, $\xi_x=g^{xx}h_{xx}$ and $\xi_y=g^{yy}h_{yy}$. Similarly to the first block, these fields form 3 physical modes, $\Phi_1$, $\Phi_2$ and $\Phi_3$ (see \eqref{eq:physicalmodeshelicityzero}). It turns out that it is more sensible to consider this fields in terms of $\Phi_\pm = \Phi_1 \pm i \Phi_2$, since they transform fundamentally under the $U(1)_3$ in the unbroken phase.

\subsubsection{Piezoelectric effect}
\label{sec:piezoeleceffect}

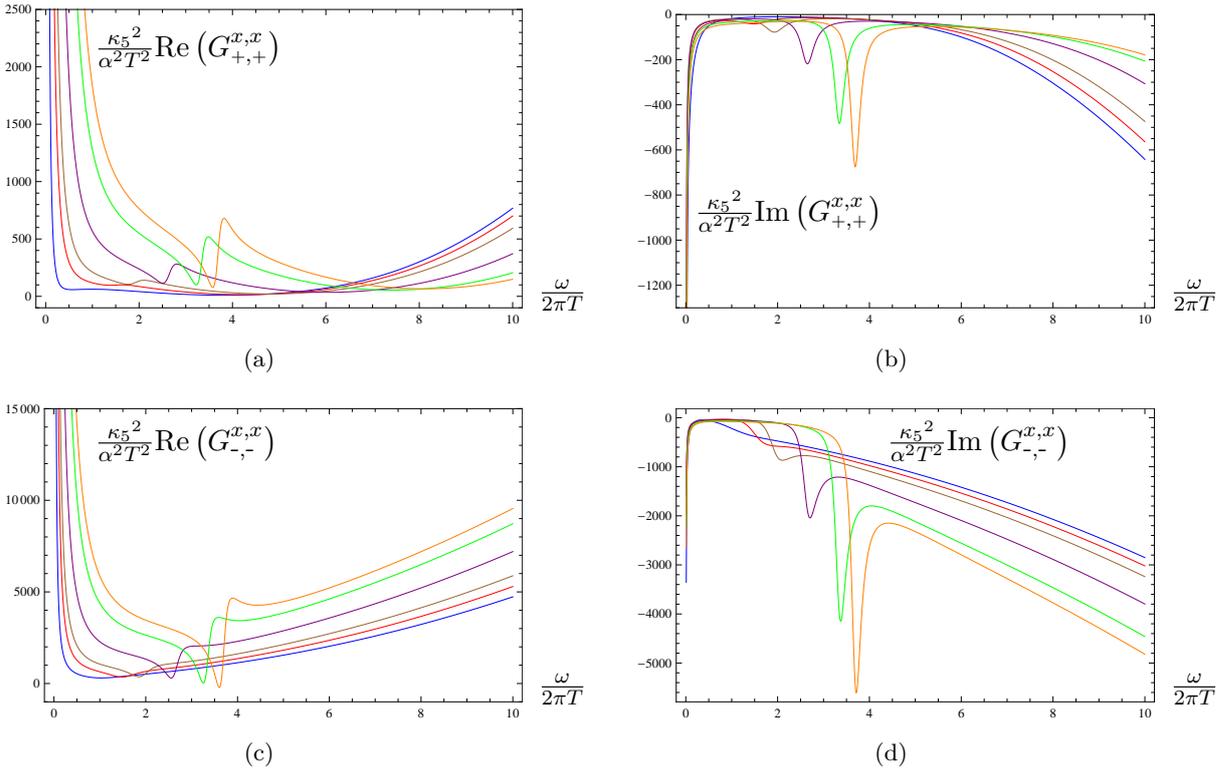
\begin{figure}
  \centering
  \subfigure[]{\label{fig:remm0316}\includegraphics[width=0.45\textwidth]{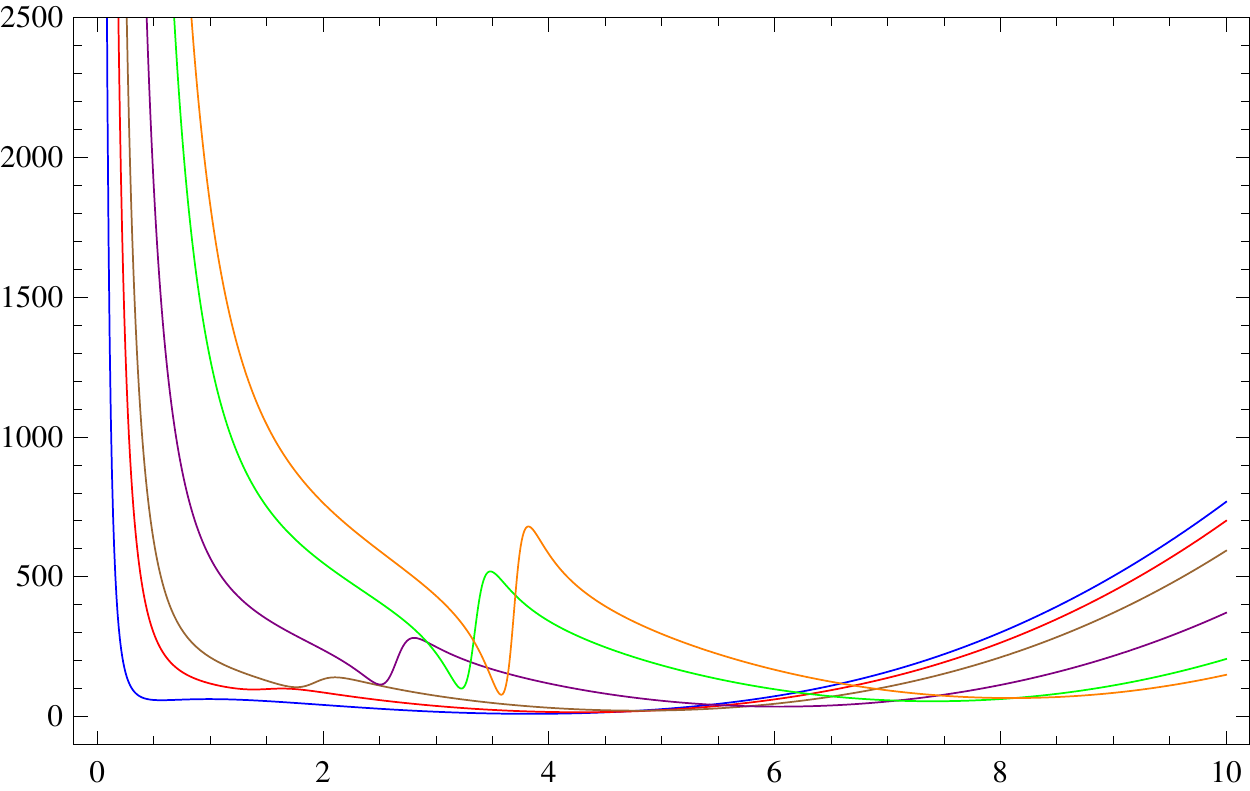}\put(-160,102){$\frac{{\kappa_5}^2}{\alpha^2 T^2} \re\left(G^{x,x}_{\p,\p}\right)$}\put(6,10){$\frac{\omega}{2\pi T}$} }
  \hfill               
  \subfigure[]{\label{fig:repp0316}\includegraphics[width=0.45\textwidth]{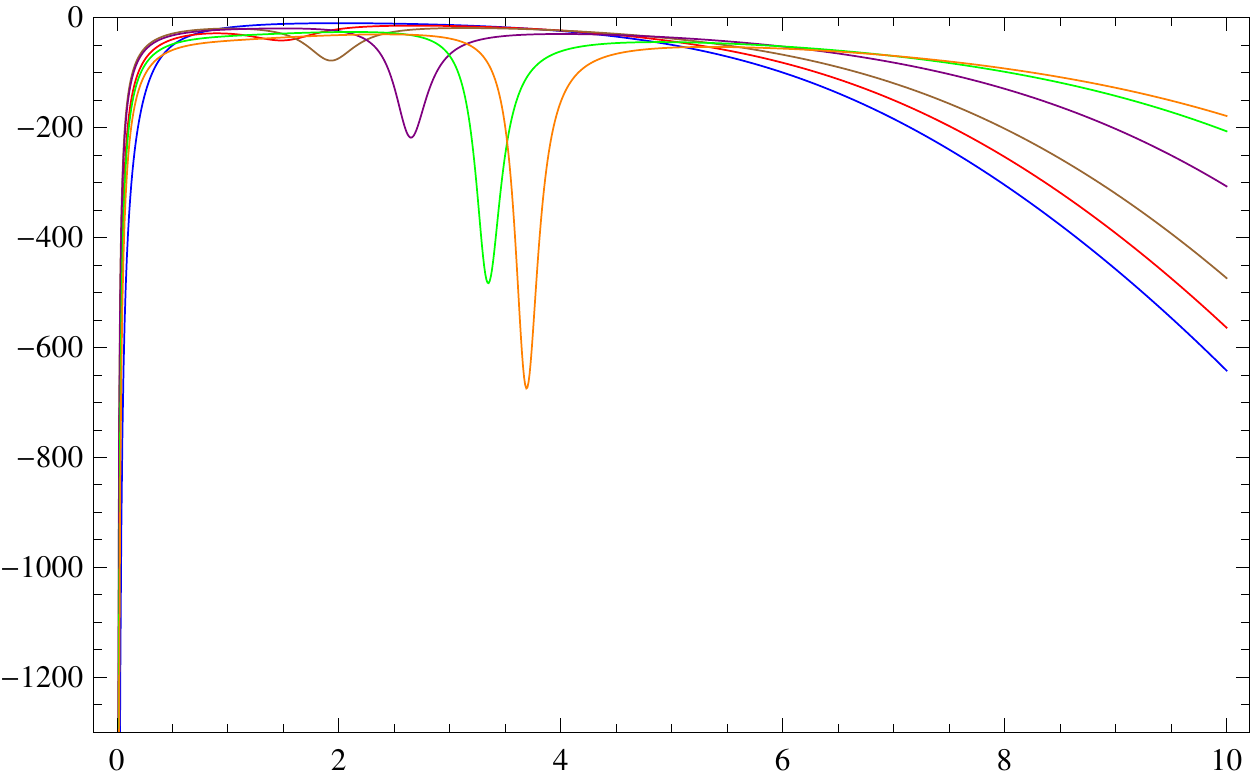}\put(-172,40){$\frac{{\kappa_5}^2}{\alpha^2 T^2} \im\left(G^{x,x}_{\p,\p}\right)$}\put(6,10){$\frac{\omega}{2\pi T}$}}
  \hfill               
  \subfigure[]{\label{fig:immm0316}\includegraphics[width=0.45\textwidth]{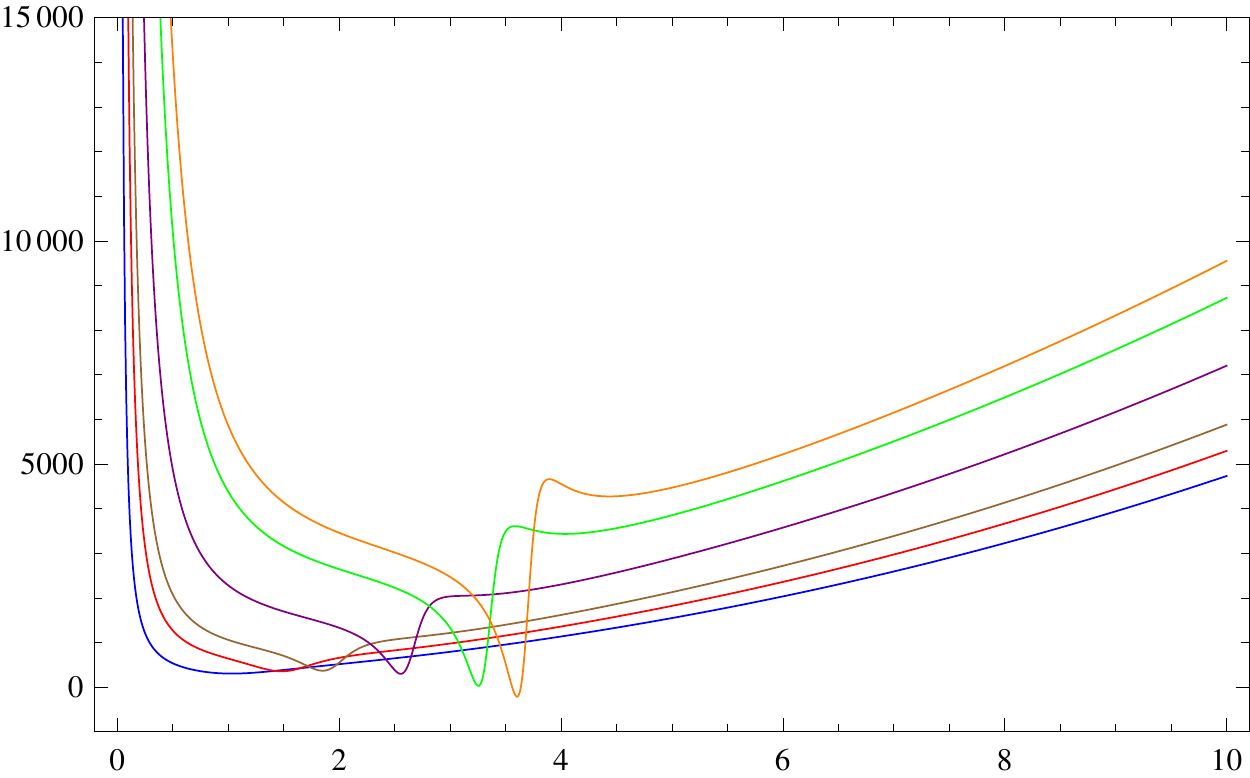}\put(-160,102){$\frac{{\kappa_5}^2}{\alpha^2 T^2} \re\left(G^{x,x}_{\m,\m}\right)$}\put(6,10){$\frac{\omega}{2\pi T}$}}
  \hfill               
  \subfigure[]{\label{fig:impp0316}\includegraphics[width=0.45\textwidth]{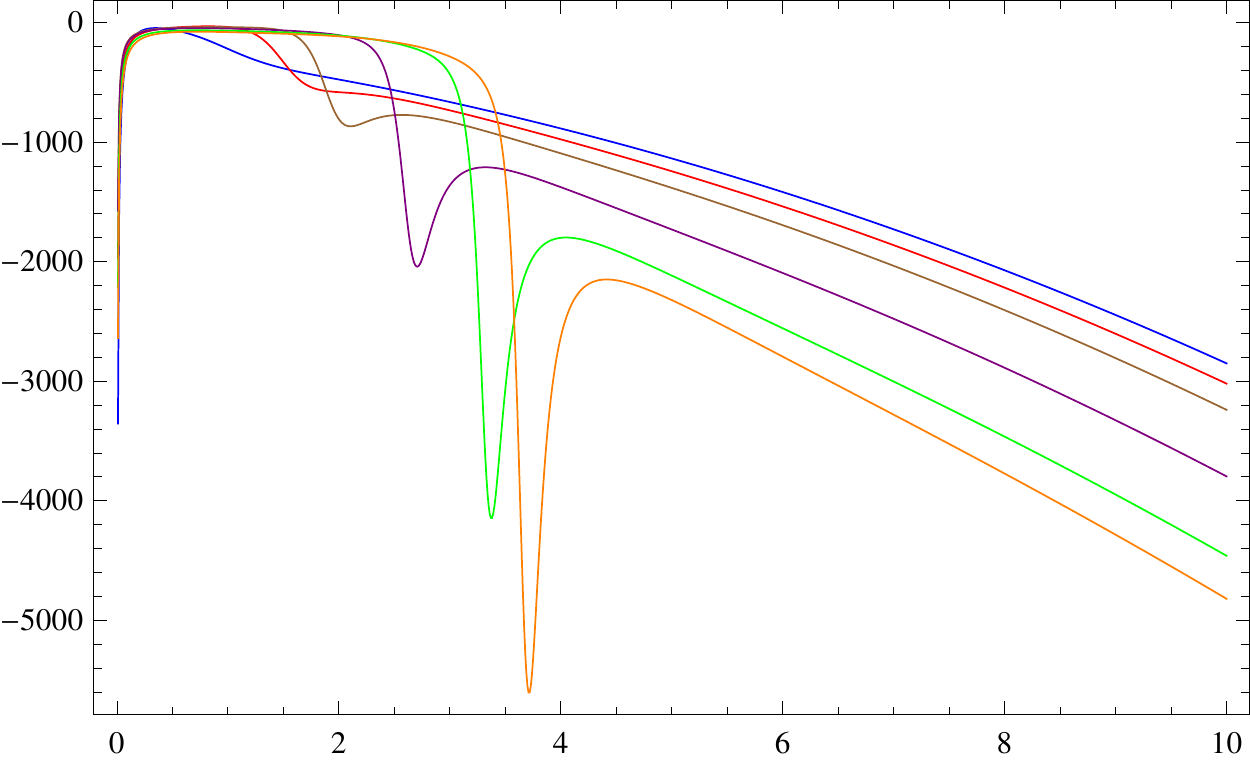}\put(-100,102){$\frac{{\kappa_5}^2}{\alpha^2 T^2} \im\left(G^{x,x}_{\m,\m}\right)$}\put(6,10){$\frac{\omega}{2\pi T}$}}
  \caption{These plots show the real and imaginary part of the correlators $G^{x,x}_{\pm,\pm}$ versus the reduced frequency $\omega/(2\pi T)$ for $\alpha=0.316$ at different temperatures: blue $T=0.98T_c$, red $T=0.88T_c$, brown $T=0.78T_c$, purple $T=0.62T_c$, green $T=0.50T_c$, orange $T=0.46T_c$.}
  \label{fig:hel0b2ppmm}
\end{figure}

The transport properties presented in this section show similarities to an effect known as piezoelectric effect\footnote{A similar effect, the flexoelectric effect, is related to the helicity one modes, see~\cite{Erdmenger:2011tj}} found in crystals~\cite{Gennes:1974lc}. This effect describes the generation of an electric current due to the squeezing and/or elongation of a crystal, or the generation of a mechanical strain due to an external electric field. A coupling between a normal stress difference and (flavour) currents that resembles this effect is found in this block. Note that the piezoelectric effect was already found in the context of black branes in \cite{Armas:2012ac}.

In simple terms, there is an interaction analogous to the one of the first block, between
\begin{align}
\begin{pmatrix}\langle J^x_\pm \rangle\\[1ex] \langle T^{xx}, T^{\perp\perp}, T^{tt} \rangle \end{pmatrix} \longleftrightarrow 
\begin{pmatrix} a^\pm_x \\[1ex]  h_{xx}, h_{\perp\perp}, h_{tt} \end{pmatrix}\,.
\end{align}
The broken phase is characterized by a condensate $\langle J^x_1\rangle$. The $a^\pm_x$ fluctuate around this background value and the system reacts by working against this perturbations by changing the diagonal stress-energy tensor components $\langle T^{xx}\rangle$, $\langle T^{\perp\perp}\rangle$ and $\langle T^{tt}\rangle$. The converse case, where we fluctuate about equilibrium values of the stress-energy tensor and look at the response of the currents $\langle J^x_1\rangle$ works in a similar fashion. Note that this is not the only response of the system to these fluctuations. However, in this section we are interested exactly in the coupling between different modes. In the field theory this may be related to an electric current being affected by, or generating, mechanical stress (Piezoelectric effect). Finally, the transport coefficients ``measure'' the strength of the response of the system, i.e.~how do the expectation values change with respect to the original values when they are perturbed.

\begin{figure}
  \centering
  \subfigure[]{\label{fig:repm0316}\includegraphics[width=0.45\textwidth]{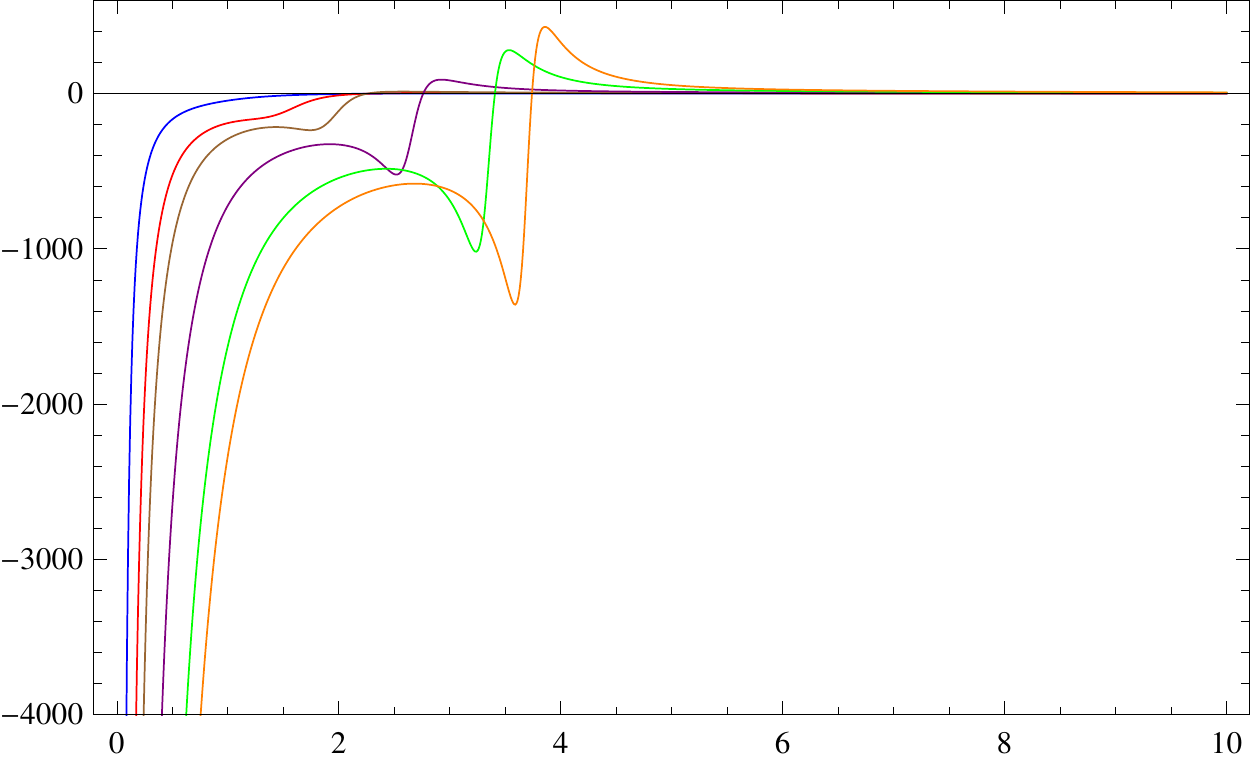}\put(-100,82){$\frac{{\kappa_5}^2}{\alpha^2 T^2} \re\left(G^{x,x}_{\mp,\pm}\right)$}\put(6,10){$\frac{\omega}{2\pi T}$}}                
  \hfill
  \subfigure[]{\label{fig:impm0316}\includegraphics[width=0.45\textwidth]{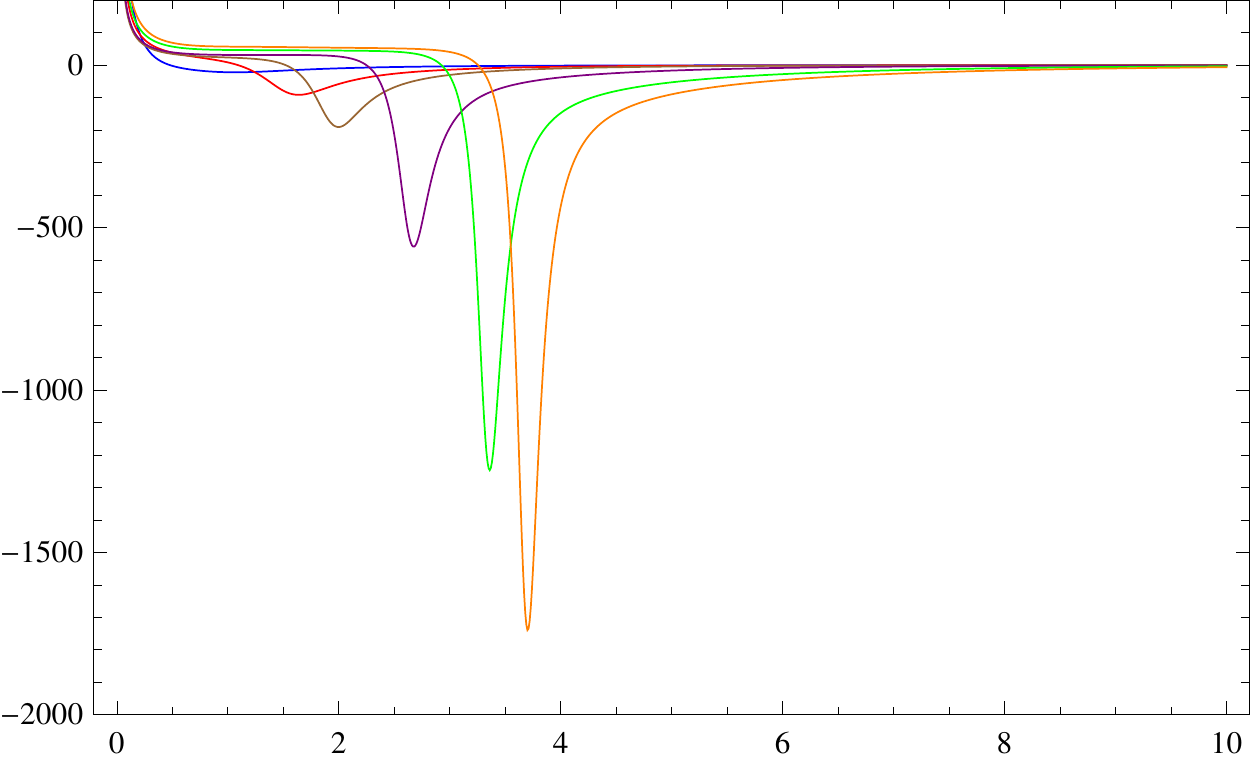}\put(-100,82){$\frac{{\kappa_5}^2}{\alpha^2 T^2} \im\left(G^{x,x}_{\mp,\pm}\right)$}\put(6,10){$\frac{\omega}{2\pi T}$}}
  \caption{These plots show the real and imaginary part of the correlator $G^{x,x}_{\p,\m}$, or equivalently $G^{x,x}_{\m,\p}$, versus the reduced frequency $\omega/(2\pi T)$ for $\alpha=0.316$ at different temperatures: blue $T=0.98T_c$, red $T=0.88T_c$, brown $T=0.78T_c$, purple $T=0.62T_c$, green $T=0.50T_c$, orange $T=0.46T_c$.}
  \label{fig:hel0b2pmmp}
\end{figure}

In figures \ref{fig:hel0b2ppmm} and \ref{fig:hel0b2pmmp} we plot the real and imaginary part of $G^{x,x}_{\pm,\pm}$ and $G^{x,x}_{\pm,\mp}$ over the reduced frequency $\omega/(2\pi T)$, for several values of the temperature, or equivalently of the chemical potential $\mu$. We find the symmetry relations
\begin{gather*}
G^{x,x}_{\m,\m}(\omega)=G^{x,x}_{\p,\p}(-\omega)^*\,, \qquad G^{x,x}_{\p,\m}(\omega)=G^{x,x}_{\m,\p}(-\omega)^*\,, \\
G^{x,x}_{\m,\m}(\omega)=G^{x,x}_{\p,\p}(-\omega)^*\,, \qquad G^{x,x}_{\p,\m}(\omega)=G^{x,x}_{\m,\p}(-\omega)^*\,,
\end{gather*}
as we did in the study of helicity one modes~\cite{Erdmenger:2011tj}. This was expected from the fact that $(\Phi_\p (\omega))^* = \Phi_\m(-\omega)$, $(\Phi_\m (\omega))^* = \Phi_\p(-\omega)$ and $\left(\Phi_3(\omega)\right)^*=\Phi_3(-\omega)^*$.

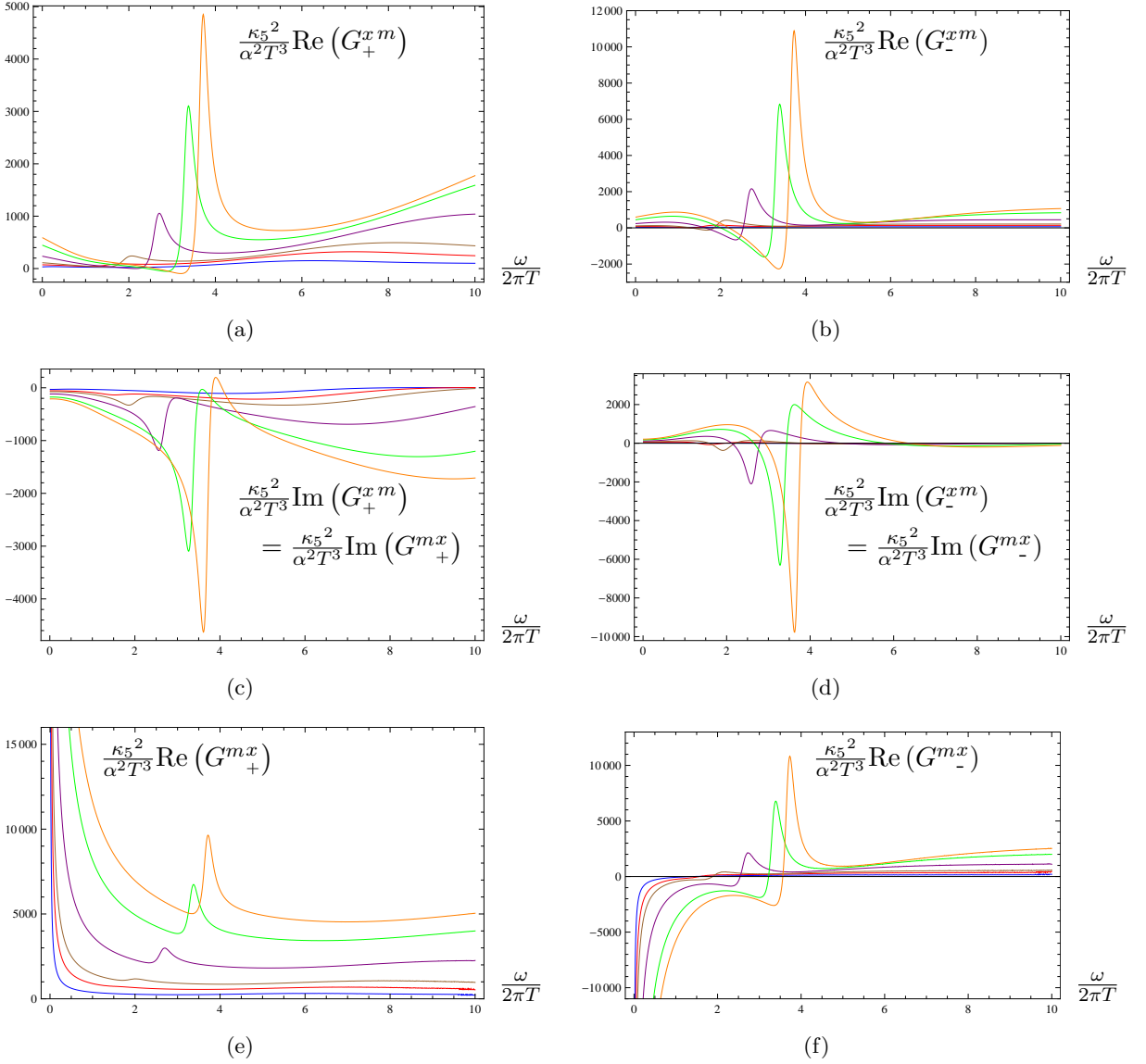
\begin{figure}
  \centering
  \subfigure[]{\label{fig:repp30316}\includegraphics[width=0.45\textwidth]{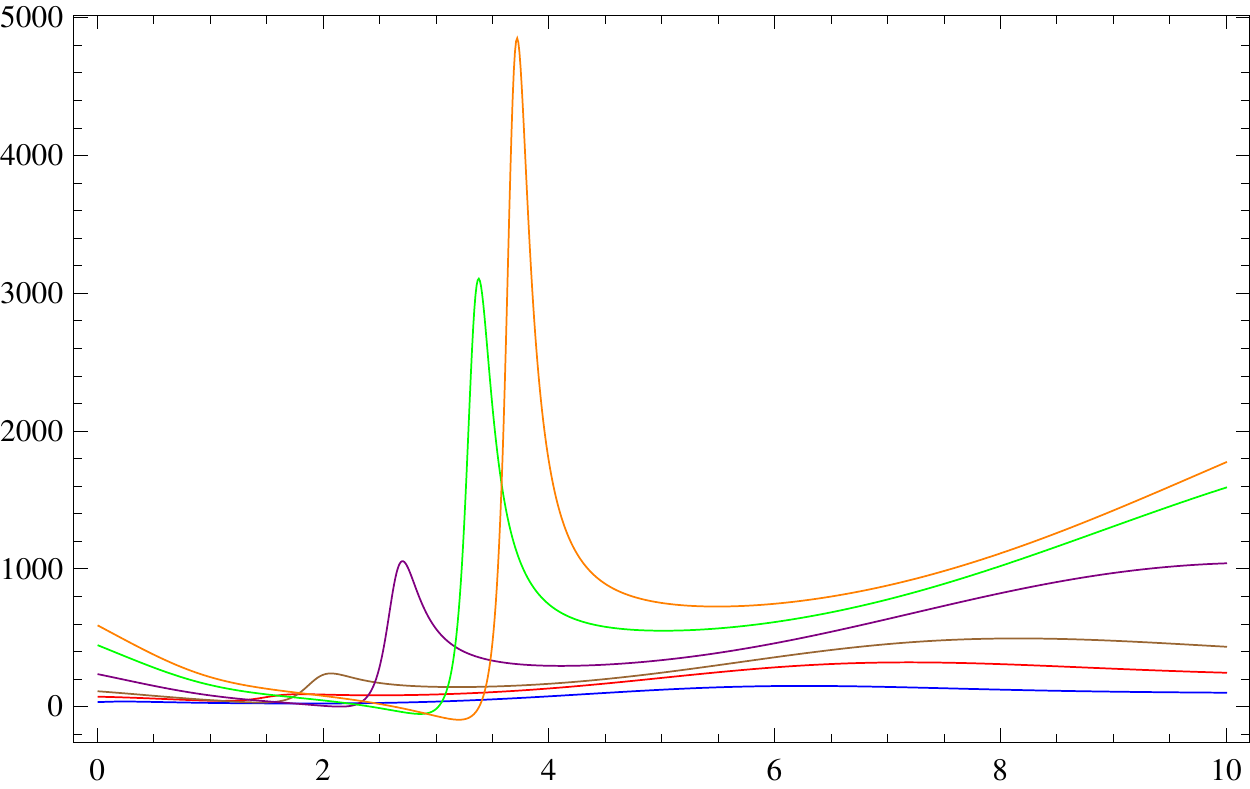}\put(-100,102){$\frac{{\kappa_5}^2}{\alpha^2 T^3} \re\left( {G_\p^x}^{m} \right)$}\put(6,10){$\frac{\omega}{2\pi T}$}}                
  \hfill
  \subfigure[]{\label{fig:rem30316}\includegraphics[width=0.45\textwidth]{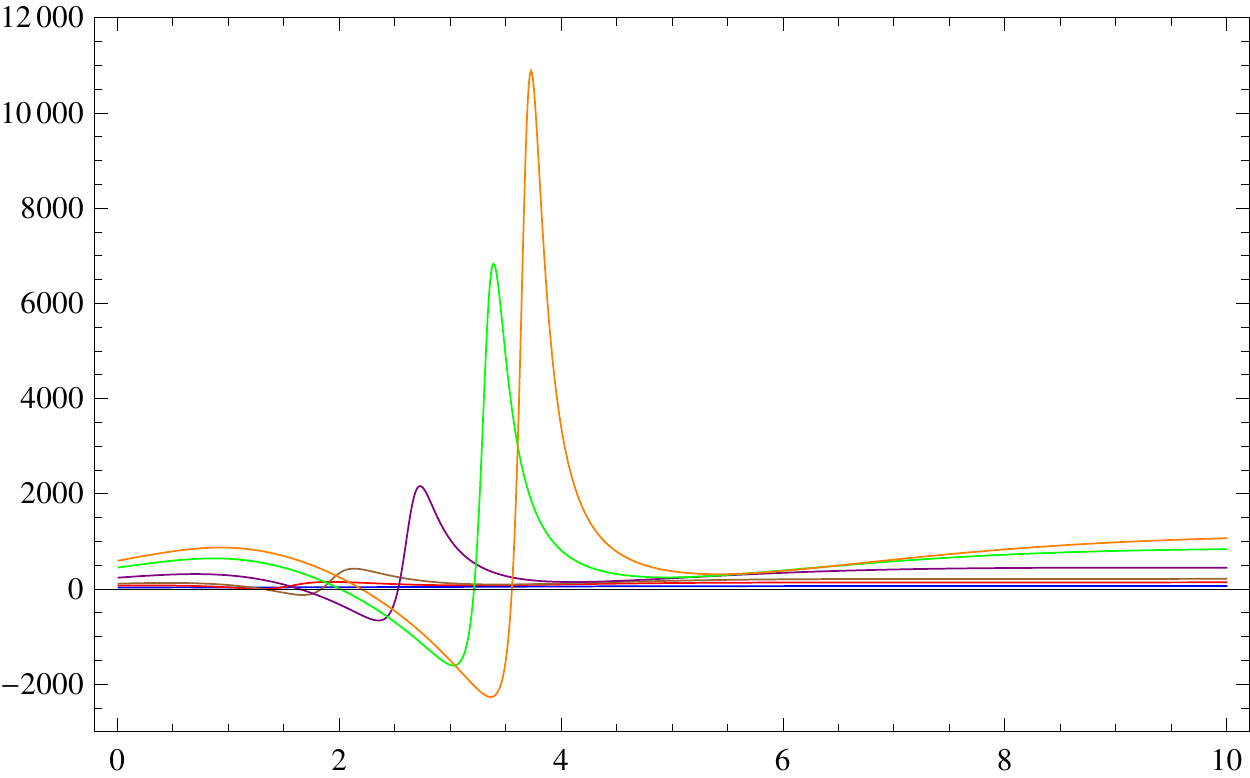}\put(-100,102){$\frac{{\kappa_5}^2}{\alpha^2 T^3} \re\left({G_\m^x}^{m}\right)$}\put(6,10){$\frac{\omega}{2\pi T}$}}                
  \hfill
  \subfigure[]{\label{fig:imp30316}\includegraphics[width=0.45\textwidth]{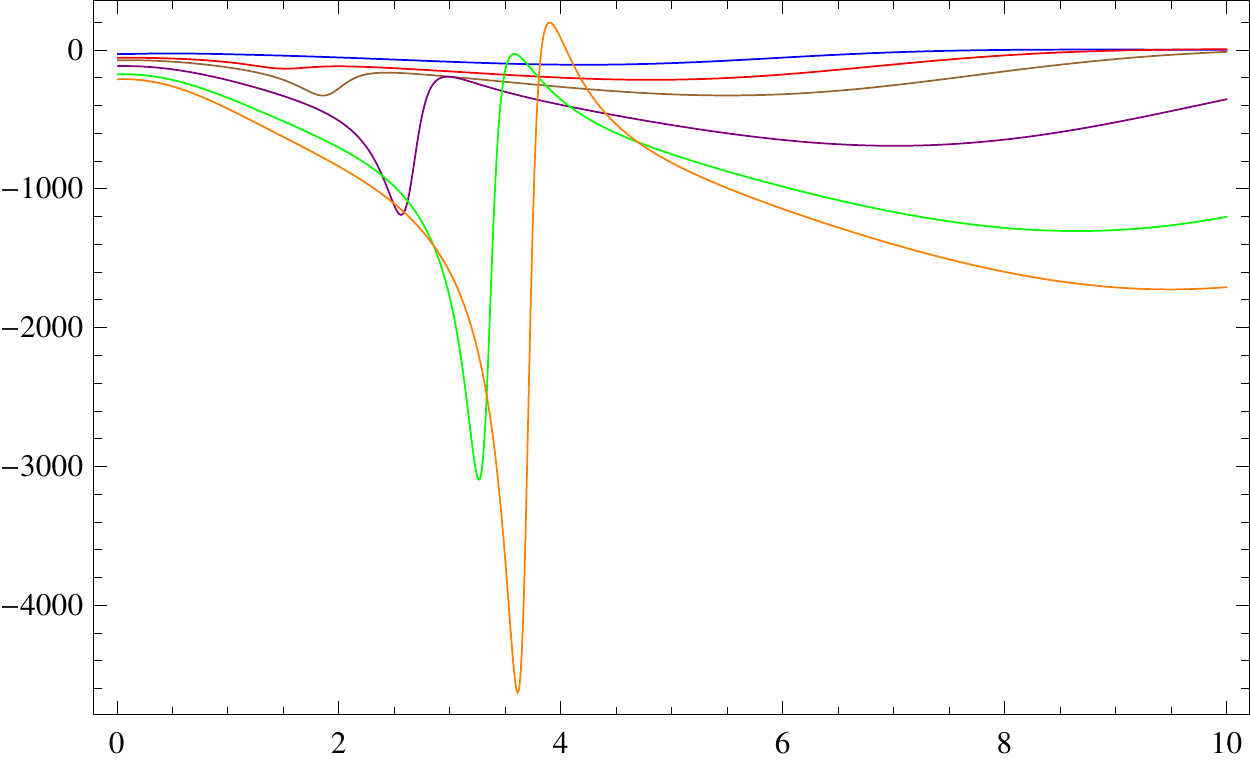}\put(-100,62){$\frac{{\kappa_5}^2}{\alpha^2 T^3} \im\left( {G_\p^x}^{m} \right)$}\put(-90,42){$=\frac{{\kappa_5}^2}{\alpha^2 T^3} \im\left( {G^{m}}^x_\p \right)$}\put(6,10){$\frac{\omega}{2\pi T}$}}                
  \hfill
  \subfigure[]{\label{fig:imm30316}\includegraphics[width=0.45\textwidth]{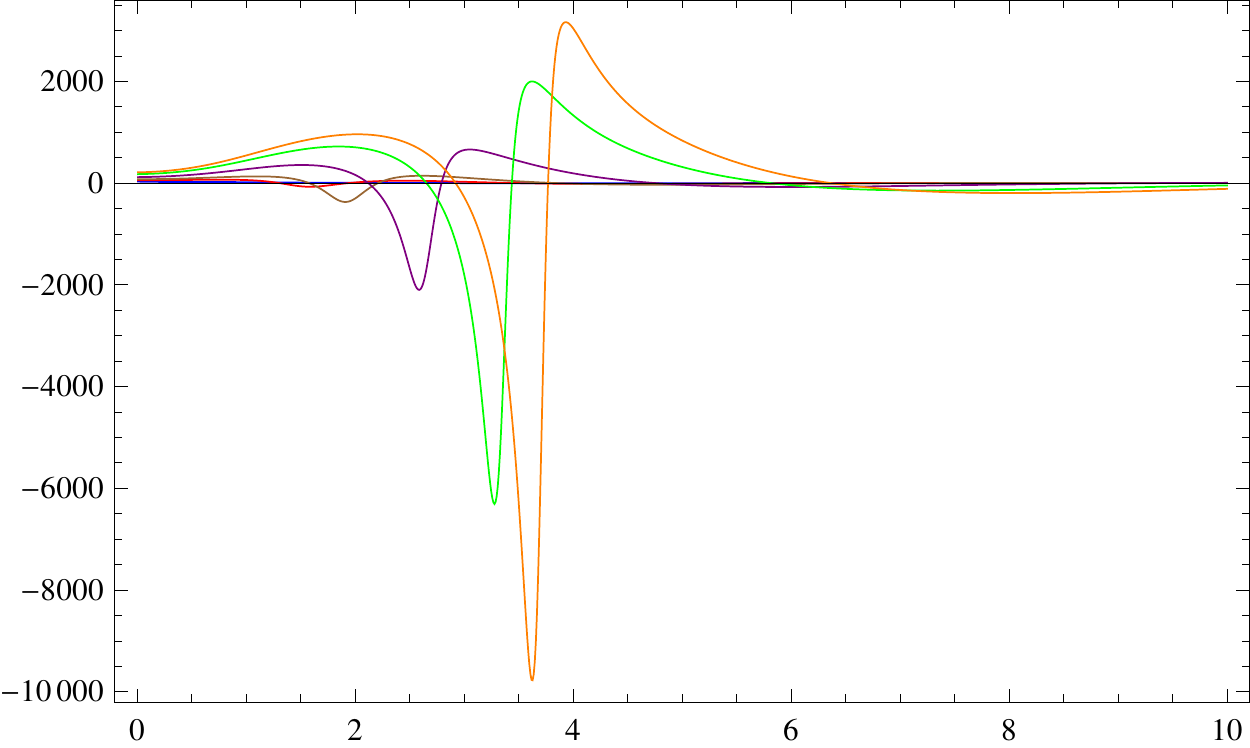}\put(-100,62){$\frac{{\kappa_5}^2}{\alpha^2 T^3} \im\left( {G_\m^x}^{m} \right)$}\put(-90,42){$=\frac{{\kappa_5}^2}{\alpha^2 T^3} \im\left( {G^{m}}^x_\m \right)$}\put(6,10){$\frac{\omega}{2\pi T}$}}
  \subfigure[]{\label{fig:rep3p0316}\includegraphics[width=0.45\textwidth]{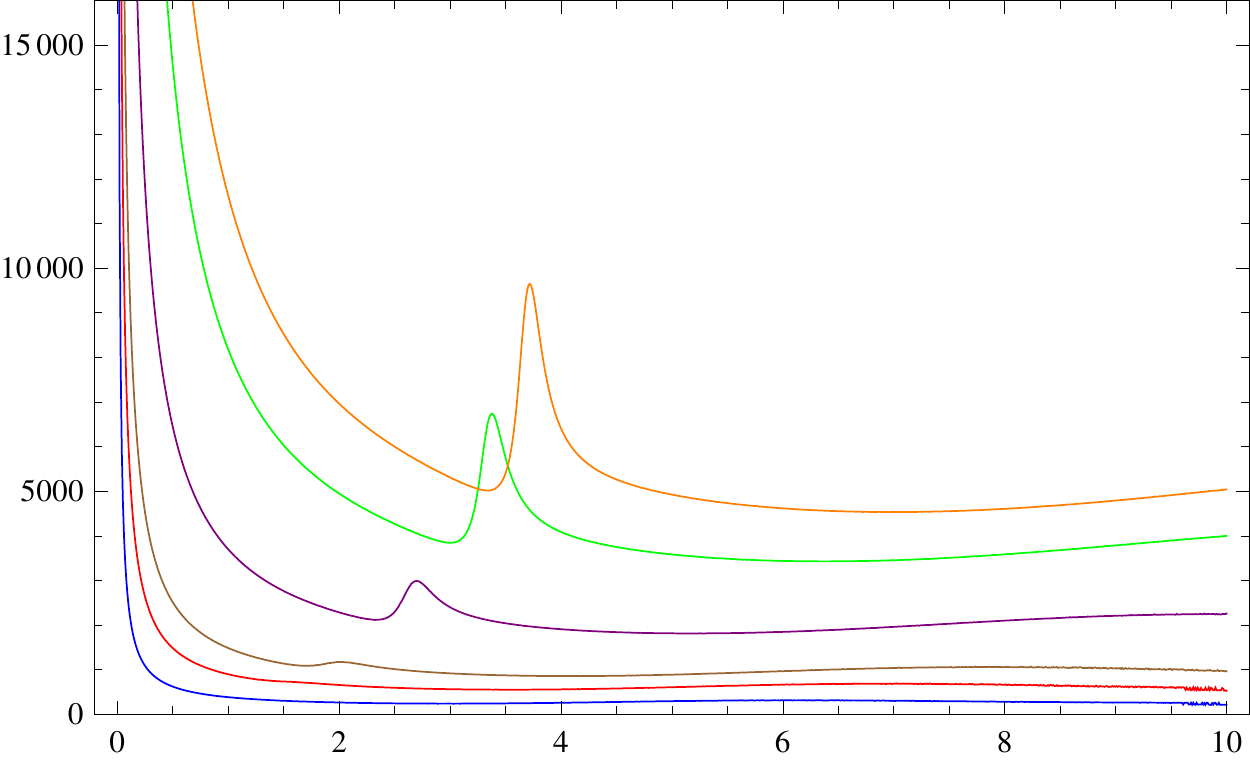}\put(-155,102){$\frac{{\kappa_5}^2}{\alpha^2 T^3} \re\left( {G^{m}}_\p^x \right)$}\put(6,10){$\frac{\omega}{2\pi T}$}}                
  \hfill
  \subfigure[]{\label{fig:re3m0316}\includegraphics[width=0.45\textwidth]{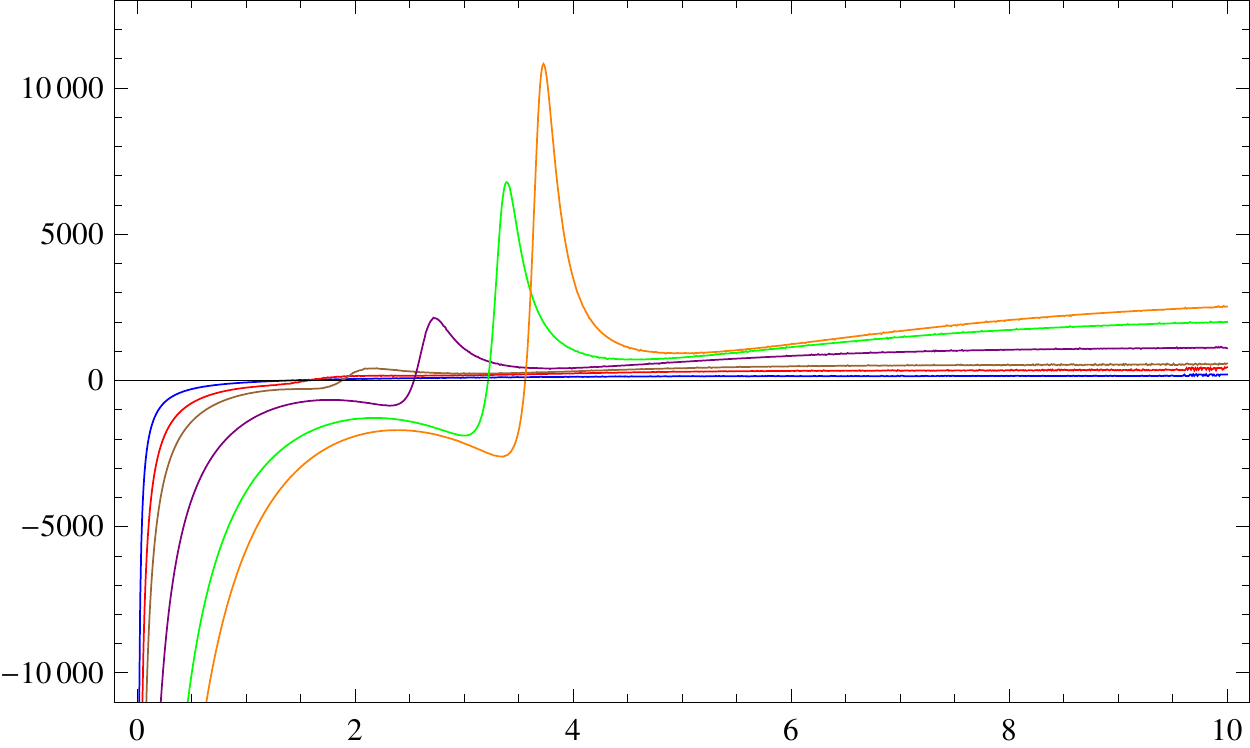}\put(-100,102){$\frac{{\kappa_5}^2}{\alpha^2 T^3} \re\left({G^{m}}_\m^x\right)$}\put(6,10){$\frac{\omega}{2\pi T}$}}                
  \hfill
  \caption{These plots show the real and imaginary part of the correlators ${G^x_\pm}^{m}$ versus the reduced frequency $\omega/(2\pi T)$ for $\alpha=0.316$ at different temperatures: blue $T=0.98T_c$, red $T=0.88T_c$, brown $T=0.78T_c$, purple $T=0.62T_c$, green $T=0.50T_c$, orange $T=0.46T_c$. We are not showing ${G^{m}}^x_\pm$ because their imaginary parts are identical. Their real parts, however, show a different low frequency behaviour.}
  \label{fig:hel0b2pmxx}
\end{figure}

In figure~\ref{fig:hel0b2pmxx} we plot the ${G^{m}}^x_\pm$, whose imaginary parts are identical to those of ${G^x_\pm}^{m}$, and whose real parts are similar, except for small frequencies compared to the temperature.

\begin{figure}[t]
\centering
\includegraphics[width=0.5\linewidth]{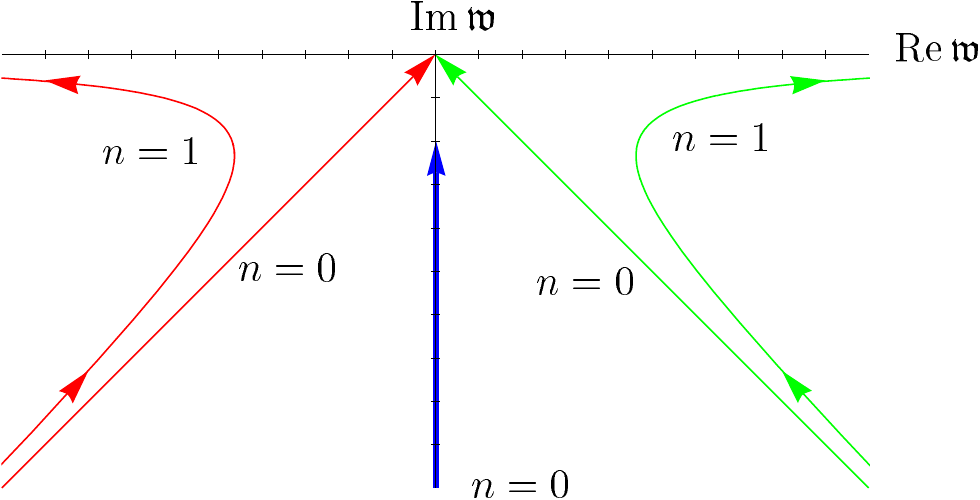}
\caption{This figure, taken from~\cite{Ammon:2008fc}, shows the different quasinormal modes in the D3/D7 system in the complex frequency plane. Here $\mathfrak{w}=\omega/(2\pi T)$. The red and green curves show the modes of the fluctuations which correspond to $\Phi_\pm$ in our setup and the blue curve corresponds to our $\Phi_4$. It is interesting to see that the backreaction and the bottom up approach we pursue in the system at hand behave in a very similar fashion as the D3/D7 probe setup. Note however, that there is one difference: due to the backreaction and consequently the rotational symmetry breaking in the superfluid phase, the $a^\pm_y$ decouple from the $\Phi_\pm$, contrary to what happens in the D3/D7 mode. Moreover, we only see the red and green modes in the $\Phi_\pm$ sector and not in the $a^\pm_y$ sector (see~\cite{Erdmenger:2011tj} for a treatment of this modes).}
\label{fig:polmov}
\end{figure}

Notice that many of the curves in figs.~\ref{fig:hel0b2ppmm}-\ref{fig:hel0b2pmxx} show a pole at $\omega=0$. To understand why, remember that the formation of $\langle J^x_1\rangle$ selects a preferred direction in flavor space, spontaneously breaking the $SO(3)$ and $U(1)_3$ symmetries. As a consequence of this, the $a^2_x$ field becomes one of the three massless Goldstone modes arising from the spontaneous symmetry breaking. This common pole reflects the formation of this Goldstone mode, since it is included in the fields $\Phi_\pm$, which are involved in all of the correlators presented here. When plotting the quasinormal modes in the complex frequency plane we also see this pole at the origin, so that for $T>T_c$ they asymptote to the origin of the frequency place $\omega=0$. In fact, although it is not apparent in figs.~\ref{fig:hel0b2pmmp} and \ref{fig:hel0b2pmxx}, the correlators vanish completely in the unbroken phase $T>T_c$, since in this limit the equations of motion of the bulk fields decouple.

Another common feature of our results is the appearance of a rich structure for the correlators in the broken phase, including the formation of a bump located on the same value of the frequency for all of them. These bumps come from higher quasinormal excitations. With decreasing temperature, they move in the direction of smaller values of the negative imaginary parts and larger real parts of the frequencies. Therefore they become more accentuated with decreasing temperature. Nevertheless the quasinormal modes stay in the lower half complex frequency plane for all the temperatures we were able to check numerically. We will leave it for future work to investigate their behavior at finite spatial momentum and zero temperature. Note that their behavior is very similar to the one found in the D3/D7 model in \cite{Ammon:2008fc} (see figure \ref{fig:polmov}). Following \cite{Ammon:2008fc}, these bumps may be interpreted as bound states, e.g. mesons. However, since we do not have a precise knowledge of the the field theory side, this interpretation should be treated with care. We cannot say much more at this stage, without having the exact formulation of the hydrodynamics dual to this gravitational setup.

The Green's functions $G^{x,x}_{\pm,\pm}$  (c.f. fig.~\ref{fig:hel0b2ppmm}) seem to have different asymptotic values, however this is just a consequence of the small frequency range displayed here. Actually, they do asymptote to the same value for all temperatures in the limit of large frequencies. However this veils the interesting details at low frequency, therefore we do not show it here. Nevertheless, the large frequency limit is proportional to $\omega^2$, in agreement with the underlying CFT.

Finally, note that the real parts of ${G^x_\pm}^{m}$ and ${G^m}^x_\pm$ are not symmetric to each other. In the latter one we see poles in the $\omega\to 0$ limit. This poles are due to the fact that in the ${G^m}^x_\pm$ case we are dividing by the boundary value of $\Phi_\pm$, which contains the (massless) Goldstone mode $a^2_x$ and therefore vanishes at $\omega=0$, whereas in the ${G^x_\pm}^{m}$ case we divide by the boundary value of $\Phi_3$, whose quasinormal mode is not located at the origin.

In the next subsection we look at $G^{m,m}$, the Green's function generated by $\Phi_3$.

\subsubsection{Transport Coefficient associated to a Normal Stress Difference}
\label{sec:new_viscosity_bound}

In the presence of anisotropy, besides the two shear viscosities $\eta_{yz}$ and $\eta_{xy}$, there are three other coefficients. But in a conformal fluid, two of them, $\zeta_x$ and $\zeta_y$, vanish due to the tracelessness condition of the energy-momentum tensor in conformal theories. The remaining nonzero component, $\lambda$, is related to the normal stress difference as discussed below.

In the $\omega \to 0$ limit the imaginary part of the two-point function of $\Phi_3$ asymptotes to a finite value different from zero, see fig.~\ref{fig:lambdaT}. We can relate this Green's function to the transport coefficient $\lambda$ (see appendix \ref{sec:General-Remarks-on}), \ie
\begin{equation}
\label{eq:kubo3}
\lambda=\lim_{\omega\to 0}\frac{3}{2\omega}\, \im\, G^{m,m}(\omega)\,. 
\end{equation}

As discussed in appendix \ref{sec:General-Remarks-on} there is no $\lambda$ in the isotropic phase. However, the corresponding transport coefficient in the unbroken phase is just the shear viscosity $\eta$ (compare equations \eqref{eq:actflucanisotropic} and \eqref{eq:actflucisotropic}). Therefore we can match $\lambda$ to $\eta$ at the phase transition. In the following we show that the Green's function of $\Phi_3$ in the isotropic case gives the correct value of $\eta$.

If we consider perturbations around a background with unbroken $SO(3)$ symmetry, \ie with zero $w(r)$, then the physical field $\Phi_3$ decouples from the other fields. Inserting the analytic solution of the $AdS$ Reissner-Nordstr\"om black hole,
\begin{equation}
\label{eq:adsrnmetric}
\dd s^2 = -N(r)\dd t^2 + \frac{1}{N(r)}\dd r^2 +r^2 \left(\dd x^2 +\dd y^2 + \dd z^2\right)\,,
\end{equation}
its equation of motion can simply be written as
\begin{equation}
\label{eq:phi3decoupled}
\frac{\omega^2 r^3}{N(r)}\Phi_3+\left[ r^3N(r)\Phi'_3 \right]' = 0\, ,
\end{equation}
where $N(r)=r^2-\frac{2m_0}{r^2}+\frac{2\alpha^2 q^2}{3r^4}$. This is the equation of motion of a minimally coupled scalar, thus we can apply the procedure developed in ~\cite{Iqbal:2008by} to derive the value of the corresponding transport coefficient. The relevant part of the boundary action in the unbroken case for this mode is 
\begin{equation}
S^{\text{on-shell}}_{\Phi_3} = \frac{r_h^4}{\kappa^2_5} \int \frac{\dd^4 k}{{(2\pi)}^4}\left[-\frac{r^5}{12}\Phi^*_3 (\omega,r) \partial_r \Phi_3(\omega,r)+ B_{33}(k,r)\Phi^*_3(\omega,r)\Phi_3(\omega,r)  \right]_{r=r_b}
\end{equation}
with $B_{33}(\omega,r)=\frac{1}{96} \left[-4\omega^2 r^2+2\omega^4 \log\left(\frac{r_h}{r}\right)-19m^b_0\right]$ and $m^b_0$ defined as in the $AdS$ Reissner-Nordstr\"om solution. Using the result of~\cite{Iqbal:2008by} and the Kubo formula~\eqref{eq:kubo3}\footnote{Just interchange $\eta$ with $\lambda$, since $\lambda$ is normalized in a way to match $\eta$ at the phase transition.}, we obtain for this particular case the viscosity coefficient
\begin{equation}
\frac{\eta}{s} = \frac{1}{4\pi}\, ,
\end{equation}
where $s$ is the entropy density. This is the expected value for the shear viscosity in the isotropic phase.

\begin{figure}[t]
\centering
\includegraphics[width=0.9\linewidth]{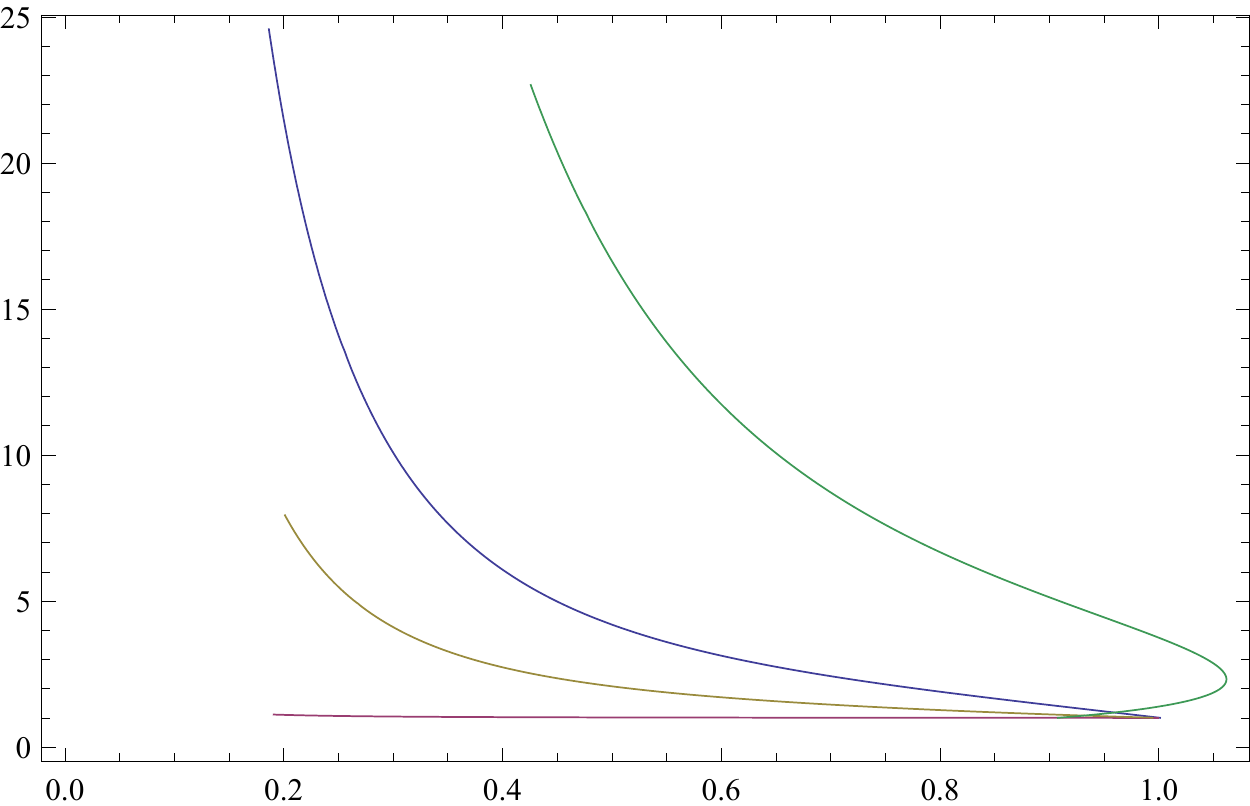}
\put(-190,210){$4\pi\frac{\lambda}{s}=\left.\frac{3}{2}\frac{1}{\omega\, s} \im\left(G^{m,m}\right) \right|_{\omega\to 0}$}\put(6,10){$\frac{T}{T_c}$}
\caption{We plot $\frac{\lambda}{s}$ over the temperature $T/T_c$ for $\alpha=0.032$ (red), $\alpha=0.224$ (yellow), $\alpha=0.316$ (blue) and $\alpha=0.447$ (green). Note that $\alpha=0.447>\alpha_c$ and therefore the phase transition is first order leading to multiple values near the phase transition. All curves tend to $1/(4\pi)$ at $T_c$, since in the unbroken phase $\lambda$ corresponds to the isotropic shear viscosity $\eta$.}
\label{fig:lambdaT}
\end{figure}

As stated above $\lambda$ is the transport coefficient related to the normal stress difference, that is, the difference of the diagonal components of the stress tensor, $\langle\frac{1}{2}\left(T^{xx}-2T^{\perp\perp}\right)\rangle$. Generically, whenever a incompressible material is squeezed between two surfaces by applying normal stresses, it will tend to expand along the directions parallel to these surfaces (e.g. normal radial squeezing on a cylinder is expected to lengthen its shape on the vertical direction).

In our setup, a positive strain difference $\left(\Phi_3\right)^b_0 \sim \left(h_{xx}-\frac{1}{2}\left(h_{yy}+h_{zz}\right)\right)^b_0$\footnote{Note that we do not differentiate between $h_{yy}$ and $h_{zz}$, since both are related via the $SO(2)$ symmetry. Therefore, we get $\frac{1}{2}\left(h_{yy}+h_{zz}\right)=h_{\perp\perp}$.} corresponds to a deformation that enlarges a direction $\vec n$, which results in the formation of a squeezing normal stress in the perpendicular direction to $\vec n$, which translates into a positive $\langle T_{xx}-2T_{\perp\perp} \rangle$. Therefore, one expects to have a positive coefficient $\lambda$. An experimental consequence of this behavior of the coefficient would be that, if the fluid is inside a recipient and a spinning rod is placed in it, the fluid would be expelled outwards more noticeably in the superfluid phase, climbing up the walls of the recipient.


\section{Conclusion}
\label{sec:Conclusion} 
We have considered a holographic p-wave superfluid within $SU(2)$ Einstein-Yang-Mills theory, in which the formation of the condensate spontaneously breaks the rotation symmetry, selecting a preferred direction $x$ and keeping transverse isotropy in the $y$, $z$-plane. This remaining $SO(2)$ symmetry group allows us to classify the perturbations about equilibrium according to their transformation properties, into three different helicity sectors. In this paper we have focused on the helicity zero states and studied their transport properties.
Due to $\mathbb{Z}_2$ parity, the helicity zero sector splits into two blocks. In the first block, we obtain the thermoelectric conductivity in the direction parallel to the condensate, whereas the second block allows us to study the piezoelectric effect and transport properties related to the normal stress difference. These are interesting new phenomena which are due to the anisotropy in our system.

We see that the thermoelectric conductivity displays some differences with respect to the transverse case (\ie the helicity one fluctuations), despite being qualitatively similar. In particular, the temperature suppression in the broken phase is much lighter in the presented longitudinal case for temperatures above around $0.5T_c$. For temperatures below the temperature suppression of the real part a finite small frequencies increases dramatically and we see a pole at $\omega=0$. This is due to a quasinormal mode traveling up the imaginary axis of the complex frequency plane. At around $0.5T_c$ this mode arrives at the origin and stays there.

On the other hand, in the parity even block we find a behavior reminiscent of the piezoelectric effect. Furthermore we see bumps in the correlators of this sector, which seem to be related to the generation of bound states.

In the zero frequency limit, we find a non-zero value for the two-point function of the diagonal metric fluctuations, which is related by a Kubo formula to a component of the viscosity tensor, denoted by $\lambda$. Since $\lambda$ has the same dimensions as a shear viscosity we investigate its behavior by taking its ratio with respect to the entropy density $s$. We find that in the broken case $\lambda/s$ is temperature dependent, whereas in the unbroken phase it turns into the isotropic shear viscosity $\eta/s=1/(4\pi)$ for all temperatures $T>T_c$. The ratio $\lambda/s$ does not fall below the $1/(4\pi)$ for all temperatures and all values of the backreaction parameter $\alpha$. The physical interpretation of this coefficient is the effect that an anisotropic strain has over the normal stress difference, $\frac{1}{2}\langle T_{xx}-2T_{\perp\perp}\rangle$.

We have determined the coefficients associated to these effects for generic values of the frequency and the temperature. Our results are valid as an effective description of the transport properties near the critical temperature $T_c$, where scale invariance is approached and simple models of AdS/CFT can be applied.

For further progress, a detailed analysis of the hydrodynamics of anisotropic superfluids is desirable to give a further interpretation to our study. In addition, it would be interesting to perform an analysis at finite spatial momentum, which would allow us to investigate the dispersion relations of the normalizable modes~\cite{Buchbinder:2008nf} and to check if there are instabilities similar to the ones found in~\cite{Nakamura:2009tf,Ooguri:2010xs}.

\section*{Acknowledgements}
We are grateful to Patrick Kerner and Amos Yarom for discussions. DF acknowledges financial support from 2009-SGR-168, MEC FPA2010-20807-C02-01, MEC FPA2010-20807-C02-02, CPAN CSD2007-00042 Consolider-Ingenio 2010, and ERC StG 306605 HoloLHC. DF is thankful to D. Mateos, J. Casalderrey-Solana, A. Buchel, C. Manuel, J. M. Pons, J. Tarr\'io and M.A. Valle for discussions, as well as to the Max Planck Institute for Physics in Munich and to Perimeter Institute for Theoretical Physics for hospitality during the initial and final stages of this work, respectively.  This work was supported in part by  {\it The Cluster of Excellence for Fundamental Physics - Origin and Structure of the Universe}.


\begin{appendix}

\section{Holographic Renormalization}
\addtocontents{toc}{\protect\setcounter{tocdepth}{1}}
\label{sec:Holographic-Renormalization}
The boundary part $S_{ct}$ of the action (\ref{eq:action}) does not have any influence on the equations of motion, but it must ensure that the action is finite on-shell. It includes the Gibbons-Hawking boundary term and some additional terms that will constitute the counterterm action $S_{ct}$, needed to cancel out any divergences that may appear. Thus, the full action is written as
  \begin{equation}
  \label{actioncomplete}
    S = \frac{1}{2\k_5^2}\, \int\!\dd^5x\,\sqrt{-g} \, \left[ R -\Lambda - \frac{\alpha^2}{2} \, F^a_{MN}
    F^{aMN} \right] + \frac{1}{\k_5^2}\, \int\!\dd^4x \,\sqrt{-\gamma} \, K +S_{ct} ,
  \end{equation}
where $K$ is the trace of the extrinsic curvature.

We will follow the lead of the references \cite{Skenderis:2002wp,Sahoo:2010sp} to perform the holographic renormalization and obtain the counterterm action.

\subsection{Asymptotic Behavior}
\label{sec:Asymptotic-Behavior}
  In this section we look at the behavior of the fluctuation fields $\{ F(r) \}$ at the horizon and at the boundary. Eventually we will want to calculate real-time retarded Green's functions \cite{Son:2002sd,Son:2000xc}, therefore at the horizon, besides regularity\footnote{The condition $\phi(r_H)=0$ at the horizon guarantees regularity. Even with all fluctuations switched on, there is no need for any further constraint.}, we have to fulfill the incoming boundary condition. For this purpose the ansatz we plug in for the behavior of the fields near the horizon is
  \begin{equation}
  \label{eq:expfluchorpwave}
   F(r)\big|_{r\rightarrow r_H} =\epsilon_h^{\beta} \sum_{i\geq0} F_i^h \epsilon_h^i \,,
  \end{equation}
where $\epsilon_h = {r / r_h} - 1$, into the equations of motion of the fluctuation fields. It turns out that, as expected, we obtain two possibilities for $\beta$, namely
  \begin{equation}
   \beta = \pm \ii \frac{\omega}{4 \pi T},
  \end{equation}
  with $T$ being the temperature defined in equation ~\eqref{eq:temperature}. As said before, we choose the solution with the ``$-$" sign which corresponds to the incoming boundary condition. The other solution represents the outgoing boundary condition.
  
  On the other hand, our ansatz at the boundary is similar to the one used for the background calculation in section~\ref{sec:Holographic-Setup-and}. However, here we have to add a logarithmic term to get a consistent solution (c.f. \cite{Skenderis:2002wp}). Therefore we use
  \begin{equation}
  \label{eq:expflucbdy}
   F(r)\big|_{r\rightarrow r_\text{bdy}} =\sum_{i\ge0}\left(F_i^b+\frac{1}{2}\hat F_i^b\ln\epsilon_b\right)\epsilon_b^i \, ,
  \end{equation}
where $\epsilon_b=\left(r_h / r \right)^2$ is the expansion parameter.

Let us now use the above expansions for the helicity zero states (the expressions for the helicity one and two states can be found in~\cite{Erdmenger:2011tj}). In this case, the equations of motion for the fluctuation fields can be distributed into two blocks. In the first block, we have 5 independent expansion coefficients at the boundary (8 free parameters from the 4 second order differential equations minus 3 free parameters due to the constraints). We choose them to be $\left(\xi_{tx}\right)_0^b,\, \left(a_t^1\right)_0^b,\, \left(a_t^2\right)_0^b,\, \left(a_x^3\right)_0^b$ and $\left(a_x^3\right)_1^b$. At the horizon, we already halved the independent parameters by choosing the incoming boundary condition. From the remaining 4 parameters, we can get rid of 3 by using the constraint equations. Therefore, we are left with just one free parameter at the horizon.
  
  We can perform similar considerations for the second block. Here we have also 3 constraints, but we are dealing with 6 fields, each with its corresponding second order differential equation. Therefore at the boundary we have $12-3=9$ independent parameters, namely $\left(\xi_y\right)_0^b,\, \left(\xi_x\right)_0^b,\, \left(\xi_t\right)_0^b,\, \left(a_x^1\right)_0^b,\, \left(a_x^2\right)_0^b,\, \left(a_t^3\right)_0^b,\, \left(a_x^1\right)_1^b,\, \left(a_x^2\right)_1^b$ and $\left(\xi_y\right)_2^b$. At the horizon, as before, we already fixed 6 free parameters by choosing the incoming boundary condition. There are $6-3=3$ free parameters that give a fully determined system.
  
  Now we will state the first few non-vanishing terms of the expansion at the boundary of the different fields, because we will need them later on to determine divergences in the on-shell action and to calculate the Green's functions. The explicit form of these expansions is
  \begin{equation}
   \label{eq:expflucbdybl2}
   \begin{split}
    \xi_y =& \left(\xi_y\right)_0^b+\omega ^2\frac{\left(\xi_y\right)_0^b-\left(\xi_x\right)_0^b}{12}\epsilon_b+\left(\left(\xi_y\right)_2^b-\frac{1}{96} \omega^4 \left[\left(\xi_y\right)_0^b-\left(\xi_x\right)_0^b \right] \log\epsilon_b \right)\epsilon_b^2 + \mathcal{O}(\epsilon_b^3)\,,\\
    \xi_x =& \left(\xi_x\right)_0^b+\omega^2\frac{\left(\xi_x\right)_0^b-\left(\xi_y\right)_0^b}{6}\epsilon_b+\left(\ldots-\frac{1}{48} \omega^4 \left[\left(\xi_x\right)_0^b-\left(\xi_y\right)_0^b \right] \log\epsilon_b \right)\epsilon_b^2+\mathcal{O}(\epsilon_b^3)\,,\\
    \xi^t =& \left(\xi_t\right)_0^b+\omega^2\frac{2 \left(\xi_y\right)_0^b+\left(\xi_x\right)_0^b}{6}\epsilon_b
+\mathcal{O}(\epsilon_b^2)\,,\\
    a_x^1 =& \left(a_x^1\right)_0^b+\left( \left(a_x^1\right)_1^b-\frac{1}{4} \left[ \left(\mu^2+\omega^2\right)\left(a_x^1\right)_0^b-2 \ii \mu  \omega \left(a_x^2\right)_0^b\right] \log\epsilon_b \right)\epsilon_b + \mathcal{O}(\epsilon_b^2)\,,\\
    a_x^2 =& \left(a_x^2\right)_0^b+\left( \left(a_x^2\right)_1^b-\frac{1}{4} \left[ \left(\mu^2+\omega^2\right)\left(a_x^2\right)_0^b+2 \ii \mu  \omega \left(a_x^1\right)_0^b\right] \log\epsilon_b \right)\epsilon_b + \mathcal{O}(\epsilon_b^2)\,,\\
    a_t^3 =& \left(a_t^3\right)_0^b+ \left( -\frac{\ii}{\omega}\left(a_x^2\right)_0^b w^b_1-\frac{1}{2} \left[2 \left(\xi_y\right)_0^b+\left(\xi_x\right)_0^b-\left(\xi_t\right)_0^b \right] \phi^b_1 \right)\epsilon_b + \mathcal{O}(\epsilon_b^2)\,;
   \end{split}
  \end{equation}
for the fields of the second block, and
  \begin{equation}
   \label{eq:expflucbdybl1}
   \begin{split}
    a_t^1 =&\left(a_t^1\right)_0^b+\frac{\left[\omega^2 \left(\xi_{tx}\right)_0^b - \left(a_x^3\right)_0^b\mu- \left(\xi_{tx}\right)_0^b\mu^2\right]  w^b_1- \left[\left(a_t^1\right)_0^b \mu+\ii\omega \left(a_t^2\right)_0^b \right] \phi^b_1 }{\omega^2-\mu^2}\epsilon_b + \mathcal{O}(\epsilon_b^2)\,,\\
    a_t^2 =& \left(a_t^2\right)_0^b+\frac{ -\ii\omega \left(a_x^3\right)_0^b w^b_1+\left[\left(a_t^2\right)_0^b \mu-\ii\omega\left(a_t^1\right)_0^b \right]\phi^b_1 }{\mu ^2-\omega ^2}\epsilon_b+\frac{\ii \omega \left(a_x^3\right)_0^b w^b_1}{8}\epsilon_b^2+\mathcal{O}(\epsilon_b^3)\,,\\
    a_x^3 =& \left(a_x^3\right)_0^b+\left(\left(a_x^3\right)_1^b-\frac{1}{4} \omega^2 \left(a_x^3\right)_0^b \log \epsilon_b\right) \epsilon_b+\mathcal{O}(\epsilon_b^2)\,,\\
    \xi_{tx} =& \left(\xi_{tx}\right)_0^b - \alpha^2 \left(a_x^3\right)_0^b \phi^b_1 \epsilon_b+ \mathcal{O}(\epsilon_b^3)\,;
   \end{split}
  \end{equation}
for the first block. Note that $\mu \equiv \phi^b_0$, $\phi^b_1$ and $w^b_1$ are the expansion coefficients of $\phi(r)$ and $w(r)$ at the boundary.

  We do not state the expansions at the horizon, since the explicit form is quite lengthy and does not provide additional information to equation (\ref{eq:expfluchorpwave}).

\subsection{Counterterms}
\label{sec:Counterterms}

By plugging the expansions~(\ref{eq:expflucbdybl1}) into~\eqref{eq:hel0actionbl1} and \eqref{eq:hel0actionbl2}, we obtain the non-renormalized on-shell action, $S_\text{on-shell} = \frac{1}{\kappa_5^2}\int\frac{\dd^4 k}{(2\pi)^4}\ \call_{r_b}$, where the integrand $\call_{r_b}$ is written in terms of the free parameters of the previous expansions as
  \begin{equation}
   \label{eq:actbdryflucpwave}
   \begin{split}
    \frac{\call_{r_b}}{r_h^4}=& \frac{\alpha^2 \mu \phi^b_1}{\omega^2-\mu^2} {\left(a_t^1\right)_0^b}^2 + \frac{\alpha^2 \mu \phi^b_1}{\omega^2-\mu^2}{\left(a_t^2\right)_0^b}^2 - \frac{\alpha^2 \omega^2}{4}\left(1+ \log\epsilon_b \right){\left(a_x^3\right)_0^b}^2 \\ &+\left(\frac{3}{2 \epsilon_b^2}-6 f^b_2\right) {\left(\xi_{tx}\right)_0^b}^2 + \frac{2 i \alpha^2 \omega  \phi^b_1}{\omega^2-\mu^2}\left(a_t^1\right)_0^b \left(a_t^2\right)_0^b + \frac{\alpha^2 \mu  w^b_1}{\omega^2-\mu^2}\left(a_t^1\right)_0^b \left(a_x^3\right)_0^b \\
&- \frac{i \alpha ^2 \omega  w^b_1}{\omega^2-\mu^2}\left(a_t^2\right)_0^b \left(a_x^3\right)_0^b +\alpha ^2 \left(a_x^3\right)_0^b \left(a_x^3\right)_1^b - \alpha ^2 \phi^b_1 \left(a_x^3\right)_0^b \left(\xi_{tx}\right)_0^b \\
& + \left(-\frac{3}{8 \epsilon_b^2}-\frac{\omega^2}{8 \epsilon_b} - \frac{7 \omega^4}{192}-f^b_2+\frac{5 m^b_0}{4}+\frac{\omega^4}{48} \log\epsilon_b \right){\left(\xi_x\right)_0^b}^2 +\left(-\frac{3}{8 \epsilon_b^2}+\frac{3 m^b_0}{4}\right){\left(\xi_t\right)_0^b}^2\\
&+ \left(\frac{\omega^2}{8 \epsilon_b}-\frac{13 \omega^4}{192}+\frac{\omega^4}{48} \log\epsilon_b \right) {\left(\xi_y\right)_0^b}^2
-\frac{\alpha^2 \left(\mu^2+\omega^2\right)}{4} \left(1+\log\epsilon_b\right) \left[{\left(a_x^1\right)_0^b}^2+{\left(a_x^2\right)_0^b}^2\right] \\
&+2 \left(\xi_y\right)_2^b \left[ \left(\xi_x\right)_0^b-\left(\xi_y\right)_0^b \right]
+\left(\frac{3}{2 \epsilon_b^2}+\frac{3 \omega^2}{4 \epsilon_b}+\frac{\omega^4}{96}+f^b_2+m^b_0 - \frac{\omega^4}{24} \log\epsilon_b\right)\left(\xi_y\right)_0^b \left(\xi_x\right)_0^b \\
&+\left(\frac{3}{2 \epsilon_b^2}-\frac{\omega^2}{4 \epsilon_b}-f^b_2-4 m^b_0\right)\left(\xi_y\right)_0^b \left(\xi_t\right)_0^b+\left(\frac{3}{4 \epsilon_b^2}-\frac{\omega^2}{8 \epsilon_b}+f^b_2-2 m^b_0\right)\left(\xi_x\right)_0^b \left(\xi_t\right)_0^b \\
&+\frac{\alpha^2 w^b_1}{2} \left(a_x^1\right)_0^b \left[ 2\left(\xi_y\right)_0^b +\left(\xi_x\right)_0^b -\left(\xi_t\right)_0^b \right] - \frac{i \alpha^2 \mu  w^b_1}{\omega}\left(a_x^2\right)_0^b \left[\left(\xi_x\right)_0^b -\left(\xi_t\right)_0^b\right]+\alpha^2 \left(a_x^1\right)_0^b  \left(a_x^1\right)_1^b \\
&+\left(1 +i \alpha^2 \mu  \omega  \log\epsilon_b\right) \left(a_x^1\right)_0^b \left(a_x^2\right)_0^b +\alpha^2  \left(a_x^2\right)_0^b \left(a_x^2\right)_1^b
-\frac{i \alpha^2 w^b_1}{\omega} \left(a_x^2\right)_0^b \left(a_t^3\right)_0^b \bigg|_{r=r_{\text{bdy}}}\, .
   \end{split}
  \end{equation}
This is evaluated at the boundary, where $\epsilon_b = (r_h/r_{\text{bdy}})^2 = 0$, so any higher order terms vanish. And we have changed to momentum space, so that in each product of expansion parameters in this expression, the first one has always to be understood as evaluated on $-\omega$ and the second on $\omega$, e.g. on the first term we have ${\big(a_t^1\big)_0^b}(-\omega){\big(a_t^1\big)_0^b}(\omega)$. Therefore, note that the order in which they are multiplied matters.

The terms that have to be considered for the counterterms are the ones in (\ref{eq:actbdryflucpwave}) with explicit $\epsilon_b$ dependence, since those are the ones responsible for the divergences
  \begin{equation}
   \begin{split}
    \frac{\call^{\text{Div}}_{r_b}}{r_h^4}=& \frac{1}{\epsilon_b^2} \left[ \frac{3}{2}{\left(\xi_{tx}\right)_0^b}^2 +\frac{3}{8} \left( -{\left(\xi_x\right)_0^b}^2-{\left(\xi_t\right)_0^b}^2  + 4\left(\xi_y\right)_0^b\left(\xi_x\right)_0^b + 4\left(\xi_y\right)_0^b\left(\xi_t\right)_0^b + 2\left(\xi_x\right)_0^b\left(\xi_t\right)_0^b \right) \right] \\
    +& \frac{1}{\epsilon_b} \frac{\omega^2}{8}\left[ {\left(\xi_y\right)_0^b}^2 - {\left(\xi_x\right)_0^b}^2 +6\left(\xi_y\right)_0^b\left(\xi_x\right)_0^b -2\left(\xi_y\right)_0^b\left(\xi_t\right)_0^b -\left(\xi_x\right)_0^b\left(\xi_t\right)_0^b \right] \\
    +& \log\epsilon_b \left[-\frac{\alpha^2 \omega^2}{4} {\left(a_x^3\right)_0^b}^2 +\frac{\omega^4}{48} \left( {\left(\xi_y\right)_0^b}^2+{\left(\xi_x\right)_0^b}^2 -2\left(\xi_y\right)_0^b\left(\xi_x\right)_0^b \right) \right. \\
    &\left. \hspace{20mm} -\frac{\alpha^2 \left( \mu^2-\omega^2 \right)}{4} \left( {\left(a_x^1\right)_0^b}^2 + {\left(a_x^2\right)_0^b}^2 \right) +\ii\alpha^2 \mu \omega \left(a_x^1\right)_0^b \left(a_x^2\right)_0^b \right] \, . \nonumber
   \end{split}
  \end{equation}
For the construction of the counterterms, first we need to define the induced metric $\gamma_{\mu\nu}$ on the $r=r_{\text{bdy}}$ plane,
    \begin{equation}
     \gamma_{\mu\nu} = \frac{\partial x^M}{\partial \tilde{x}^\mu}\frac{\partial x^N}{\partial \tilde{x}^\nu}g_{MN}(r)\bigg|_{r=r_\text{bdy}},
    \end{equation}
    resulting in
    \begin{equation}
     \dd s^2_{r_\text{bdy}} = -N(r_{\text{bdy}})\sigma(r_\text{bdy})^2 \dd t^2 + \frac{r_\text{bdy}^2}{f(r_\text{bdy})^4}\dd x^2 + r_\text{bdy}^2 f(r_\text{bdy})^2(\dd y^2 + \dd z^2).
    \end{equation}
    Note that the expansion of the determinant of the induced metric for $r \gg 1$ is divergent and is given by
    \begin{equation}
     \begin{split}
      \sqrt{-\gamma}\bigg|_{r \gg 1} &= r^4 \left[ \frac{1}{2} {\left(\xi_{tx}\right)_0^b}^2 - \frac{1}{8} \left({\left(\xi_x\right)_0^b}^2+{\left(\xi_t\right)_0^b}^2 -4 \left(\xi_y\right)_0^b \left(\xi_x\right)_0^b -4 \left(\xi_y\right)_0^b\left(\xi_t\right)_0^b -2 \left(\xi_x\right)_0^b \left(\xi_t\right)_0^b\right) \right] \\
&+\frac{\omega ^2 r^2}{24} \left[ 2 {\left(\xi_y\right)_0^b}^2 -{\left(\xi_x\right)_0^b}^2 +8 \left(\xi_y\right)_0^b \left(\xi_x\right)_0^b -2 \left(\xi_y\right)_0^b \left(\xi_t\right)_0^b -\left(\xi_x\right)_0^b \left(\xi_t\right)_0^b \right] \\
&+\frac{\omega^4}{48}\log \frac{1}{r} \left[ {\left(\xi_y\right)_0^b}^2 +{\left(\xi_x\right)_0^b}^2 -2 \left(\xi_y\right)_0^b \left(\xi_x\right)_0^b \right] \\
&+\left(\xi_y\right)_2^b \left(\left(\xi_x\right)_0^b-\left(\xi_y\right)_0^b \right)-\frac{\omega^4}{288} \left(7 {\left(\xi_y\right)_0^b}^2+4 {\left(\xi_x\right)_0^b}^2-2 \left(\xi_y\right)_0^b \left(\xi_x\right)_0^b\right) \\
&- 2f^b_2 {\left(\xi_{tx}\right)_0^b}^2- f^b_2 \left({\left(\xi_x\right)_0^b}^2-\left(\xi_y\right)_0^b \left(\xi_x\right)_0^b +\left(\xi_y\right)_0^b \left(\xi_t\right)_0^b -\left(\xi_x\right)_0^b \left(\xi_t\right)_0^b \right) \\
&+\frac{m^b_0}{2} {\left(\xi_{tx}\right)_0^b}^2+\frac{m^b_0}{8} \left(5 {\left(\xi_x\right)_0^b}^2 +{\left(\xi_t\right)_0^b}^2 +4 \left(\xi_y\right)_0^b \left(\xi_x\right)_0^b -12 \left(\xi_y\right)_0^b \left(\xi_t\right)_0^b -6 \left(\xi_x\right)_0^b \left(\xi_t\right)_0^b \right) \\
&- \alpha^2 \phi^b_1 \left(a_x^3\right)_0^b \left(\xi_{tx}\right)_0^b +\frac{\alpha^2 w^b_1}{2}\left(\left(\xi_x\right)_0^b-\left(\xi_t\right)_0^b\right)\left( \left(a_x^1\right)_0^b-\frac{\ii \mu }{\omega }  \left(a_x^2\right)_0^b\right) \, .
     \end{split}
    \end{equation}
We will use these divergences to cancel out the ones we find in the non-renormalized action, together with other counterterms that have to be considered. It is not necessary to rigorously derive the covariant counterterms here in this work. By looking at the ones that B. Sahoo and H.-U. Yee calculated in \cite{Sahoo:2010sp}, we get an idea of how they should look like; namely, some combinations of $R[\gamma],\, R_{\mu\nu}[\gamma]$ and $F^a_{\mu\nu}$ (i.e. the Ricci scalar and Ricci tensor on the induced surface, and the field strength tensor on that surface). Possible covariant combinations of the three terms are $\sqrt{-\gamma}$, $\sqrt{-\gamma}R[\gamma]$, $\sqrt{-\gamma}R[\gamma]^2$, $\sqrt{-\gamma}R^{\mu\nu}[\gamma]R_{\mu\nu}[\gamma]$ and $\sqrt{-\gamma}F^a_{\mu\nu}F^{a\mu\nu}$. The coefficients in front of them can be guessed by requiring the divergences to vanish in the complete action. Their expansions for $r \gg 1$ are
    \begin{equation}
     \begin{split}
      \sqrt{-\gamma}R[\gamma]\bigg|_{r \gg 1} =& -\frac{r^2 \omega^2}{2}\left[{\left(\xi_y\right)_0^b}^2+2 \left(\xi_y\right)_0^b \left(\xi_x\right)_0^b\right] +\frac{\omega ^4}{12} \left({\left(\xi_y\right)_0^b}^2+{\left(\xi_x\right)_0^b}^2-2 \left(\xi_y\right)_0^b\left(\xi_x\right)_0^b\right)\, ,\\
      \sqrt{-\gamma}R[\gamma]^2\bigg|_{r \gg 1} =& \omega ^4\left(4 {\left(\xi_y\right)_0^b}^2 + {\left(\xi_x\right)_0^b}^2+4 \left(\xi_y\right)_0^b\left(\xi_x\right)_0^b\right) \, ,\\
      \sqrt{-\gamma}R^{\mu\nu}[\gamma]R_{\mu\nu}[\gamma]\bigg|_{r \gg 1} =& \frac{\omega^4}{2} \left(3 {\left(\xi_y\right)_0^b}^2+{\left(\xi_x\right)_0^b}^2 + 2\left(\xi_y\right)_0^b\left(\xi_x\right)_0^b\right)\, ,\\
      \sqrt{-\gamma}F^a_{\mu\nu}F^{a\mu\nu}\bigg|_{r \gg 1} =& -2\omega^2 {\left(a_x^3\right)_0^b}^2 -2 \left(\mu^2+\omega^2\right) \left({\left(a_x^1\right)_0^b}^2+{\left(a_x^2\right)_0^b}^2\right)+8 \ii \mu  \omega  \left(a_x^1\right)_0^b \left(a_x^2\right)_0^b \, .
     \end{split}
    \end{equation}
    It can be checked that by adding the real space action
    \begin{equation}
     S_\text{ct} = -\frac{1}{\kappa_5^2}\int \dd^4x\ \sqrt{-\gamma}\ \left(3+\frac{1}{4}R[\gamma]+\left[\frac{1}{48}R[\gamma]^2-\frac{1}{16}R^{\mu\nu}[\gamma]R_{\mu\nu}[\gamma]+\frac{\alpha^2}{8}F^a_{\mu\nu}F^{a\mu\nu} \right]\log\epsilon_b\right)\bigg|_{r=r_{\text{bdy}}}
    \end{equation}
    to the action $S_\text{on-shell}$~\eqref{eq:action} we get a divergence-free theory (up to second order in the fluctuations) for $r_{\text{bdy}}\gg 1$, i.e. also the real time Green's functions are divergence-free. The renormalized $r_{\text{bdy}}\gg 1$ on-shell action of the helicity 0 modes are
\begin{equation}
\begin{split}
S^{\text{on-shell}}_{\text{hel.0, bl.1}} =& \frac{r_h^4}{\kappa^2_5} \int \frac{\dd^4 k}{{(2\pi)}^4} \left\lbrace \frac{\alpha^2 \mu \phi^b_1}{\omega^2-\mu^2} {\left(a_t^1\right)_0^b}^2 + \frac{\alpha^2 \mu \phi^b_1}{\omega^2-\mu^2}{\left(a_t^2\right)_0^b}^2 - \frac{\alpha^2 \omega^2}{4}{\left(a_x^3\right)_0^b}^2 \right.\\
& -\frac{3}{2} m^b_0 {\left(\xi_{tx}\right)_0^b}^2 + \frac{2 \ii \alpha^2 \omega  \phi^b_1}{\omega^2-\mu^2}\left(a_t^1\right)_0^b \left(a_t^2\right)_0^b + \frac{\alpha^2 \mu  w^b_1}{\omega^2-\mu^2}\left(a_t^1\right)_0^b \left(a_x^3\right)_0^b \\
& \left. - \frac{\ii \alpha ^2 \omega  w^b_1}{\omega^2-\mu^2}\left(a_t^2\right)_0^b \left(a_x^3\right)_0^b +\alpha ^2 \left(a_x^3\right)_0^b \left(a_x^3\right)_1^b +2 \alpha ^2 \phi^b_1 \left(a_x^3\right)_0^b \left(\xi_{tx}\right)_0^b  \right\rbrace \,
\end{split}
\end{equation}
and
  \begin{equation}
   \label{eq:hel0actionrenbl2}
   \begin{split}
S^{\text{on-shell}}_{\text{hel.0, bl.2}} =& \frac{r_h^4}{\kappa^2_5} \int \frac{\dd^4 k}{{(2\pi)}^4} \left\lbrace -\frac{\omega^4}{64} {\left(\xi_y\right)_0^b}^2 - \left( \frac{\omega^4}{64}-2f^b_2+\frac{5 m^b_0}{8} \right){\left(\xi_x\right)_0^b}^2 +\frac{3 m^b_0}{8}{\left(\xi_t\right)_0^b}^2 \right. \\
&+\left(\frac{\omega^4}{32}-2f^b_2-\frac{m^b_0}{2} \right)\left(\xi_y\right)_0^b \left(\xi_x\right)_0^b +\left(2f^b_2+ \frac{m^b_0}{2}\right)\left(\xi_y\right)_0^b \left(\xi_t\right)_0^b \\
& - \left(2f^b_2 -\frac{m^b_0}{4}\right)\left(\xi_x\right)_0^b \left(\xi_t\right)_0^b + \left(\xi_y\right)_2^b \left[ \left(\xi_y\right)_0^b-\left(\xi_x\right)_0^b \right]\\
& -\frac{\alpha^2 \left(\mu^2+\omega^2\right)}{4} \left[{\left(a_x^1\right)_0^b}^2+{\left(a_x^2\right)_0^b}^2\right] +\ii\alpha^2\mu\omega\left(a_x^1\right)_0^b \left(a_x^2\right)_0^b \\
& -\frac{\ii \alpha^2 w^b_1}{\omega} \left(a_x^2\right)_0^b \left(a_t^3\right)_0^b +\alpha^2 \left[ \left(a_x^1\right)_0^b  \left(a_x^1\right)_1^b + \left(a_x^2\right)_0^b \left(a_x^2\right)_1^b \right]\\
&\left. +\alpha^2 w^b_1 \left(a_x^1\right)_0^b \left[ \left(\xi_y\right)_0^b -\left(\xi_x\right)_0^b +\left(\xi_t\right)_0^b \right] + \frac{\ii \alpha^2 \mu  w^b_1}{2\omega}\left(a_x^2\right)_0^b \left[\left(\xi_x\right)_0^b -\left(\xi_t\right)_0^b\right] \right\rbrace .
   \end{split}
  \end{equation}
Since we have 6 fields determined by second order differential equations and 3 constraints, we end up with $12-3=9$ undetermined coefficients of the boundary expansion, in terms of which the expression above is written. They are
\begin{equation}
\left\{ \left(a_t^3\right)_0^b, \left(a_x^2\right)_0^b, \left(a_x^2\right)_1^b, \left(a_x^1\right)_0^b, \left(a_x^1\right)_1^b, \left(\xi_t\right)_0^b, \left(\xi_y\right)_0^b, \left(\xi_y\right)_2^b, \left(\xi_x\right)_0^b \right\}\, .
\end{equation}

\section{Constructing the Gauge Invariant Fields}
\label{sec:Constructing-the-Gauge}

The gauge group of the $SU(2)$ Einstein-Yang-Mills theory can be a subject of formal studies, as outlined in~\cite{Pons:1999xu}. It is shown that diffeomorphism-induced transformations of the metric functions and pure Yang-Mills transformations of the Yang-Mills fields ought not to be considered separately. On general grounds, we must look for the most general combination, which can be written as
    \begin{equation}
\label{eq:gaugefull}
     \begin{split}
      \delta \mathcal{N} &= \partial_t \Sigma^t + \Sigma^i \partial_i \mathcal{N} - \mathcal{N}^i \partial_i \Sigma^t,\\
      \delta N^i &= \partial_t \Sigma^i - \mathcal{N} \partial^i \Sigma^t + \Sigma^t \partial^i \mathcal{N} + \Sigma^j \partial_j \mathcal{N}^i - \mathcal{N}^j \partial_j \Sigma^i, \\
      \delta g_{ij} &= \frac{\Sigma^t}{\mathcal{N}}\partial_t g_{ij} + \left( \Sigma^k - \frac{\Sigma^t \mathcal{N}^k}{\mathcal{N}} \right)\partial_k g_{ij} + 2 g_{k(i} \partial_{j)} \Sigma^k - 2 \frac{\Sigma^t g_{k(i} \partial_{j)} \mathcal{N}^k }{\mathcal{N}} , \\
      \delta A^a_t &= A^a_i \partial_t \Sigma^i + \Sigma^i \partial_i A^a_t + F^a_{ti}\frac{\Sigma^t \mathcal{N}^i}{\mathcal{N}} + \partial_t \Lambda^a + \epsilon^{abc} \Lambda^b A^c_t , \\
      \delta A^a_i &= F^a_{ti} \frac{\Sigma^t}{\mathcal{N}} + F^a_{ij} \frac{\Sigma^t \mathcal{N}^j}{\mathcal{N}} + A^a_j \partial_i \Sigma^j + \Sigma^j \partial_j A^a_i + \partial_i \Lambda^a + \epsilon^{abc} \Lambda^b A^c_i \,;
     \end{split}
    \end{equation}
where the $i, j, \ldots$ indices denote the spatial coordinates $\{x, y, z, r\}$ of our spacetime. The metric $g_{ij}$ is the spatial metric and $g^{ij}$ is its inverse, and the functions $\mathcal{N}$ and $\mathcal{N}^i$ (called lapse and shift vector respectively) are defined as
    \begin{equation}
\dd s^2 = g_{MN} \dd x^M \dd x^N = -\mathcal{N}^2 \dd t^2 + g_{ij}(\dd x^i + \mathcal{N}^i \dd t)(\dd x^j + \mathcal{N}^j \dd t) .
    \end{equation}
A general infinitesimal gauge transformation acting on a perturbed solution is given in terms of the 8 descriptors $\{ \Sigma^M, \Lambda^a \}$. We define
    \begin{equation}
     \begin{split}
      \hat{g}_{MN} &= g_{MN} + h_{MN},\\
      \hat{A}^a_M &= A^a_M + a^a_M.
     \end{split}
    \end{equation}
where $g_{MN}$ and $A^a_M$ are the background fields of the hairy black hole solution that is considered in Sec.~\ref{sec:Holographic-Setup-and}. This part of the fields is therefore fixed, and the fluctuations $h_{MN}$ and $a^a_M$ are our dynamical variables. Thus, for instance the variation of the fluctuation field defined as $\xi_t = g^{tt}h_{tt}$ is given by $\delta \xi_t = g^{tt}\delta h_{tt} = g^{tt}\delta \hat{g}_{tt}$.

Furthermore, since they are considered as perturbations, they will be of the same order as the parameters $\Sigma^M$ and $\Lambda^a$. This allows us to give simple expressions to their variations, which will be approximated to lowest order.
    
\subsection{Residual Gauge Transformations}
\label{sec:Diffeomorphism-Invariance}
In Sec. \ref{sec:Characterization-of-Fluctuations}, we decided to choose a gauge where $a_r^a\equiv 0$ and $h_{Mr}\equiv 0$. This kind of gauge fixing is allowed as long as, for any given configuration, there exists a gauge transformation such that it makes these components vanish. Since there are 8 functions that categorize each possible transformation, in principle this is feasible.
Here we will see that this is justified, however the gauge is not completely fixed by these choices.

We begin by defining the background metric, which corresponds to the ansatz use in section \ref{sec:Holographic-Setup-and}, so it is of the form
    \begin{equation}
     \dd s^2 = g_{MN}\dd x^M \dd  y^N = -{c_1(r)}^2 \dd t^2 + {c_2(r)}^2 \dd x^2 + {c_3(r)}^2 (\dd y^2+\dd z^2) + {c_4(r)}^2 \dd r^2,
    \end{equation}
the only non-zero components of the background Yang-Mills field are $A^3_t=\phi(r)$ and $A^1_x=w(r)$,
and we will be working in momentum space, i.e.
    \begin{equation}
     \begin{split}
     \Sigma^M(t,x,r) &= \int \dd^4 x\ \ee^{ik_\mu x^\mu} \Sigma^M(\omega,k,r), \\
     \Lambda^a(t,x,r) &= \int \dd^4 x\ \ee^{ik_\mu x^\mu} \Lambda^a(\omega,k,r),
     \end{split}
    \end{equation}
where $k^\mu = (\omega,k,0,0)$, since in the case we are studying the rotational symmetry $SO(2)$ is preserved so that the fluctuations can be classified.

With these assumptions, we look at the variations of the $h_{Mr}$ components of the metric and the $a_r^a$ components of the Yang-Mills field, under an infinitesimal gauge transformation~\eqref{eq:gaugefull} acting on a perturbed background solution. To first order, these are
    \begin{subequations}
     \begin{align}
      \delta h_{tr} &= -i \omega {c_4}^2 \Sigma^r + {c_1}' \Sigma^t - c_1 \partial_r \Sigma^t,\\
      \delta h_{xr} &= i k {c_4}^2 \Sigma^r + {c_2}^2 \partial_r \Sigma^x,\\
      \delta h_{yr} &= {c_3}^2 \partial_r \Sigma^y,\\
      \delta h_{zr} &= {c_3}^2 \partial_r \Sigma^z,\\
      \delta h_{rr} &= 2 c_4 \left( {c_4}' \Sigma^r + c_4 \partial_r \Sigma^r \right),\\
      \delta a_r^1 &= w\partial_r \Sigma^x + \partial_r \Lambda^1,\\
      \delta a_r^2 &= \partial_r \Lambda^2,\\
      \delta a_r^3 &= -\frac{\Sigma^t}{c_1}\partial_r \phi + \partial_r \Lambda^3.
     \end{align}
    \end{subequations}
It is easy to convince oneself that by choosing carefully the $\Sigma_M$ and $\Lambda^a$ functions, one could make the $h_{Mr}$ and $a_r^a$ vanish. Now the residual gauge freedom would correspond to any further transformation that, while keeping these components null, changes the rest of the dynamical variables. We will find the most general form of a residual gauge transformation.
The solutions to $\delta h_{Mr}=0, \delta a_r^a=0$ can be written in terms of 8 constants $\{K_M, \Lambda^a_0 \}$ as
    \begin{equation}
\label{eq:transfdeltar}
     \begin{split}
      \Sigma^t(\omega,k,r) &= -K_t c_1 - i\omega K_r c_1 A, \quad \text{ with } A = \int \dd r \frac{c_4}{{c_1}^2};\\
      \Sigma^x(\omega,k,r) &= K_x - i k K_r B, \quad \text{ with } B = \int \dd r \frac{c_4}{{c_2}^2};\\
      \Sigma^y(\omega,k,r) &= K_y,\\
      \Sigma^z(\omega,k,r) &= K_z,\\
      \Sigma^r(\omega,k,r) &= \frac{K_r}{c_4},\\
      \Lambda^1(\omega,k,r) &= ik K_r C_w + \Lambda^1_0, \quad \text{ with } C_w = \int \dd r \frac{c_4 w}{{c_2}^2};\\
      \Lambda^2(\omega,k,r) &= \Lambda^2_0,\\
      \Lambda^3(\omega,k,r) &= -K_t \phi-i\omega K_r \left( \phi A-C_\phi \right) + \Lambda^3_0, \quad \text{ with } C_\phi = \int \dd r \frac{c_4 \phi}{{c_1}^2}.
     \end{split}
    \end{equation}
The physics ought to be invariant under any gauge transformation. Therefore, those dynamical fields affected by these residual gauge transformations must be unphysical. Those linear combinations with the property of being invariant constitute the physical fields.

\subsection{The Physical Fields}
\label{sec:Physical-Fields}
The helicity two fluctuations, $\Xi=g^{yy}h_{yz}$ and $h_{yy}-h_{zz}$ are already invariant, that is,
    \begin{equation}
     \begin{split}
      \delta\Xi = g^{yy}\delta h_{yz} &= 0,\\
      \delta(h_{yy}-h_{zz}) &= 0;
     \end{split}
    \end{equation}
therefore they are already physical modes. The helicity one fluctuations transform as
    \begin{equation}
     \begin{split}
      \delta h_{xy} &= \ii k {c_3}^2 K_y,\\
      \delta h_{ty} &= -\ii\omega{c_3}^2 K_y,\\
      \delta a^a_y &= 0;
     \end{split}
    \end{equation}
so that the $a^a_y$ are physical, and the invariant combination of the other two gives the physical mode $\Psi=g^{yy}(\omega h_{xy}+k h_{ty})$.
Note that the same applies to the $z$ components, which behave exactly the same as the $y$ components.

Now, for the helicity zero fields\footnote{Where we had defined $\xi_y = g^{yy}h_{yy}$, $\xi_x = g^{xx}h_{xx}$, $\xi_t = g^{tt}h_{tt}$ and $\xi_{tx} = g^{xx}h_{tx}$ in (\ref{eq:defxi}).} $\xi_{tx},\, \xi_{t},\, \xi_{x},\, \xi_{y}, a^a_x$ and $a^a_t$, we arrange any possible physical mode $\Phi$ as a linear combination of them given by some $r$-dependent coefficients $\tau_n$, so that its invariance translates into
    \begin{equation}
     \delta \Phi = \sum_{a=1}^3 (\tau_a\delta a^a_x + \tau_{3+a}\delta a^a_t) + \tau_7 \delta \xi_{tx} + \tau_8\delta \xi_{t}+\tau_9\delta \xi_{x}+\tau_{10}\delta \xi_{y} = 0.
\label{eq:physfieldinv}
    \end{equation}
Each of the variations in this expression are given by
    \begin{equation}
\label{eq:gengaugetr}
     \begin{split}
      \delta \xi_{tx} &= -\ii\omega K_x+ \ii k\frac{{c_1}^2}{{c_2}^2}K_t-\omega k \left( B+\frac{{c_1}^2}{{c_2}^2}A \right)K_r,\\
      \delta  \xi_t &= 2\ii\omega K_t + \left( \frac{2{c_1}'}{c_1 c_4} -2\omega^2 A \right)K_r,\\
      \delta  \xi_x &= 2\ii k K_x + \left( \frac{2{c_2}'}{c_2 c_4} + 2k^2 B \right)K_r,\\
      \delta  \xi_y &= \frac{2{c_3}'}{c_3 c_4}K_r,\\
      \delta a^1_x &= \ii k\Lambda^1_0 +\ii k w K_x+\left(\frac{w'}{c_4}+k^2 \left(w B-C_w \right)\right)K_r,\\
      \delta a^1_t &= -\ii\omega\Lambda^1_0-\phi \Lambda^2_0 - \ii\omega  w K_x - \omega  k \left(w B - C_w \right)K_r,\\
      \delta a^2_x &= \ii k\Lambda^2_0-w\Lambda^3_0 - \ii\omega  w C_\phi K_r,\\
      \delta a^2_t &= -\ii\omega \Lambda^2_0+\phi\Lambda^1_0+\ii k \phi C_w K_r,\\
      \delta a^3_x &= \ii k \Lambda^3_0+w\Lambda^2_0 - \ii k\phi K_t + \omega  k \left(\phi A - C_\phi \right) K_r,\\
      \delta a^3_t &= -\ii\omega \Lambda^3_0 + \ii\omega \phi K_t+\left(\frac{\phi'}{c_4}-\omega^2 \left(\phi A-C_\phi \right)\right)K_r.
     \end{split}
    \end{equation}
Plugging everything into equation (\ref{eq:physfieldinv}) results in 6 algebraic equations, due to the fact that the variation of the physical mode must vanish for any residual transformation, that is, for any $K_t$, $K_x$, $K_r$, $\Lambda^1_0$, $\Lambda^2_0$, $\Lambda^3_0$. Thus, we can solve for 6 of the $\tau_n$ coefficients in terms of the other four. The solution gives the most general gauge invariant combination and it turns out to be independent of the $\{ A,B,C_w,C_\phi \}$ functions.

What we call the four physical fields, $\Phi_i$ ($i$: $1,\ldots,4$), are chosen as a set of independent fields that generate that invariant combination. There is more than one choice, but the one we have taken is
\begin{equation}
\label{eq:physicalfields}
     \begin{split}
      \Phi_1 =&a^1_x-\frac{\ii k}{\phi}a^2_t+\frac{k^2}{w \phi}a^3_t+\frac{k \omega}{w \phi}a^3_x+\frac{k w}{\omega}\xi_{tx}-\\
&-\frac{k^2 f^4 N w \sigma^2}{2 r^2 \omega ^2}\xi_t+\frac{k^2 f^5 w^2 \sigma \phi \left(\sigma N'+2 N \sigma' \right)-2 r^2 \omega ^2 f\left(w \phi w'+k^2 \phi ' \right)}{4 r \omega ^2 w \phi \left(f+r f'\right)}\xi_y,\\
      \Phi_2 =&a^2_x+\frac{\ii \left(-k^2+w^2\right)}{\omega  w}a^3_t-\\
&-\frac{\ii k}{w}a^3_x-\frac{\ii w \phi}{2 \omega }\xi_t+\frac{\ii r f \left(w^2 \phi \left (\sigma N' + 2 N\sigma' \right) + 
 2 N \left (k^2 - w^2 \right)\sigma \phi'\right)}{4 \omega  N w \sigma \left(f+r f'\right)}\xi_y,\\
      \Phi_3 =&\xi_x+\frac{2 k}{\omega }\xi_{tx}-\frac{k^2 f^4 N\sigma^2}{r^2 \omega ^2}\xi_t+\frac{4 r^2 \omega ^2 f'-2 r \omega ^2 f+k^2 f^5 \sigma \left(\sigma N'+2 N \sigma '\right)}{2 r \omega ^2 \left(f+r f'\right)}\xi_y,\\
      \Phi_4 =&a^3_x+\frac{k}{\omega }a^3_t-\frac{w \phi}{\omega ^2-\phi^2}a^1_t-\frac{\ii \omega w}{\omega ^2-\phi^2}a^2_t+\frac{w^2 \phi}{\omega ^2-\phi^2}\xi_{tx}-\\
&-\frac{k f^4 N w^2 \sigma^2 \phi}{2 r^2 \omega \left(\omega^2- \phi^2\right)}\xi_t+\frac{k f \left(f^4 w^2 \sigma \phi \left(\sigma N'+2 N \sigma '\right)+2 r^2 \left(-\omega ^2+\phi^2\right) \phi '\right)}{4 r \omega  \left(\omega ^2-\phi^2\right) \left(f+r f'\right)}\xi_y.
     \end{split}
    \end{equation}

\section{Numerical Evaluation of Green's Functions}
\label{sec:Numerical-Evaluation-of}
Here we review and generalize the algorithm to evaluate Green's functions in cases when there is operator mixing \cite{Kaminski:2009dh}. The starting point of the algorithm would be a general bilinear bulk action for some fields $\Phi_\tmI(x^\mu,r)$ given by
   \begin{equation}
\label{eq:KLMSTansatz}
S=\int d^dx~dr\left[\partial_\mu\Phi_\tmI \;\mathcal{A}_{\tmI\tmJ}(x,r)\;\partial^\mu\Phi_\tmJ + \Phi_\tmI\; \mathcal{B}^\mu_{\tmI\tmJ}(x,r)\;\partial_\mu \Phi_\tmJ+\Phi_\tmI \;\mathcal{C}_{\tmI\tmJ}(x,r)\;\Phi_\tmJ\right]
   \end{equation}
In principle, one could be considering a perturbed background solution, as in the problem discussed in this paper, and it may be possible that there is some gauge freedom associated with those perturbation fields. In our case, that would be given by the transformation~\eqref{eq:gengaugetr}. But of course, gauge symmetry implies that the only relevant fields are the gauge-invariant combinations of the perturbations~\eqref{eq:physicalmodeshelicityzero}. Therefore, the most sensible strategy would be to write the action in terms of these physical degrees of freedom $\Phi_\tmI$, and proceed from there.

\subsection{Writing action in the correct basis}
\label{sec:Action-with-physical}

Even though the action is constituted as a gauge-invariant itself, it may not be possible to express it in terms of the physical fields only. It depends on the number of fluctuation fields and the extent of the gauge freedom. Let's say, for instance, that after whatever gauge fixing, the perturbed background is described by $N$ fields $\varphi_i(x)$ and we are left with a residual gauge freedom parametrized by $M$ constants. Then, the set of the possible gauge invariant linear combinations of $\varphi_i(x)$ is generated by $N-M$ independent physical fields $\Phi_\tmI(x)$. But the part of the action that is quadratic in perturbations may be of the form
   \begin{equation}
\label{eq:KLMSTansatz2}
S=\int d^dx~dr\left[\partial_\mu\varphi_i \; a_{ij}(x,r)\;\partial^\mu\varphi_j + \varphi_i \;b^\mu_{ij}(x,r)\;\partial_\mu \varphi_j+\varphi_i \;c_{ij}(x,r)\;\varphi_j\right] \, ,
   \end{equation}
with, assuming for simplicity dependence on $r$ only, $(2N+1)N$ coefficients $\{a_{ij},b_{ij},c_{ij}\}$ (Note that $a$ and $c$ form symmetric matrices). The only requirement upon this action is that it be invariant under any gauge transformation. This gives $2NM$ equations (one for every field or derivative of field, and for every transformation), from which some coefficients are determined, leaving $(1+2 N-2M)N$ undetermined coefficients that one is free to choose. On the other hand, an action written using only physical modes $\Phi_\tmI$ is constructed using $2(N-M)^2+(N-M)=(1+2N-2M)(N-M)$ coefficients. Thus, the freedom in writing a gauge-invariant action is always greater than what the $\Phi_\tmI$ allow for.

We conclude that in general the action will not be expressible as in~\eqref{eq:KLMSTansatz}. Not with the $\Phi_i$ being physical, gauge-invariant fields. The generalization of the algorithm consists in getting as close as possible to an expression of that kind, as we explain below. Our starting point, for now, will consist in taking the complete quadratic action~\eqref{eq:KLMSTansatz2} and forgetting about the gauge symmetry issues. Varying this action, one can obtain the equations of motion for the perturbation fields, integrate the Lagrangian by parts, insert the equations of motion to obtain the action evaluated on-shell and add the proper counterterms to cancel out any divergences. Finally, this expression can be transformed carefully into an integration in Fourier space (see \cite{Kaminski:2009dh} for a description of the procedure).

The physical fields obey a set of coupled equations of motion of their own, and each particular solution gives a vector of functions $\{\Phi_\tmI(x,r)\}$, such that as we approach the boundary, it asymptotes to some boundary values. It is in general possible to normalize the physical modes and to parametrize the perturbations in such a way that each boundary value of a physical mode coincides with the boundary value of one of the fluctuations,
   \begin{equation}
\Phi_\tmI(k,r) \xrightarrow[r\to\infty]{} \varphi_\tmI(k).
   \end{equation}
So we can make an association one to one between the physical modes and $n$ of the $N$ fluctuation fields. At any other distance $r$, each physical mode will of course depend on the values of all the other fluctuations that are involved in its definition.

Then, the first of the instructions would be to normalize the physical modes and to choose the appropriate fluctuation modes in order to be able to make this association on the boundary. The second step is to invert the definitions of the physical modes and solve for the fluctuations $\varphi_\tmI(k)$ that enter in the association. The idea is to replace these fields\footnote{From this point on, indices $I,J,\ldots$ denote the $n=N-M$ physical modes, while indices $i,j,\ldots$ denote the $N-n$ fluctuation modes that have not been replaced.} by inserting that solution into the on-shell action. In doing so, one obtains a contribution that involves only the physical modes, another one with couplings between the physical modes and the remaining fluctuations, and finally some terms given in terms of these remaining fluctuations only. That is, $S_{\text{o.s.}}=S_{\text{o.s.1}}+S_{\text{o.s.2}}$ where\footnote{We do not state it explicitly, but each term in the next actions includes the product of a field evaluated in $k$ and another evaluated in $-k$. This is natural for a quadratic Lagrangian written in Fourier space.}
    \begin{gather*}
S_{\text{o.s.1}}  =\int d^dk\left[\Phi_\tmI \;\mathfrak{A}(k,r)_{\tmI\tmJ}\;\partial_r \Phi_\tmJ+\Phi_\tmI\; \mathfrak{B}(k,r)_{\tmI\tmJ}\;\Phi_\tmJ  \right]_{r=r_b} \,,\\
S_{\text{o.s.2}} =\int d^dk \left[ \Phi_\tmI \mathfrak{a}(k,r)_{\tmI j}\partial_r \varphi_j+ \varphi_i \mathfrak{b}(k,r)_{i \tmJ}\partial_r \Phi_\tmJ+\Phi_\tmI \mathfrak{c}(k,r)_{\tmI j}\varphi_j  +\varphi_i \mathfrak{d}_{ij}(k,r)\partial_r \varphi_j+\varphi_i \mathfrak{e}_{ij}(k,r)\varphi_j \right]_{r=r_b} \,,
    \end{gather*}
The associated $\varphi_\tmI(x)$ fields no longer enter in the action. Now let's assume that we cannot find an analytic solution to the $n$ coupled equations of motion, which is expected except for some simple cases. Nevertheless, since this action is evaluated on the boundary $r_b$, a possible analytic approach would be to solve for the equations of motion on the limit $r\gg r_h$ and obtain the asymptotic expansions of the fields. As shown in the expansions of Sec.~\ref{sec:Asymptotic-Behavior}, the expanded solutions are usually not determined by the boundary values $\varphi^b_0$ only. There are also some undetermined coefficients $\varphi^b_p$ which can only be fixed by supplying initial conditions at a given point from which integration starts. Since these coefficients depend on the whole integration up to the boundary, they will not be solved for analytically. It is for this reason that some Green's functions can only be evaluated numerically.

A convenient position to start the integration is the horizon of the bulk geometry, because the initial conditions can be made easily at that point by demanding incoming solutions. This condition is related to the fact that we will ultimately be calculating retarded Green's functions. For convenience, let us refrain here what has been stated elsewhere in the text: The condition at the horizon halves the number of degrees of freedom and from a basis of $2N$ solutions ($N$ fluctuations under second order differential equations), we end up with just $N$ solutions. Furthermore, there are $M$ constraints coming from the equations of motion of the gauge-fixed fields which reduce these solutions to the $n=N-M$ degrees of freedom that manifest themselves through the physical solutions and can be found by requiring invariance under residual gauge transformations. However, the analytic expanded solution obtained around the boundary knows nothing about the incoming condition at the horizon. We have only the constraints, so accordingly we are dealing with a basis of $2N-M=N+n$ solutions - that is, $N+n$ undetermined coefficients. $N$ of them can be taken to be the boundary values $\varphi^b_0$. Therefore, the number of undetermined coefficients $\varphi^b_p$ is expected to be precisely $n$, the same as the number of physical modes. So a numerical integration of the equations of motion of the physical fields starting at the boundary is sufficient to fix them, since we are implicitly setting $n$ initial conditions.

The expanded on-shell action can be then divided into two terms, $S_{\text{o.s.}}=S_{\text{o.s.(I)}}+S_{\text{o.s.(II)}}$ where
   \begin{equation}
\label{eq:generalseparation}
    \begin{split}
S_{\text{o.s.(I)}} &= \int d^dx \left[\alpha_{\tmI \tmJ} \left(\Phi_\tmI\right)^b_0 \left(\Phi_\tmJ\right)^b_0 + \beta_{\tmI \tmJ} \left(\Phi_\tmI\right)^b_0 \left(\varphi_\tmJ\right)^b_p + \zeta_{\tmI \tmJ} \left(\varphi_\tmI\right)^b_p \left(\varphi_\tmJ\right)^b_p \right] \, ,\\
S_{\text{o.s.(II)}} &= \int d^dx \left[\kappa_{\tmI j} \left(\Phi_\tmI\right)^b_0 \left(\varphi_j\right)^b_0 + \lambda_{ij} \left(\varphi_i\right)^b_0 \left(\varphi_j\right)^b_0 \right]\, , 
\end{split}
   \end{equation}
provided that the expansions are arranged in such a way that the $n$ $\left(\varphi_\tmI\right)^b_p$ do not cross with the $N-n$ $\left(\varphi_i\right)^b_0$, which in general can be done because the $\varphi^b_p$ are fixed by integrating the equations of motion of the physical fields, which are obtained by varying the action with respect to the physical fields, and therefore the part of the action which contains the remaining fluctuations $\varphi_i$ is irrelevant to them.

In this expression, remember that $\left(\Phi_\tmI\right)^b_0=\left(\varphi_\tmI\right)^b_0$. Now, to obtain the Green's functions, the AdS/CFT prescription instructs us to take the functional derivative of the action with respect to the boundary values of the fields~\cite{Son:2002sd}.
The Green's functions of the fluctuations $\varphi_i$ can be easily extracted from~\eqref{eq:generalseparation} and read
\begin{equation}
\label{eq:restgr}
G^R_{ij}(k) = -\lambda_{ij} , \;\;\;\;\;\; G^R_{\tmI j}(k) = -\kappa_{\tmI j}
\end{equation} 
On the other hand, the Green's functions $G^R_{\tmI \tmJ}(k)$, associated to the boundary values of the physical modes cannot be extracted directly, nor can they be expressed in an analytic way. These are the most interesting Green's functions because they have physical meaning, as discussed in Sec.~\ref{sec:Transport-Properties}. We have computed them numerically following the method presented in the next section.

\subsection{Prescription for numerical solutions}
\label{sec:numerical-prescription}

It is tempting to establish the identifications $S_{\text{o.s.(I)}}=S_{\text{o.s.1}}$ and $S_{\text{o.s.(II)}}=S_{\text{o.s.2}}$. However, it is important to note that this is incorrect, because the equations of motion of the fluctuations are all coupled, and consequently the expansions of the $\varphi_i(k,r)$ may depend on boundary values of the replaced $\varphi_\tmI(k,r)$. Thus, expanding $S_{\text{o.s.2}}$ it is possible to produce terms like the ones found in $S_{\text{o.s.(I)}}$.

In order to deal with this, an effective action $S_{\text{eff}}$ can be constructed, using physical fields only, such that its expansion near the boundary reproduces exactly those undesirable terms. This effective action is to be subtracted from $S_{\text{o.s.2}}$ to cancel them out, and at the same time added to $S_{\text{o.s.1}}$ producing
\begin{equation}
\label{eq:effectivefornum}
S_{\text{eff}}+S_{\text{o.s.1}} = \int d^dk\left[\Phi_\tmI (-k,r) A(k,r)_{\tmI\tmJ}\partial_r \Phi_\tmJ(k,r)+\Phi_\tmI(-k,r) B(k,r)_{\tmI\tmJ}\Phi_\tmJ(k,r)  \right]_{r=r_b} \, .
\end{equation}
The matrices $A, B$ are obtained analytically, but only their expression at the boundary is necessary. We refer to the action as ``effective" because its contribution matches exactly $S_{\text{o.s.(I)}}$, but it is just an artifact - we do not derive any equation of motion from it. The fields that will be inserted are solutions of the equation of motions derived from the original action.

These solutions are obtained by numerical integration, starting from some selected values of the horizon $\left(\Phi_\tmI\right)^h_0$. In fact, this set of values determines completely the coefficients of the expansion at the boundary. We may choose $n$ linearly independent sets $\left(\Phi_\tmI\right)^{h (\tmJ)}_0 = e^{(\tmJ)}_\tmI$, in order to obtain $n$ linearly independent sets of boundary values. In particular, a possible choice is 
   \begin{equation}
    e^{(1)}_\tmI = (1,0,0,\ldots)\,,\quad     e^{(2)}_\tmI = (0,1,0,\ldots)\,,\quad  \ldots\,,\quad   e^{(n)}_\tmI = (\ldots,0,0,1)\,.
   \end{equation}
Alternate choices are possible. This is just the one we used because we got good numerical results (with less noise). A numerical integration can be performed for each set in order to obtain $n$ independent solutions  $\{\Phi^{(\tmJ)}_\tmI(k,r)\}_\tmJ$ extended in the bulk, which can be arranged in a matrix $H(k,r)$, with entries
   \begin{equation}
    H_{\tmI \tmJ} (k,r) = \Phi^{(\tmJ)}_\tmI(k,r) .
   \end{equation}
Thus, the $J^{\text{th}}$ solution appears as the $J^{\text{th}}$ column.
On the other hand, we know that when each physical field approaches the boundary, it asymptotes to the value of its associated perturbation, $\left(\varphi_\tmI\right)^b_0(k)$. At any other distance or scale $r$, since the system of differential equations is coupled, they will in general evaluate to a linear combination of all the $\{\left(\varphi_\tmJ\right)^b_0\}_\tmJ$, so that the set of functions can be written as
   \begin{equation}
\label{eq:solutionmatrix}
\Phi_\tmI(k,r) = F_{\tmI\tmJ}(k,r)\left(\varphi_\tmJ\right)^b_0(k) \, .
   \end{equation}
In this way, all the dynamics of the fields is encoded in the solution matrix $F_{\tmI\tmJ}(k,r)$, which has the nice property of becoming the identity at the boundary, $F_{\tmI\tmJ}(k,r_b)=\delta_{\tmI\tmJ}$.

Any complete set of $n$ independent solutions to the equations of motion is enough to build the matrix $F$, because any solution (any one that satisfies the incoming condition at the horizon) can be written as a linear combination of them. \emph{In particular}, the matrix $F(k,r)$ must be linearly related to $H(k,r)$ because each $I^{\text{th}}$ column of $F$ is composed by a set of solutions that asymptotes to $\left(\varphi_\tmJ\right)^b_0=0$ for all $J$, except for $\left(\varphi_\tmI\right)^b_0=1$. Since at the boundary, by definition, $F$ is the identity, the linear relation must be given by
   \begin{equation}
F(k,r)=H(k,r) \cdot H^{-1}(k,r_b)\,.
   \end{equation}
This result enables us to calculate the solution matrix, which encodes the dynamics, from $n$ numerically integrated solutions. Then, by inserting~\eqref{eq:solutionmatrix} into our on-shell action~\eqref{eq:effectivefornum}, we obtain
   \begin{equation}
S_{\text{eff}+\text{s.o.1}}=\int d^dk \left. \Phi_\tmI(-k,r)  \mathcal{F}_{\tmI \tmJ} (k,r) \Phi_\tmJ(k,r) \right|_{r=r_b}\,,
   \end{equation}
where $\mathcal{F}= F^\dagger~A~\partial_r F + F^\dagger~B~F$. But, since this is evaluated at the boundary, where the matrix $F$ becomes the identity and the physical fields coincide with their associated fluctuations, we might as well write
   \begin{equation}
S_{\text{eff}+\text{s.o.1}}=\int d^dk  \left(\varphi_\tmI\right)^b_0 (-k) \left[ A_{\tmI \tmK}(k,r_b)~\partial_r F_{\tmK \tmJ}(k,r_b) + B_{\tmI \tmJ}(k,r_b) \right] \left(\varphi_\tmJ\right)^b_0 (k)\,.
   \end{equation}
The Green's functions can now be directly extracted from this expression using the AdS/CFT correspondence prescription, to give
   \begin{equation}
G^R_{\tmI \tmJ}(k) = -A(k,r_b) H'(k,r_b) H^{-1}(k,r_b) -B(k,r_b) \; .
   \end{equation}
Notice that this formula reproduces the well known result for the Green's function of a decoupled equation. Here, instead of the derivative of the field, there is the matrix of derivatives. And instead of taking the ratio with the boundary value of the field, a factor given by the inverse of the matrix of solutions is included.

This completes~\eqref{eq:restgr}, giving the way of calculating all the Green's functions of the problem. However, as opposed to the ones given in~\eqref{eq:restgr}, these Green's functions are not determined by the background only, the solution to the equations of motion of the perturbed degrees of freedom enters through $H$. Their physical meaning is clearer and more important, since they correspond holographically to the correlators of the dual operators.

\section{General Remarks on Viscosity in Anisotropic Fluids}
\label{sec:General-Remarks-on}
The concept of viscosity is linked to the internal motion of a system that causes dissipation of energy \cite{Landau:fluid}. In general, we may define a general dissipation function $\Xi$, such that the dissipative forces that describe the internal motion are obtained from it as velocity derivatives. Typically, frictional forces are linear in velocities $u^\mu$, which suggests that the general form of this function be quadratic in velocities.

But, for an internal motion which describes a general translation or a general rotation, the dissipation is zero. Since it describes dissipative processes only, $\Xi$ ought to vanish for these configurations of velocities. Because of this argument, the function must depend on the velocities through the combination of gradients of velocities $u_{\mu\nu}=\frac{1}{2}\left(\nabla_\mu u_\nu+\nabla_\nu u_\mu\right)$, rather than on the velocities themselves directly. Thus, the general form is given by the sum $\Xi=\frac{1}{2}\eta^{\mu\nu\lambda\rho}u_{\mu\nu}u_{\lambda\rho}$, where the coefficients $\eta^{\mu\nu\lambda\rho}$ define the viscosity tensor \cite{Landau:1959te}, whose symmetries are given by
\begin{equation}
\label{eq:symeta}
\eta^{\mu\nu\lambda\rho}=\eta^{\nu\mu\lambda\rho}=\eta^{\mu\nu\rho\lambda}=\eta^{\lambda\rho\mu\nu}\,.
\end{equation}
The part of the stress tensor which is dissipative 
due to viscosity is defined by
\begin{equation}
\label{eq:stress}
\Pi^{\mu\nu}=-\frac{\del \Xi}{\del u_{\mu\nu}}=-\eta^{\mu\nu\lambda\rho}u_{\lambda\rho}\,.
\end{equation}
In the case of a fluid in the rest frame $u^t=1$, and in order to satisfy the Landau frame condition $u_\mu \Pi^{\mu\nu}=0$, the stress energy tensor (and the viscosity tensor, correspondingly) must have non-zero components only in the spatial directions $i,j,\ldots=\{x,y,z\}$. In general, only 21 independent components of $\eta_{ijkl}$ appear in the expressions above.

For the particular case of an isotropic fluid, the tensor can be written using only 2 independent components, which are usually parametrized by the shear viscosity $\eta$ and the bulk viscosity $\zeta$, so that the dissipative part of the stress tensor can be expressed as $\Pi^{ij}=-2\eta(u^{ij}-\frac{1}{3}\delta^{ij}u_{l}^l)-\zeta u_{l}^l\delta^{ij}$, which is a well-known result.

For a transversely isotropic fluid, there are 5 independent components in the tensor $\eta^{ijkl}$. Without loss of generality, we choose the symmetry axis to be the $x$-axis. The non-zero components are parametrized by
\begin{equation}
\label{eq:etatensorpwave}
\begin{aligned}
&\eta^{xxxx}=\zeta_x+\frac{4}{3}\lambda\,,\qquad
&&\eta^{yyyy}=\eta^{zzzz}=\zeta_y+\frac{\lambda}{3}+\eta_{yz}\,,\\
&\eta^{xxyy}=\eta^{xxzz}=-\frac{2}{3}\lambda\,,
&&\eta^{yyzz}=\zeta_y+\frac{\lambda}{3}-\eta_{yz}\,,\\
&\eta^{yzyz}=\eta_{yz}\,,
&&\eta^{xyxy}=\eta^{xzxz}=\eta_{xy}\,.
\end{aligned}
\end{equation}
So that the non-zero off-diagonal components of the stress tensor are
\begin{equation}
\label{eq:off-diagonalstressanisotroptic}
\begin{split}
&\Pi^{xy}=-2\eta_{xy} u_{xy}\,,\qquad \Pi^{xz}=-2\eta_{xy} u_{xz}\,,\\
&\Pi^{yz}=-2\eta_{yz} u_{yz}\,.
\end{split}
\end{equation}

In this consideration we are including only the contribution to the stress tensor due to the dissipation via viscosity, and we find the terms in the constitutive equation which depend on the velocity of the normal fluid $u_\mu$. But in general, there would also be terms depending on the derivatives of the Nambu-Goldstone boson fields $v_\mu=\del_\mu \varphi$ on the superfluid velocity and on the velocity of the director, which may contribute to the dissipative part of the stress tensor (the director is the vector pointing in the preferred direction).

However, these terms do not contribute to the off-diagonal components of the energy-momentum tensor because (1) a shear viscosity due to the superfluid velocity leads to a non-positive divergence of the entropy current \cite{Landau:fluid,Pujol:2002na}, and (2) no rank two tensor can be formed out of degrees of freedom of the director if the gradients of the director vanish \cite{LESLIE01011966}. In our case, the second argument is fulfilled since the condensate is homogeneous and the fluctuations depend on time only. Even though these degrees of freedom will generate additional transport coefficients, they do not change the shear viscosities, so we can write Kubo formulae which give the shear viscosities in terms of the stress energy correlation functions.

Let us consider a conformal fluid, so that $\zeta_x=\zeta_y=0$ (this can easily be shown using the tracelessness condition of the stress-energy tensor,\ie $\Pi^a_{~a}=0$, with $a=x,y,z$). The usual way to perturb a system in thermal equilibrium is to look at small perturbations of the background fields and add these sources to the action. Here we are interested in the metric fluctuations about the flat Minkowski metric,\ie the terms of interest here that we add to the action are
\begin{equation}
\label{eq:actflucanisotropic}
\Pi^{xx}h_{xx}+\Pi^{yy}h_{yy}+\Pi^{zz}h_{zz} = +i\omega\frac{2}{3}\lambda \left(h_{xx}-\frac{1}{2}\left(h_{yy}+h_{zz}\right)\right)^2 +i\omega\frac{\eta_{yz}}{2}\left(h_{yy}-h_{zz}\right)^2\,.
\end{equation}
To derive the left side we use equations \eqref{eq:stress} and \eqref{eq:etatensorpwave}, as well as $u_{aa}=-\frac{i\omega}{2}h_{aa}$\footnote{Here also used the fact, that we are in the rest frame of the fluid and therefore $u^\mu=(1,0,0,0)$}, with $a=x,y,z$. Applying the same calculation to the isotropic case we obtain
\begin{equation}
\label{eq:actflucisotropic}
\begin{split}
\Pi^{xx}h_{xx}+\Pi^{yy}h_{yy}+\Pi^{zz}h_{zz} &=+\frac{i\omega}{3}\eta\left(\left(h_{xx}-h_{yy}\right)^2+\left(h_{xx}-h_{zz}\right)^2+\left(h_{yy}-h_{zz}\right)^2\right)\\
					     &=+i\omega\frac{2}{3}\eta \left(h_{xx}-\frac{1}{2}\left(h_{yy}+h_{zz}\right)\right)^2 + i\omega\frac{\eta}{2}\left(h_{yy}-h_{zz}\right)^2\,.
\end{split}
\end{equation}
We only have one shear viscosity $\eta$ in the isotropic case. The purpose of the rewriting of the latter case is to show the connection to the transversely isotropic case. This rewriting shows that at the phase transition $\lambda$ turns into the isotropic shear viscosity $\eta$, explaining the behavior we see in figure \ref{fig:lambdaT}. Note that this is a computation taking place on the field theory side. Therefore the metric we need to lower and raise indices is the flat Minkowski metric.

By plugging in the components of $\Pi^{aa}$ it is easy to show that the left hand side of equation \eqref{eq:actflucanisotropic} is equivalent to
\begin{equation}
\label{eq:rewritactfluc}
\begin{split}
 \Pi^{xx}h_{xx}+\Pi^{yy}h_{yy}+\Pi^{zz}h_{zz} &= \frac{1}{2}\left(\Pi^{xx}-\left(\Pi^{yy}+\Pi^{zz}\right)\right)\left(h_{xx}-\frac{1}{2}\left(h_{yy}+h_{zz}\right)\right)\\ &+\frac{1}{2}\left(\Pi^{xx}+\left(\Pi^{yy}+\Pi^{zz}\right)\right)\left(h_{xx}+\frac{1}{2}\left(h_{yy}+h_{zz}\right)\right)\\
 &+ \frac{1}{2}\left(\Pi^{yy}-\Pi^{zz}\right)\left(h_{yy}-h_{zz}\right)\,.
\end{split}
\end{equation}
Applying linear response theory we obtain the Green's function for the first term in the equation above, which can be related to $G^{m,m}$ on the gravity side
\begin{equation}
\begin{split}
&G^{m,m}(\omega) = \lim_{|\vec k|\to 0}\int \dd t\,\dd^3 x e^{-\ii k_\mu x^\mu} \theta(t)\times\\
  &\left\langle\left[\frac{1}{2}\left(T^{xx}(t,\vec x)-\left(T^{yy}(t,\vec x)+T^{zz}(t,\vec x)\right)\right),\frac{1}{2}\left(T^{xx}(0,0)-\left(T^{yy}(0,0)+T^{zz}(0,0)\right)\right)\right]\right\rangle\,.
\end{split}
\end{equation}
 The Kubo formula that gives the $\lambda$ viscosity is
\begin{equation}
\lambda=\lim_{\omega\to 0}\frac{3}{2\omega}\, \im\, G^{m,m}(\omega)\,.
\end{equation}
Note that the Green's function of the dissipative part of the second term of \eqref{eq:rewritactfluc} is zero therefore we only get background fields for this components of the Green's function \eqref{eq:greenmathel0b2}. Finally the last term of \eqref{eq:rewritactfluc} corresponds to the helicity two mode and is related to the shear viscosity $\eta_{yz}$.

\end{appendix}

\bibliographystyle{JHEP}
\bibliography{library}

\end{document}